\documentclass[%
aps,
prx,
reprint,
superscriptaddress,
nofootinbib,
amsmath,
amssymb
]{revtex4-2}
\usepackage{
    physics,
    graphicx,
    multirow,
    tabularx,
    mathtools, %for \begin{rcases} \end{rcases}
    placeins, %To include \FloatBarrier
    nicefrac, %For inline fraction
    hyperref, %To use put in url, enable links in bibiliography
    threeparttable, % For table footnotes
    siunitx,
    enumitem,
    makecell
}

\usepackage{mathtools}
\DeclarePairedDelimiter\ceil{\lceil}{\rceil}

\usepackage[capitalize]{cleveref}
\Crefname{section}{Sec.}{Secs.}

\usepackage[dvipsnames]{xcolor}
\hypersetup{
    colorlinks,
    linkcolor={blue!50!black},
    citecolor={blue!50!black},
    urlcolor={blue!80!black}
}% remove the box around citation link and bibilio link

\usepackage[caption=false]{subfig} %The caption part is not compatible with revtex
\begin{document}

%\preprint{APS/123-QED}

\title{{\Large Snakes on a Plane:\\} Mobile, low dimensional logical qubits on a 2D surface}

%\title{Snakes on a Plane}

\newcommand{\qmaddress}{\affiliation{Quantum Motion, 9 Sterling Way, London N7 9HJ, United Kingdom}}
\newcommand{\oxddress}{\affiliation{Department of Materials, University of Oxford, Parks Road, Oxford OX1 3PH, United Kingdom}}
\newcommand{\uclddress}{\affiliation{Department of Physics and Astronomy, University College London, Gower St, London WC1E 6BT, United Kingdom}}

\newcommand{\scb}[1]{\textcolor{purple}{#1}}

\author{Adam Siegel}
\email{adam@quantummotion.tech}
\qmaddress
\oxddress

\author{Zhenyu Cai}
\qmaddress
\oxddress

\author{Hamza Jnane}
\qmaddress
\oxddress

\author{Balint Koczor}
\qmaddress
\affiliation{Mathematical Institute, University of Oxford, Woodstock Road, Oxford OX2 6GG, United Kingdom}

\author{Shaun Pexton}
\qmaddress

\author{Armands Strikis}
\qmaddress
\uclddress

\author{Simon Benjamin}
\qmaddress
\oxddress

\date{\today}

\begin{abstract}
Recent demonstrations indicate that silicon-spin QPUs will be able to shuttle physical qubits rapidly and with high fidelity -- a desirable feature for maximising logical connectivity, supporting new codes, and routing around damage. However it may seem that shuttling at the logical level is unwise: static defects in the device may `scratch' a logical qubit as it passes, causing correlated errors to which the code is highly vulnerable. Here we explore an architecture where logical qubits are 1D strings (`snakes') which can be moved freely over a planar latticework. Possible scratch events are inferred via monitor qubits and the complimentary gap; if deemed a risk, remarkably the shuttle process can be undone in a way that negates any corruption. Interaction between logical snakes is facilitated by a semi-transversal method. We obtain encouraging estimates for the tolerable levels of shuttling-related imperfections. 
\end{abstract}

\maketitle

\section{Introduction}
Quantum computing promises to outperform classical machines provided sufficiently low logical rates are achieved so that deep quantum algorithms can be run. Such low error rates can be obtained by assembling a significant number of qubits together, in so-called error correcting codes \cite{Fowler_2012, resource_estimation_shor_algo}. The choice of the code is often determined by the constraints that the platform imposes, for example, the surface code is well suited for devices that have 2D qubit layouts with nearest-neighbour connectivity~\cite{Bravyi_Kitaev_1998, Kitaev_1997}.
Yet, even apparently simple 2D layouts may prove challenging to  engineer as systems scale: a dense grid of qubits can quickly become unmanageable, due to the supporting infrastructure that is necessary for each qubit's control and measurement. Recent studies have advocated for alternative and more experimental-friendly architectures that are enabled by qubit shuttling~ \cite{li_crossbar_2018, cai_looped_pipelines_2023, kunne_spinbus_2024, siegel_two_by_n_2024, pataki_surface_code_crossbar_2024}. Indeed when a platform supports the shuttling of physical qubits, there are potential advantages beyond the increased flexility of design. Examples can include support for higher connectivity LDPC codes, logical qubit connectivity, and routing around damage.  

A recent work proposed and explored an extremely simplified architecture: a 2$\times$N array of qubits whose first row can be collectively shuttled with respect to the second row\,\cite{siegel_two_by_n_2024}. While several qubit platforms have been identified as suitable for implementing shuttling with high fidelity \cite{Bluvstein_2023, Pino_2021}, silicon spins may be the most promising one for the implementation of a 2$\times$N array \cite{langrockBlueprintScalableSpin2023}. Although the linear architecture imposed various constraints, it was shown that powerful code classes could still efficiently be embedded in such an paradigm. Full universal quantum computation was shown to be possible using a surface code encoding, making this approach a feasible candidate for early fault-tolerance.

This architecture however suffered two main drawbacks. Firstly, the highly-restricted dimensionality induced connectivity lockups at the logical level, meaning that in general two-qubit logical gates could not be implemented in parallel, thus drastically increasing the runtime. Secondly, the proposed architecture was not defect-resistant. If a qubit site were to become inoperable sometime during the computation, preventing qubits from shuttling through it, the device would simply be split into two disconnected halves.

\begin{figure*}
    \centering
    \includegraphics[width=0.9\linewidth]{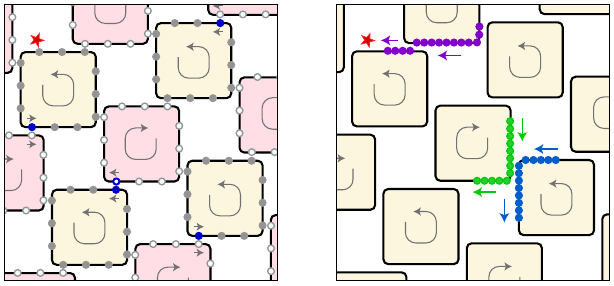}   
    \caption{Two paradigms for fault tolerant QC in a 2D latticework, exploiting shuttling of physical along linear arrays (black lines) which in places run parallel and permit interactions. The left panel illustrates the older `looped pipeline' approach (c.f. Fig. 10 of Ref.\,\cite{cai_looped_pipelines_2023}). Open circles are ancilla qubits, filled circles are data qubits, and the four blue filled circles are data qubits belonging to a given logical qubit. Each logical qubit is therefore spread over many loops, as summarised in the main text.  A very different paradigm is the `snakes on a plane' approach on the right. Here a logical qubit is a 1D chain of data qubits, local on the scale of the latticework structure. Stabilisers are evaluated by small to-and-fro shuttling motions with respect to fixed ancilla and measurement elements (not shown). In the figure, the purple snake is moving from one loop to another (e.g. via SWAP operations in the parallel-array region) while the blue and green snakes are performing a semi-transversal logical gate. A key difference in the two approaches is the impact of a phase-shifting defect in the device (red asterisk). In the LP-paradigm this damages only individual data or ancilla qubits (albeit afflicting multiple logical qubits) whereas in the snakes model there is a risk of `scratching' an entire logical qubit as it moves past, thus inducing a logical error. The purple snake will imminently encounter this risk.  }
    \label{fig:LPversusSOAP}
\end{figure*}

In this paper, we present an approach for quantum computing on a 2D {\it latticework} formed of low-dimensional array strands that occasionally meet at junctions or run parallel. The model is tailored to silicon spin qubits, although other platforms which support shuttling may also be compatible with the concepts. There are a wide range of possible latticework structures; the strands can be very long, creating large void areas to accommodate support systems including measurement devices. Logical qubits are snake-like objects: trains of physical qubits that move through the latticework. We analyse the advantages and challenges of this approach, and estimate the required shuttling properties. 

Latticework structures of this general kind have been considered in Ref.\,\cite{cai_looped_pipelines_2023} but the paradigm for fault tolerant computation was quite different (see Fig.\,\ref{fig:LPversusSOAP}, left panel). In that prior approach, a logical qubit is spread thinly, so to speak, over the latticework; each local region (a square shaped loop in the figure) hosts a single data qubit or ancilla qubit of a given logical qubit. The highlighted blue circles correspond to four data qubits and one ancilla; we observe that as the physical qubits move according to the arrows, the ancilla will encounter each of the data qubits in turn, thus permitting a stabiliser measurement via measurement of the ancilla, using devices in the void spaces (not shown). As Ref.\,\cite{cai_looped_pipelines_2023} explains in detail, by hosting multiple physical qubits in each loop the architecture can effectively `stack' logical qubits. An advantage of the approach is that damage to the structure corresponds to damage to only one point in a logical qubit (e.g. a missing or corrupt data qubit or ancilla), albeit this damage would be replicated for all the logical qubits `stacked' in that region. 

In the present `snakes on a plane' model, instead the logical qubits are strings of physical qubits local to a given region of the latticework. Shuttling, combined with junctions for re-routing the snakes, allows for the global movement of the logical qubits at the device scale and the exclusion of potentially defective links (see Fig.\,\ref{fig:LPversusSOAP}, left panel). To stabiliser the logical snakes, we adopt ideas from the  2$\times$N paradigm\,\cite{siegel_two_by_n_2024}. Although the 2D latticework is only an incrementally more challenging technological scenario, the resulting approach profoundly superior in terms of connectivity and damage tolerance. However, in comparison to prior models there is now a significant new problem to address: by shuttling logical qubits through a solid state environment there is the risk that static defects in the device will `scratch' the logical qubit, i.e. create correlated strings of errors against which the code is vulnerable. We introduce a means to mitigate this risk. 

In Section \ref{sec:general_idea}, we present in more depth the general ideas and layouts underlying the \textit{snakes on a plane} proposal. The following Section \ref{sec:noise_protection} is the largest part of the paper: here we review the noise processes that dominate in silicon spin qubits, and describe a means to mitigate their impact on shuttled logical qubits. We give particular attention to charge noise, which originates from the random fluctuation of charges between defects spread across the device and effectively induces uncontrolled variation in the magnetic field landscape. These can be responsible for catastrophic dephasing errors affecting entire logical qubits shuttled near the defect, but we describe a protocol that is highly effective in mitigating the impact of such events. The approach is enabled by two defect detection schemes: one is based on the continual evaluation of the effective magnetic field landscape via the use of {\it monitor qubits}; the other analyses the stabiliser measurements signature, more precisely by computing the syndrome's complementary gap. The capability to infer likely defect-related events enables an robust shuttling protocol which we call \textit{snake surgery}, allowing one effectively to reverse time and retrieve the logical qubit's state before the defect occurred. The joint use of these methods enables the shuttling of logical snakes across long distances without accumulating errors. 

In Section \ref{sec:univ_quant_comput}, we show that universal quantum computation is possible and efficient on our proposed architecture. In particular, the long-distance-shuttling ability enables an all-to-all connectivity at the logical level, where two-qubit gates are efficiently implementable transversally, or \textit{semi-transversally}. The latter protocol, which we describe in Section \ref{sec:semi_transversal_gates}, promises to reduce logical error rates if accepting higher time costs than for usual transversal gates, in case the noise level is prohibitively high.

\section{General idea} \label{sec:general_idea}

In this section, we present the general layout of the snakes on a plane architecture and highlight in general terms where its advantage arises.

\subsection{Error correction with a 2$\times$N array of qubits}

\begin{figure}
    \centering
    \includegraphics[width=.8\linewidth]{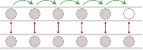}
    \caption{Representation of the 2$\times$N architecture. The device is characterised by two parallel rails of evenly-spaced qubits (gray circles with gray disks inside). Adjacent qubits from different rails are allowed to interact via two-qubit gate operations (red vertical lines with disks on their ends). Finally, the qubits of the first row are allowed to shuttle along their rail (green arrows). Some empty locations are kept at the end of the device to facilitate movement (empty gray circle).}
    \label{fig:2xN_device_layout}
\end{figure}

\begin{figure}
    \centering
    \includegraphics[width=\linewidth]{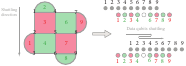}
    \caption{Two-way mapping for the rotated surface code. Left: canonical 2D layout. Right: schematic for the layout on a 2$\times$N architecture. Here, red and green colours indicate $X$ and $Z$ stabilisers respectively. On the 2$\times$N architecture, interactions are enabled by shuttling the data qubits' row back and forth and interacting the qubits with corresponding nearest-neighbour ancillas. As an example, the top right (resp. bottom right) panel depicts an interaction between an ancilla and its South-East (resp.North-West) data qubit.}
    \label{fig:surface_code_2xN}
\end{figure}

We here review our previously-published work \cite{siegel_two_by_n_2024}, which is incorporated into the current paradigm. In this paper, we analysed the suitability of 2$\times$N arrays of qubits equipped with shuttling for embedding error correcting codes. Such 2$\times$N arrays, simply consisting of two parallel rows (or `rails') of physical qubits, are the building block of our current architecture and are represented in Fig. \ref{fig:2xN_device_layout}. In this figure, the first row of qubits can be collectively shuttled while the second row is kept static. Two-qubit gates can be implemented between neighbouring qubits across the rails. By placing data qubits in the mobile rail and ancilla qubits in the static one, all-to-connectivity is virtually obtained between data and ancilla qubits thanks to shuttling: this theoretically permits the implementation of any code. Nonetheless, as shuttling is a finite-time and noisy process, one should aim at minimising the overall shuttling distance over a stabiliser cycle. In \cite{siegel_two_by_n_2024}, we demonstrated that an appropriate choice of qubit labelling along with the right gate and shuttling sequencing leads to advantageous performances for certain code classes. The rotated surface code is one of them and is the code we will focus on for the rest of this paper. A schematic representation of a 3$\times$3 surface code embedded on a 2$\times$N array is given in Fig. \ref{fig:surface_code_2xN}.

As noted in\,\cite{siegel_two_by_n_2024}, the use of a 1D embodiment for the canonically 2D surface code logical qubits has the effect that asymptotically the code cannot scale: for any  given finite shuttling speed there is some logical qubit so large that a stabiliser cycles cannot be completed before the accumulated noise is too great. However, the asymptotic performance may not be relevant to the practical utility: attainable speeds and fidelities for shuttling mean that this problem need not arise for logical qubit sizes of practical interest, even those with distance 30 or more. This is explored further in later sections.

An important characteristic of the rotated surface code embedded in the 2$\times$N architecture is that stabiliser cycles can be implemented in only four shuttles of total length $2d+2$ (where $d$ is the code distance). This is visible in Fig. \ref{fig:surface_code_2xN}: South-East and North-West data qubits around an ancilla are separated by $d+1$ increments (top right and bottom right panels). Shuttling back and forth along these $d+1$ increments allows for all interactions within the stabiliser cycle and for the data qubits to return to their initial position. We also showed in \cite{siegel_two_by_n_2024} that this implementation protects against hook errors.

\subsection{Snakes on a plane architecture}

In \cite{siegel_two_by_n_2024} we gave an end-to-end protocol for universal quantum computation using one long 2$\times$N array where all logical qubits were appended next to each other. Long-range interactions between logical qubits were then implemented via the use of a logical ancilla bus rather than by physically bringing the logical qubits closer together. While promising for the early fault-tolerant regime, this proposal suffered two drawbacks presented in the introduction: a limited connectivity between logical qubits and a high sensitivity to single-qubit failure.

\begin{figure*}
    \centering
    \subfloat{
        \centering
        \includegraphics[width=0.45\linewidth]{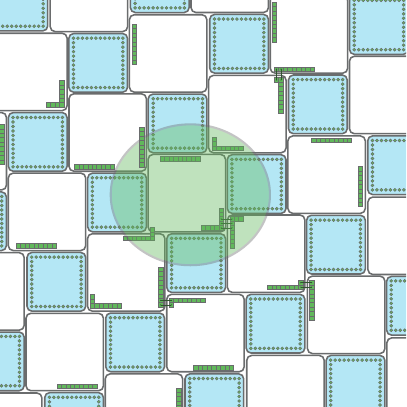}}
    % \hspace{0.1\linewidth}
    \hfill
    \subfloat{
        \centering
        \includegraphics[width=0.45\linewidth]{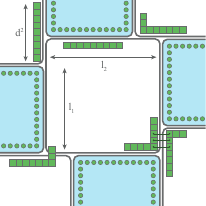}}
    \caption{An example layout for the snakes on a plane architecture. Each edge is a 2$\times$N filament as in Fig. \ref{fig:2xN_device_layout}. Data qubits (green squares) can be shuttled along the edges of a white square, while ancillas (green circles) are static and placed along the edges of a blue square. Logical qubits are built from multiple data qubits, creating a snake that can be shuttled across the device. The left panel is a zoomed out schematics of the device. The right panel is a close-up of the region inside the green faded oval. The two bottom right logical qubits are interacting through two-qubit transversal gates, while the three other logical qubits are normally being stabilised and shuttled.}
    \label{fig:general_layout}
\end{figure*}

In this paper, we consider a new paradigm, where logical qubits can be displaced across a latticework structure of 2$\times$N filaments (Fig.~\ref{fig:general_layout}) meeting at occasional three-way junctions. These filaments form loops of two kinds: small loops (of size $l_1$) holding ancilla qubits and larger loops (of size $l_2$) holding data qubits. The sizes are expressed in the number of quantum dots. Additionally, one can distinguish two kinds of edges on the lattice: data/ancilla and data/data edges. These will respectively be denoted as \textit{stabiliser edges} and \textit{interaction edges}. Indeed, in the former, data qubits are facing ancilla qubits, which can be utilised to measure stabilisers. In contrast, in the latter, while stabilisers cannot be measured, physical qubits from two distinct logical qubits can meet and interact across the ridge. In our architecture, data qubits are allowed to shuttle (while ancilla qubits stay fixed). Therefore, for the interaction edges, shuttling is permitted on both rails. The size of the interaction edges is $l_{int}=l_2-l_1$. 

\begin{figure}
    \centering
    \includegraphics[width=0.9\linewidth]{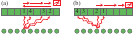}
    \caption{Stabiliser measurements while shuttling a snake. Green squares (resp. circles) represent data (resp. ancilla) qubits. Wiggly arrows connect a given ancilla with the qubits it must entangle with for a subsequent stabiliser measurement; this occurs when the qubits become aligned. Straight arrows show the snake movement necessary to implement a stabiliser cycle, which ends with the measurement of an ancilla qubit. (a) Stabilising a snake with no net shuttling. (b) Stabilising a snake with a global forward movement.}
    \label{fig:snake_movement}
\end{figure}

\begin{figure}
    \centering
    \includegraphics[width=0.5\linewidth]{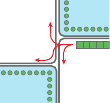}
    \caption{Representation of a junction. Red arrows show the three potential directions a snake can take.}
    \label{fig:junction}
\end{figure}

In this representation, each logical qubit is thus embodied as a snake that can travel long distances. Denoting the code distance as $d$, each snake has length $d^2$. When no logical operation needs to be implemented, a snake can be shuttled back and forth along a stabiliser edge in order to measure stabilisers in the manner of \cite{siegel_two_by_n_2024} (Fig. \ref{fig:snake_movement}(a)). When the snake must be moved to a specific location of the chip \textit{e.g.} to interact with another snake, it can simply be moved forward (Fig. \ref{fig:snake_movement}(b)). Note that doing so may however introduce hook errors. Therefore, it may be preferable to implement such forward motion separately from stabiliser cycles. For instance, if shuttling is low-noise and fast compared to ancilla qubit measurement times, the information can be stabilised via the sequence of Fig. \ref{fig:snake_movement}(a), and the data qubits moved forward while ancillas are being measured. When a snake eventually reaches a junction, the user can  decide on the path it will follow, either keeping it the current loop or transferring it to the neighbouring one (Fig. \ref{fig:junction}). Note that the implementation of junctions that snakes can jump across can be circumvented, in case it represents a significant experimental challenge. An alternative solution yielding the same net result is to bring a \textit{blank} snake initialised in, say, the $\ket{0}$ state, to the other side of a junction and swap it with the current snake (similarly to the bottom right scenario of Fig. \ref{fig:general_layout}, right panel).

\begin{figure}
    \centering
    \includegraphics[width=0.9\linewidth]{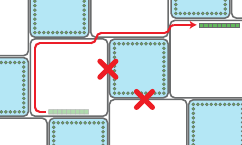}
    \caption{Rerouting of a snake in the presence of defects. The initial and target positions are indicated by a faded and brighter snake respectively. The red arrow represents a possible path between these positions and that avoids the defective shuttling links (red crosses)}
    \label{fig:rerouting}
\end{figure}

In general terms, one can easily understand where the strength of this architecture comes from. Assuming fast and high-fidelity shuttling, the snake-like logical qubits can reliably be moved across the device, enabling all-to-all connectivity at the logical level. Additionally, transversal gates rather than lattice surgery may be carried out through the interaction edges, thus accelerating the implementation of two-qubit gates. Finally, the device is resilient against defective edges, as long as they can be detected (see Section \ref{sec:noise_protection}): if a chosen path has been deemed as faulty, snakes can simply be rerouted and travel through a different path (Fig. \ref{fig:rerouting})

Note that other latticework structures are conceivable for our proposal. A hexagonal lattice and a rectangular lattice are for instance sketched in Appendix \ref{app:alternative_latticework}.

\subsection{Choice of a platform: silicon spins} \label{sec:silicon_spin_qubits}

One qubit platform is particularly well suited to the implementation of all the operations described above: silicon spins qubits.

From a general point of view, these qubits appear as an excellent candidate for the implementation of a quantum computer, due to their high coherence times \cite{Struck_2020}, scalability \cite{gonzalez-zalbaScalingSiliconbasedQuantum2021}, compatibility with advanced manufacturing techniques \cite{Maurand_2016, Zwerver_2022} and cryogenic classical electronics \cite{XueXiao2021Ccco, ruffino2021integrated}. Small silicon spin qubit processors have already been engineered \cite{Philips_2022} and small error correcting codes have been implemented \cite{Takeda_2022}. Single- and two-qubit gates with fidelity above $99\%$ have been demonstrated \cite{Xue_2022, Noiri_2022, Mills_2022} even in commercially-sourced chips\,\cite{steinacker2024300mmfoundrysilicon} and fidelities beyond $99.9\%$ have been reported \cite{yonedaQuantumdotSpinQubit2018}). Among the implementable two-qubit gates, we find the $CZ$ and $CROT$ gates, which are both natural operations for stabiliser measurements.

More specifically to our proposed architecture, 2$\times$N arrays of silicon spin qubits have already been demonstrated \cite{hutin2019gate}. Moreover, recent experiments succeeded in shuttling electrons across a silicon spin qubit chip with $99.99\%$ fidelity per shuttling increment and at a high speed \cite{desmet2024highfidelitysinglespinshuttlingsilicon}. This rapid development compared to previously demonstrated shuttling capabilities \cite{yonedaCoherentSpinQubit2021, seidlerConveyormodeSingleelectronShuttling2021} promises to bring even better fidelities in the near future. We are therefore confident that silicon spin qubits are a promising candidate for the implementation of the shuttling-based fault-tolerant protocol we advocate for in this piece of research.

In the rest of the paper, we will base all numerical parameters on state-of-the-art silicon spin qubit technologies. Namely, while two-qubit gates require around $0.1\mu$s \cite{MAUNEB.M2012Csoi, Xue_2022, Noiri_2022, Mills_2022}), initialisation, measurement and single-qubit gates can last up to a few microseconds \cite{Philips_2022, Zheng_2019, Takeda_2024}. The modes of shuttling described in \cref{fig:snake_movement} require $2d$ and $d$ shuttling increments respectively. Supposing $d\sim30$ for relevant applications \cite{Gidney_2021}, an interdot distance $l_{dd}=100$nm and a shuttling speed $v=10$m/s \cite{langrockBlueprintScalableSpin2023}, shuttling would take up 300 to 600ns in a stabiliser cycle. Adding this contribution to the initialisation, gates and measurement times, one can realistically expect a stabiliser cycle time of $3\mu$s.

\section{Protection against charge noise} \label{sec:noise_protection}

As the impact of nuclear spins can be reduced by isotopically purifying devices \cite{stano_review_2022}, the most detrimental source of noise is charge noise \cite{kepa_2023, elsayed_low_2024, yoneda_noise_correlation_2023, jnane_ab_initio_2024, shehata_2023, yonedaQuantumdotSpinQubit2018, culcer_2009, rojas_spatial_correlations_2023}. If not dealt with properly, it can lead to out-of-calibration drifts or catastrophic events that would hinder the performance of our error correcting protocols. In this section, we first give a description of the noise processes we envision, before providing four distinct protocols that can be used jointly so as to reach exponentially low logical errors rates. The first one relies on the choice of a qubit encoding that is resilient against phase errors. The second method is a fault-tolerant protocol allowing one to, in effect, \textit{reverse time} and go back to the state of the system before a catastrophic phase event occurred, if such an event is suspected. Detection is enabled by the last two schemes, which respectively exploit additional monitor qubits and the analysis of the syndrome signature to infer the presence of a defect.

\subsection{Noise description} \label{sec:noise_description}

While the physical origin of charge noise remains a matter of debate and is device-dependent, models based on two-level fluctuators have shown good agreement with experiments \cite{elsayed_low_2024}. These fluctuations are typically characterised by a $1/f$ noise spectrum which will lead to a slow drift of the $g$-factor landscape. Whether the electrons are shuttled or not, this drift needs to be characterised, or the electrons would progressively acquire some relative phase with respect to each other, leading to dephasing noise \cite{mehmandoost_decoherence_2024, kepa_2023, shehata_2023}. A problematic aspect of charge noise is the dynamic nature: a defect can arise during computation and persist for a short or long period, or be permanent. 

A highly damaging situation could occur if a defect appears near the shuttling tracks and its local impact cannot be entirely encapsulated in the above description. Rather, we will suppose that it will lead to some phase-like error being applied to all qubits flowing near the defect. To encompass the most problematic cases, we will suppose that the phase shift imparted by such effects is arbitrary, potentially way above the surface code threshold, and can occur abruptly. One can thus easily see the dramatic impact these defects might have when shuttling an entire logical qubit next to them, as they can effectively \textit{scratch} all the physical qubits of the code. Indeed, if the imparted phase shift were close to $\pi$, the scratch event would directly give rise to a logical phase flip. This process is sudden and could lead to orbital excitations, and thus to dephasing due to the difference in $g$-factor between the orbital states. Fortunately, given that the difference in $g$-factor between the states is small and the fast decay rate of the excited states, we expect that this effect will not be the most common source of dephasing. Nevertheless even rare events of this kind could create a floor in the achievable logical error rates; a scalable fault tolerant architecture must actively remediate the issue.

The general description we will adopt is thus the following: each location in the device will be associated with a phase channel, parametrised by the local $g$-factor. Charge noise, as described above, can lead to fluctuations of these channels of two types. The first model gives rise to slow and small variations of the magnetic field felt by the electrons, while the second more local description can lead to sudden and severe phase errors. Our aim is to mitigate the impact of both these contributions. In the remainder of this section we characterise the severity of the more ubiquitous slow and small phase variations.

By definition, the rate of appearance of the fluctuators yielding the phase variations is at least of the order of $T_2^*$, whose demonstrated values reach up to 20$\mu$s \cite{yonedaQuantumdotSpinQubit2018, stano_review_2022}. This is an order of magnitude higher than stabiliser cycles times (a few microseconds, see Section \ref{sec:silicon_spin_qubits}): we thus expect few variations at this time scale. Of course, variations that are slower than the whole computation time will be accommodated by calibration beforehand and application of unitary corrections during the computation. Alternatively, one could use a dynamical-decoupling-based scheme allowing a snake to pass through the same phase channel multiple times under a different basis to cancel out the noise. Therefore, the only timescale that is relevant to study is that of phase fluctuations that may arise in the middle of the computation or evolve before its end.

\subsection{Encoding the qubits in a decoherence-free subspace}

When a single qubit is represented by a single spin, one obtains the so-called Loss-DiVincenzo (LD) encoding. Although this is the original and the most natural encoding \cite{lossQuantumComputationQuantum1998}, quantum information can also be stored over multiple spins \cite{burkardSemiconductorSpinQubits2021}. This provides interesting features, such as additional protection through the use of a decoherence-free subspace when using singlet-triplet (ST) encoding \cite{levy_st_2002}, and full electrical control when using exchange-only (EO) qubits \cite{divincenzo_universal_2000}, which are two- and three-spin encodings respectively.

As explained before, fluctuations of the $g$-factor landscape due to charge noise represents an important source of errors for silicon-based quantum computers. For conventional quantum processor architectures, these fluctuations affect individual physical qubits. However, as long as these effects remain below a certain threshold, error correction will suppress their impact on the computation. Since different encodings behave differently in fluctuating environments, our goal is to determine which encoding is better suited for our architecture. While noting that a carefully chosen encoding can mitigate the effect of a sudden apparition of a charge near the shuttling track, we focus here on mitigating the effect of the slowly varying $g$-factor landscape. The techniques presented in the next subsections will be targeted towards reducing the impact of catastrophic \emph{scratching} events.
Here, we focus on the comparison between LD and ST qubits as performing gates on EO qubits remains challenging \cite{burkardSemiconductorSpinQubits2021}. 

Recently, the authors of Ref.~\cite{mokeev2024modelingdecoherencefidelityenhancement} explored this comparison under the assumption that the environmental changes induce a randomly fluctuating magnetic field in space and time. 
Their model does have limitations; it does not directly represent charge noise as it does not have a $1/f$ spectrum and it does not account for the valley degree of freedom. Nevertheless, since the main point of this section is to convey the idea that the choice of encoding can drastically reduce the effect of a fluctuating magnetic field, and as the valley splitting is believed to be large in Si/SiO$_2$ devices, we will also adopt this model. It allows us to account for slow variations of the $g$-factor landscape while being simple enough to analytically derive the impact of the environment on shuttled qubits.
We leave the inclusion of charge noise and the valley degree of freedom for future work. 

More precisely, the authors of Ref.~\cite{mokeev2024modelingdecoherencefidelityenhancement} represent the fluctuating field $B(x,t)$ as a zero-mean Ornstein–Uhlenbeck (OU) sheet. As it is a Gaussian random sheet with a defined mean, its behaviour is fully determined by its covariance matrix,
\begin{align}
    C((x_1,t_1), (x_2,t_2)) = \frac{\sigma^2}{4\kappa_x\kappa_t}e^{-\kappa_x|x_1-x_2|-\kappa_t|t_1-t_2|},
    \label{eq:noise_model}
\end{align}
where $\sigma$ is the process' standard deviation, and $\kappa_x = 1/\lambda, \kappa_t = 1/\tau$ are respectively the inverse correlation length and inverse correlation time. Note that we suppose that the field only depends on a single spatial dimension corresponding to the shuttling path. 
Here, we set $\lambda = 100$ nm, $\tau = 20 ~\mu$s and $\sigma^2/4\kappa_x\kappa_t = \sqrt{2}\kappa_t$ \cite{mokeev2024modelingdecoherencefidelityenhancement}.

While shuttling through such an environment, a system of $N$ non-interacting qubits will experience the following Hamiltonian \cite{mokeev2024modelingdecoherencefidelityenhancement},
\begin{align}
    H(t) = \frac{g_0\mu_B\hbar}{2}\sum_{i=1}^{N}B(x_i(t),t)\sigma_z,
    \label{eq:hamiltonian_random_field}
\end{align}
where $g_0$ is an averaged $g$-factor, $\mu_B$ is the Bohr magneton, $x_i(t)$ is the time-dependent position of the centre of the wavepacket describing the $i^{th}$ electron, and $\sigma_z$ is the Pauli $Z$ matrix. 

\begin{figure}
    \centering
    \includegraphics[width=\linewidth]{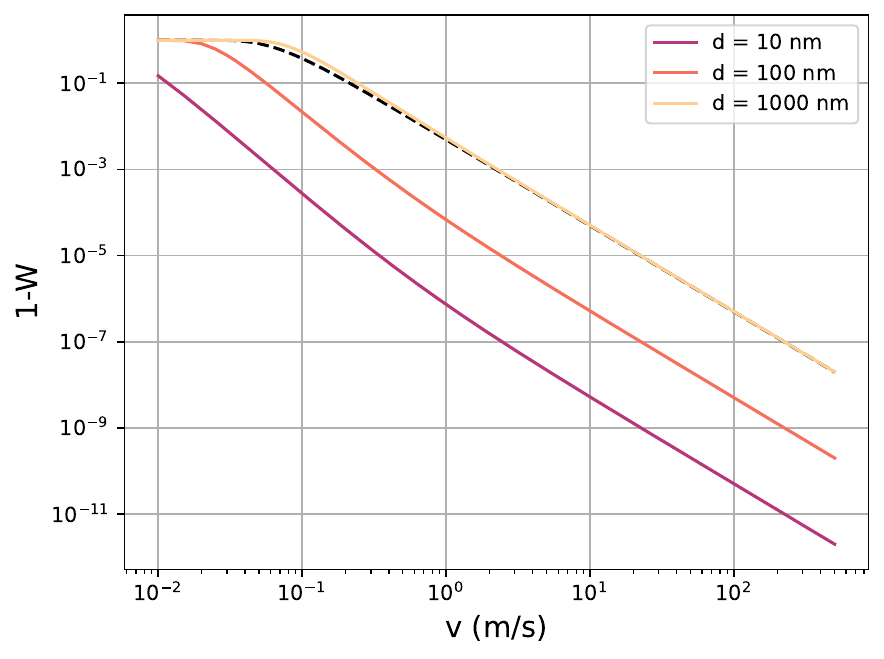}
    \caption{Dephasing error as a function of the shuttling speed for different qubit encodings. The qubit is shuttled for a distance $L = 10 \; \mu$m and a time $T = L/v$. As the distance $d$ between the electrons composing the ST qubit increases (solid lines), the probability for them to experience a different field at a given point increases, leading to a higher infidelity. Within this noise model, the ST always outperforms the LD for reasonable values of $d$. One would require a distance $d > 1 \; \mu $m for the LD (black dashed line) encoding to be better. These curves are obtained using the analytical expressions derived in \cite{mokeev2024modelingdecoherencefidelityenhancement}.}
    \label{fig:dephasing_error}
\end{figure}

To compare different encodings, we can focus on characterising the random phase acquired when the qubits are shuttled. For simplicity, consider a single qubit shuttled for a time $T$. The phase it will acquire under the Hamiltonian described in \cref{eq:hamiltonian_random_field} is given by, 
\begin{align}
    \phi(T) = \frac{g_0\mu_B\hbar}{2}\int_0^{T} B(x_1(t),t)dt.
\end{align}
The density matrix of the electron at time $T$ thus reads,
\begin{align}
    \rho(T) = \begin{pmatrix}
            \rho_{00}(0) & \rho_{01}(0)\exp(-i\phi(T)) \\
             \rho_{10}(0)\exp(i\phi(T)) & \rho_{11}(0)\\
    \end{pmatrix}   
\end{align}
with $\rho_{ij}(0)$ representing the coefficients of the initial density matrix. Note that only the off-diagonal elements are affected by the shuttling channel. 
This phase will vary between experiments and could lead to the complete dephasing of the state when averaging the results. The quantity of interest is thus the dephasing factor $W(T)$ obtained by averaging the phase $\exp(-i\phi(T))$ over the random sheet $B$. This factor appears in the averaged density matrix as follows, 
\begin{align}
    \mathbb{E}\left[\rho(T)\right] = \begin{pmatrix}
            \rho_{00}(0) & \rho_{01}(0)W(T) \\
             \rho_{10}(0)W^{*}(T) & \rho_{11}(0)\\
    \end{pmatrix} 
\end{align}
where $W(T) = \mathbb{E}\left[ \exp\left( -i\phi(T)\right)\right]$.
Ref. \cite{mokeev2024modelingdecoherencefidelityenhancement} generalises this derivation to more than one electron and gives analytic expressions for $W$ for both the LD and ST encodings. 
Using these expressions, we plot in \cref{fig:dephasing_error} the evolution of $1-W(T)$ -- which can be understood as a measure of the shuttling infidelity --  for both encodings. Here, we considered a shuttling length $L = 10 \; \mu$m and a time $T = L/v$ with $v$ the shuttling speed.

While \cite{mokeev2024modelingdecoherencefidelityenhancement} chose to explore the situation in which both electrons are shuttled one after the other with a fixed delay, here we consider that the electrons are separated by a fixed distance $d$ and move together in the spirit of conveyor-belt shuttling. They will thus experience the field with a delay $\tau_d = d/v$, which depends on the shuttling speed. 

As the speed increases in \cref{fig:dephasing_error}, a qubit spends less time shuttling and therefore experiences less magnetic field fluctuation. This explains why, irrespective of the encoding, the infidelity decreases monotonically with speed. The fact that the ST encoding (solid orange and purple lines) outperforms the LD encoding (black dashed lines) for practical values of $d$ can be explained as follows. When using a ST qubit, the state is protected against situations in which both electrons experience the same change in magnetic field. In \cref{fig:dephasing_error}, we benefit from this protection and the remaining infidelity comes from the fact that the two electrons are separated by a small distance, and thus feel slightly different $B(x,t)$. As $d$ is reduced, the difference in the magnetic field experienced by both electrons decreases and so does the infidelity. However, as $d$ gets larger, there comes a point where both electrons are too far apart to experience a similar field, leading to a larger infidelity. For distances $d = 1 \; \mu $m (solid yellow line), we even find that the LD encoding performs better. In \cref{app:shuttling_infidelity}, we report the evolution of $1-W$ with speed for different values of the correlation length $\lambda$ and correlation time $\tau$, which confirms our understanding. 

Within this noise model, the infidelity keeps decreasing with speed. However, it does not account for charge noise, spin-orbit coupling, or valley effects which are known to be detrimental to shuttling at high speed \cite{langrockBlueprintScalableSpin2023}. Therefore one needs to be careful before making definitive conclusions. 
Nevertheless, since spin-orbit-induced effects and valley state excitations are believed to be negligible at slow speeds, this study suggests that ST qubits are better suited in this regime. For larger speeds, a more realistic noise model would be required to conclude.

The remaining dephasing error when using the ST encoding stems from the fact that the two electrons cannot exactly experience the same field as one will always be in front of the other. While further suppression may be possible with additional steps, for the present paper we simply note that the achievable dephasing error is very low at the likely shuttling speeds of 10 to 50\,ms$^{-1}$. Consequently this aspect of shuttling noise, while ubiquitous in the device, will not be a dominant noise source in well-designed stabiliser cycles involving modest shuttling distances (as employed here). In essence, the decoder handles these errors along with sources such as gate infidelity. We therefore now turn to consider the more rare, but more problematic case of defects that can `scratch' logical qubits.

\subsection{Snake surgery} \label{sec:snake_surgery}

As described in Section \ref{sec:noise_description}, we assume here that the appearance of charge defects gives rise to unwanted phase effects. If a charge defect materialises near a shuttling track, a sudden and severe phase error can affect the data qubits that are shuttled through it, alarmingly increasing the logical error probability. While the ST qubit encoding presented in the previous section should reduce the impact of the resulting error surge, additional protection may be necessary to bring it to the extremely low levels required to run deep quantum algorithms. In this section, we suppose that catastrophic events are phase-like and can be detected (we give in Sections \ref{sec:monitor} and \ref{sec:complementary_gap} protocols that are able to do so). Importantly, while detection needs to be highly {\it sensitive}, \textit{i.e.} able to flag genuine error events reliably, it need not be highly {\it specific}: it is not problematic if the majority of flagged cases are in fact error-free. With these assumptions, we present a fault-tolerant scheme allowing one to retrieve the logical qubit state prior to any corruption that may have occurred, removing its impact regardless of how strong its effect was.

Let us assume that a localised `pin' charge did appear near a shuttling track. Two different cases need to be distinguished, depending on whether a given snake is merely being stabilised in place with modest back-and-forth motion or shuttled across a long distance (Fig. \ref{fig:snake_movement}(a) and (b)). The first situation can easily be dealt with by noticing that during the stabiliser cycles, a snake is only being shuttled back and forth by $d+1$ increments (as in Figs. \ref{fig:surface_code_2xN} and \ref{fig:snake_movement}(a)). Assuming that the shuttling direction is, say, along the columns of a surface code, a pin defect would only \textit{scratch} one single column of the code. By choosing to orient the $Z$ logical operator perpendicularly (along the rows), the code would naturally be protected against such adversarial phase errors. Note however that, while exponential error suppression is still guaranteed, the logical error rate and threshold may be moderately affected by the presence of defect-induced columns of $Z$ errors, albeit perpendicular to the $Z$ logical operator. Indeed, these strings offer more opportunities for actual $Z$ logical errors to form, both by slightly reducing the code distance and increasing the number of error paths.

Consequently, the scenario one should carefully scrutinise is that of Fig. \ref{fig:snake_movement}(b), where a snake is shuttled far away from its original position. This kind of movement could indeed lead to errors affecting all data qubits, potentially beyond the error threshold of the surface code. More specifically, we want to protect a snake from a defective shuttling link that the monitoring techniques (see Sections \ref{sec:monitor} and \ref{sec:complementary_gap}) detected only \textit{after} the snake shuttled through it. If a defect is detected before the snake reaches the intended shuttling path, it can be simply rerouted.

\begin{figure}
    \centering
    \includegraphics[width=.9\linewidth]{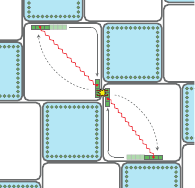}
    \caption{Snake surgery protocol. During long shuttles, snakes are prone to \textit{scratch} events, \textit{i.e.} catastrophic phase changes along the shuttling tracks due to the appearance of point-like particles. Fault-tolerance is guaranteed by only shuttling the head of a snake (solid arrow) while the tail is stabilised in place. Entanglement is maintained between them throughout the shuttling process (red wiggly line). If a catastrophic event was detected by the monitor qubits or stabiliser measurements (yellow star), the information can be teleported back to the tail (dashed arrow).}
    \label{fig:snake_surgery_overview}
\end{figure}

This kind of situation can be tackled by the protocol depicted in Fig. \ref{fig:snake_surgery_overview}, which we call \textit{snake surgery}. It relies on the transfer of the logical information contained in one snake onto two distinct but entangled snakes, called \textit{head} and \textit{tail}. These are obtained by successively growing and cutting the original snake via lattice surgery \cite{Horsman_2012}. While the tail remains static and retains full error correction capabilities, the head is shuttled across long distances. If no defect is \textit{a posteriori} detected, the tail is destroyed, projecting the delocalised logical state onto the head: this process thus correctly transfers the information to the remote location in the device. On the contrary, if a defect was suspected, the head is measured out, which will project the logical state back to the tail. Importantly, this will succeed regardless of whether a defect event in fact occurred. The protocol can then be restarted through another shuttling route or delayed until the defect has disappeared or become stable.

Further, this analysis not only holds when shuttling across long distances, but also when two distinct logical qubits interact. Specifically, one would follow the same protocol, \textit{i.e.} stabilising the tails of both logical qubits in place, while shuttling the heads so as to bring them to both sides of an interacting edge. They can then be interacted through transversal CNOTs (or via the semi-transversal protocol of Section \ref{sec:semi_transversal_gates}). As such interaction requires shuttling across the interaction edge, defect detection protocols must assess if this edge became defective or not throughout the process. Depending on the outcome of such monitoring, the heads or tails are measured out as per the above protocol.

\begin{figure}
    \centering
    \includegraphics[width=\linewidth]{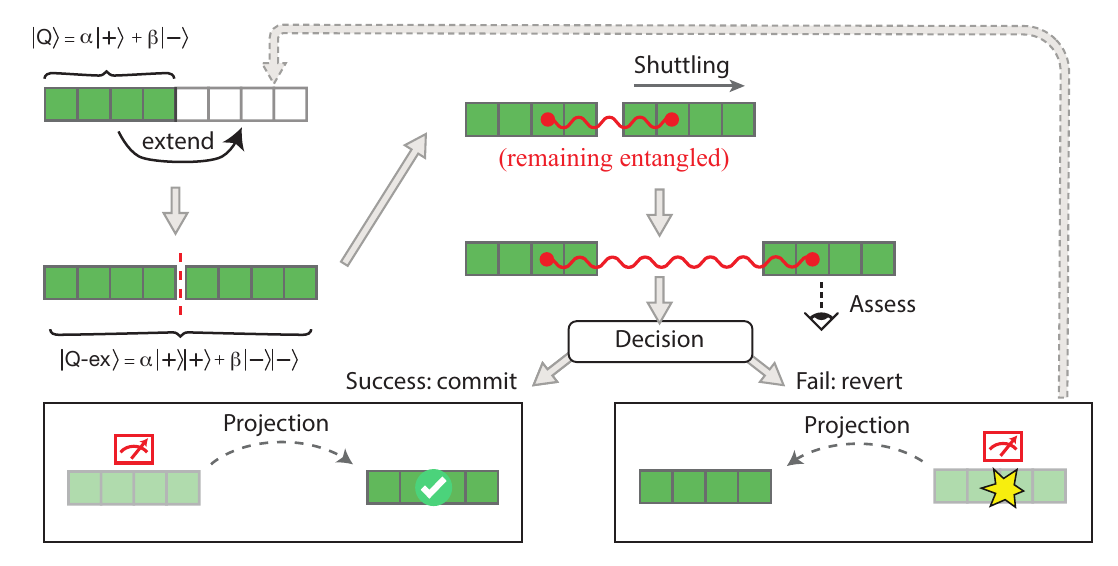}
    \caption{Detailed steps of the snake surgery protocol. $(i)$ The length of a snake is doubled by initialising new data qubits in the $\ket{0}$ state and performing an $XX$ lattice surgery. $(ii)$ The double-length snake is split in halves by performing another $XX$ lattice surgery. This separates the head from the tail but generates entanglement between the two. $(iii)$ The tail is stabilised in place, thereby protecting the logical information, while the head is shuttled across a long distance, making it sensitive to \textit{scratch} events. Entanglement between the head and tail is maintained throughout the process. $(iv)$ At the end of the head's shuttling, its integrity is assessed by analysing its syndrome signature and making use of the monitor qubits (Sections \ref{sec:monitor} and \ref{sec:complementary_gap}). $(v)$ If no catastrophic phase change was detected, the tail is measured out, projecting the logical information onto the head. Otherwise, the head is destroyed, projecting the information back to the tail. The procedure has to be restarted from step $(i)$}
    \label{fig:snake_surgery_details}
\end{figure}

A more detailed description of the scheme can be found below and in Fig. \ref{fig:snake_surgery_details}, in the case of shuttling one snake across long distances. The case of two interacting snakes is left to Appendix \ref{app:snake_surgery}.

\paragraph*{Step 1}
Double the length of the logical snake. This is performed by initialising additional data qubits in the $\ket{0}$ state and performing an $XX$ lattice surgery between the original snake and a logical snake prepared in the $\ket{0_L}$ state. Note that because $X_L$ is parallel to the direction of the shuttling, this $XX$ lattice surgery can be performed efficiently, \textit{i.e.} in $O(d)$ shuttles (in Fig. \ref{fig:snake_surgery_details}(a) the last $d$ qubits of the first snake and the first $d$ qubits of the second snakes are both $X_L$ logical operators). Starting from the initial state $\ket{\psi}$ encoded in a $d\times d$ surface code, this step therefore encodes the same state $\ket{\psi}$ in a $2d\times d$ surface code (where $2d$ is the length of the $Z$ logical operator):
\begin{equation}
    \ket{\psi}_{d\times d} \rightarrow \ket{\psi}_{2d\times d}
\end{equation}

\paragraph*{Step 2}
Split the extended surface code into two separate logical qubits, the \textit{head} and the \textit{tail}. The split operation is again lattice-surgery-based and along the same boundary. Assuming without loss of generality that the previously measured $X_LX_L$ was $+1$, this results in:
\begin{equation}
    \ket{\psi}_{2d\times d} = \alpha\ket{+} + \beta\ket{-}
    \rightarrow \alpha\ket{++} + \beta\ket{--}
\end{equation}
where the resulting states are $d\times d$ surface codes.

\paragraph*{Step 3}
Stabilise the tail in place while shuttling the head across a long distance. Throughout this process, the physical qubits of the latter can acquire some $Z$ errors, which we denote as $\mathcal{D}_\varphi[.]$:
\begin{equation}
    \alpha\ket{++} + \beta\ket{--} \rightarrow \alpha\ket{+}\mathcal{D}_\varphi[\ket{+}] + \beta\ket{-}\mathcal{D}_\varphi[\ket{-}]
\end{equation}

\paragraph*{Step 4}
At the end of the head's shuttling, detection techniques reassess the viability of the shuttling route. If no defect was detected ($\mathcal{D}_\varphi[.]$ is close to the identity), the tail is measured out in the $Z$ basis (which amounts to measuring all its data qubits in the 0/1 basis). Depending on the measurement outcome $m$, the head will be projected onto:
\begin{equation}
    \alpha\ket{++} + \beta\ket{--} \rightarrow \alpha\ket{+} + ({-1})^m\beta\ket{-}
\end{equation}
One can then apply an $X$ gate when $m=1$, thereby correctly transferring the initial state $\ket{\psi}=\alpha\ket{+} + \beta\ket{-}$ to the head.

If a defect was detected however, the head must be measured in the $Z$ basis to reverse the process and teleport the information back to the tail. As $\mathcal{D}_\varphi[.]$ commutes with such $Z$ measurement, the calculation is identical to the previous case.\\
\\
\indent The above protocol therefore promises to fully suppress the impact of catastrophic events, provided they are pure $Z$ errors and can be detected with high probability by the defect-detection schemes of Sections \ref{sec:monitor} and \ref{sec:complementary_gap}. If this is the case, the probability of correctly performing each step of our protocol (snake growth and split and teleportation of the information to the head or the tail) is simply given by the logical error rate of a $d\times d$ surface code. Indeed, all these steps are fault-tolerant, \textit{e.g.} based on lattice surgery.

The time overhead induced by this protocol is simply that of the lattice surgery of steps 1 and 2 \textit{i.e.} $d$ stabiliser cycles for the merge and one for the split.

Note that there exists a natural extension of the scheme we here presented. Instead of generating a single head and tail per snake, one could envision initialising a multi-headed snake (which one might call a hydra), in the state 
\begin{equation} \label{eq:hydra}
    \ket{\psi} = \alpha\ket{+}\ket{+\ldots +} + \beta\ket{-}\ket{-\ldots-}
\end{equation}
where the first qubit is the tail's state, and the other qubits correspond to the heads. While the (single) tail would still be stabilised in place to preserve the target error rate, heads can be shuttled separately. This has a dual use. First, in the case of a high defect rate, all heads could be brought to the same location in the device, yet through different routes, thereby multiplying the chances that at least one path was stable. Second, if the defect rate is instead relatively low, one could send the heads to different locations and exploit their entanglement (Eq. \ref{eq:hydra}) to optimise quantum circuits by quantum fan-out \cite{fan_out}.

\subsection{Monitor qubits} \label{sec:monitor}

\subsubsection{Generalities}

In the previous section, we demonstrated that snakes are resilient against phase-like defects, no matter how strong they are, as long as they can be detected. In the present section, we give a first protocol enabling such detection. More specifically, we detail the use of monitor qubits that are employed to validate in real-time the quality of shuttling links and to build an accurate model of their characteristics over the course of repeated shuttling events. These will be used jointly with stabiliser measurements, whose signature can as well inform the user about errors accumulated by shuttled data qubits (Section \ref{sec:complementary_gap}). Note however that the sole use of stabilisers cannot be sufficient, as a $\pi$ rotation on all data qubits of a snake (induced by a pin defect) would yield a logical error with no stabiliser signature.

\begin{figure}
    \centering
    \includegraphics[width=\linewidth]{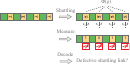}
    \caption{Schematic representation of the monitor-qubit-based scheme for task $(i)$ \textit{i.e.} inferring the presence of a defect. Monitor qubits (light orange squares) are interleaved in between data qubits (green squares) so as to probe the phase environment the logical snake evolves in. Step 1: monitor qubits are initialised in the $\ket{+}$ state. Step 2: The logical snake is shuttled. Every qubit picks up a phase along the way, modelled by the channel $\Phi$. Step 3: monitor qubits are measured in the $X$ basis. Step 4: the measurement outcomes are used to infer if a defect occurred. The workflow is similar for task $(ii)$ \textit{i.e.} precisely estimating the phase channel $\Phi$.}
    \label{fig:monitor_qubits_workflow}
\end{figure}

For this purpose we assume that a set of $N$ monitor qubits are interlaced with the data qubits and that before a shuttling event, they are prepared in an initial quantum state $\rho$. The qubits are then shuttled which ideally should act as an identity channel on the monitor qubits but due to imperfections, this channel will be non-trivial which we model via a CPTP map $\Phi$. Our aim is to learn characteristics of this channel in real time by applying a decoding operation on the state $\Phi(\rho)$ and measuring in the standard basis. The measurement basis obtained via the decoding operation, and the estimator used, crucially affect the performance of  the approach. The overall procedure is represented in Fig. \ref{fig:monitor_qubits_workflow}.

As motivated by Section \ref{sec:noise_description}, we will first assume that the noise $\Phi$ acts as a pure unitary single-qubit $Z$ rotation of angle $\omega$ on all shuttled qubits. Monitor qubits must respond to two tasks: $(i)$ detect with high precision when $\omega$ is larger than a given value $\omega_{\max}$ (this is a decision problem); and $(ii)$ estimate the value of $\omega$ with low uncertainty. The first task is the most crucial as it can inform the user of the presence of catastrophic events due to pin defects, which can induce logical errors. The second task is a refinement of the previous one, yet is less vital. It can be utilised as a means to update our knowledge of a given shuttling link for subsequent calibration. As an example of these two tasks, let us suppose that a defect appears at a given time and that a snake with interlaced monitor qubits is shuttled in its vicinity. This defect must be detected with high probability by task $(i)$, informing the user that the shuttling event should be reversed (using the protocol of Section \ref{sec:snake_surgery}). The value of the phase shift induced by the defect does not matter at this stage, as one only needs to know whether a defect occurred. If further monitoring confirms the presence of a defect, one can temporarily deactivate the corresponding shuttling link, thereby forbidding any snake from using this route.
Eventually, if later monitoring shows that the defect has become stable, inducing a constant known phase shift (task $(ii)$), then qubits can again be routed through this link.

We detail in \cref{app:monitor} that for the considered case of unitary $Z$ noise, then a simple yet optimal strategy to estimate if a given rotation is above a certain threshold (task $(i)$) is to initialise the state $\rho$ of the monitor qubits as $\ket{+}^{\otimes N}$ and measure them in the $X$ basis. An estimator $\hat{\omega}$ for the true angle $\omega$ can then be computed based on the number of times the ``1'' measurement outcome is observed. When $\hat{\omega}$ is above a given threshold, one can assume that a defect occurred. Nonetheless, this procedure cannot distinguish between angles of opposite signs (rotations by $\omega$ and $-\omega$ render on average the same number of non-trivial measurement outcomes). Therefore, this scheme must be refined for task $(ii)$ \textit{i.e.} for the precise estimation of the rotation angle $\omega\in[-\pi,\pi]$.
For this task, the strategy rather is to initialise the state $\rho$ of the monitor qubits as $\ket{-\pi/4}^{\otimes N}$ and measure half of them in the eigenbasis of the Pauli $X$ operator and the other half in the $Y$ operator. We devise an estimator $\hat{\omega}$ in the appendix to optimally sense the rotation angle $\pi$ which is detrimental to the present surface code construction (as it has no stabiliser signature). In both cases, we show in Appendix \ref{app:monitor} that the monitoring strategy saturates the fundamental, asymptotic Cramér-Rao bound, and that it shows remarkable noise robustness. Namely, we demonstrate that even in the presence of dephasing and measurement noise, this theoretical bound is only marginally affected, by a factor $(1-\frac{5}{2} \lambda)^{-1/2}$, where $\lambda$ is the total noise strength. Given a sub-surface-code threshold noise $\lambda$ of the order of $10^{-3}$, the noisy output is only $0.1\%$ larger than the noiseless one. In the following, we will assume $\lambda=0.2\%$.

\subsubsection{Task $(i)$: defect detection}

For task $(i)$, we first define $\omega_{\max}$ as the maximum value of $\omega$ a logical snake can tolerate when it is shuttled near a defect, without inducing a dramatic increase of the logical error rate. We set $\omega_{\max}=\omega_{th}/2$, where $\omega_{th}=9\%$ is the error correction threshold for such a scenario (assuming circuit-level noise of strength $p=0.1\%$ with one round of additional dephasing at a rate $q=\sin^2(\omega/2)$, see Appendix \ref{app:defect_threshold} for details). When $\omega < \omega_{\max}$, a strong exponential suppression of the error is guaranteed, and by an appropriate choice of the code distance $d$, the logical error rate can be kept low enough to run deep quantum algorithms. In contrast, if $\omega \geq \omega_{\max}$, the initially chosen code distance may fail to bring the error rate sufficiently low. Note that we here ignore ameliorations provided by the complementary gap scheme of Section \ref{sec:complementary_gap}. Besides, we set a detection threshold $\hat{\omega}_{\max}$, defined as the value of the estimator $\hat{\omega}$ above which a defect is declared. This parameter controls the rate of false positives (deeming a non-defective shuttling link as defective) and false negatives (failing to detect a defect in a defective shuttling link). In our case, we want to minimise the latter, \textit{i.e.} guarantee that almost every defect is detected, so as to minimise the logical error rate. This comes at the cost of a higher rate of false positives, meaning that a finite proportion of the shuttling links may be deemed defective even when they are not. While this has no impact on the logical error rate, it can lead to an increase in the computation time if the snake surgery protocol of Section \ref{sec:snake_surgery} has to be used repeatedly. We will thus aim to keep this false positive rate within 5-10$\%$, while bringing the false negative rate as close to 0 as possible.

We illustrate the performance of the monitor qubits for task $(i)$ in the left panel of \cref{fig:monitor_qubits}. Namely,
starting from a uniform distribution of the true angle $\omega$ before monitoring, we plot its distribution $p_{\text{mon}}(\omega)$ after the monitor-qubit-based filtering. The selection rule we choose is $\hat{\omega} < \hat{\omega}_{\max} = \omega_{\max}/4$, which leads to a rejection rate of the shuttling links of approximately $5\%$ (see Appendix \ref{app:monitor}).
This amounts to invoking the reversion option in snake surgery protocol approximately $5\%$ of the time.
The vertical dashed lines indicate the value of $\pm\omega_{\max}$. One must minimise the probability $P_m$ that a qubit is shuttled through a post-selected shuttling link whose true angle $\omega$ is outside the window defined by the dashed lines. This is computed as:
\begin{equation} \label{Pm}
    P_m = 2 \int_{\omega_{\max}}^{\pi} p_{\text{mon}}(\omega)\mathrm{d}\omega
\end{equation}
For 900 monitor qubits, \textit{i.e.} interlacing one monitor per data qubit in a $30\times30$ surface code, $P_m$ is as low as $10^{-8}$. More detailed plots can be found in Appendix \ref{app:monitor}.

\subsubsection{Task $(ii)$: angle estimation}

For task $(ii)$, we simply plot the RMS error $\Delta \omega$ of the estimator $\hat{\omega}$ with respect to $\omega\in[-\pi,\pi]$. The dashed lines represent the corresponding Cramér-Rao bounds as $1/\sqrt{N}$ and confirm that even for a relatively low number of shots these asymptotic bounds are practically saturated near the rotation angles $\pm\pi/2$, $0$ and $\pm \pi$. In other words, we are optimally sensitive to these rotation angles. Dominant noise contributions \textit{i.e.} dephasing and readout errors merely shift the dashed lines above by a fraction of a percentage. In contrast, in the worst case the estimator has a factor of $\sqrt{2}$ larger RMS error (at angles $\pm 3\pi/4, \pm \pi/4$) given we split the sensing states into two batches and use them to estimate different (but not independent) regions of the unit circle.
In conclusion, using $N=800$ monitor qubits we can achieve an RMS error
$\Delta \omega \leq 0.05$, which one order of magnitude below the maximum tolerable angle $\omega_{\max}=0.3$.

\begin{figure*}
    \centering
    \subfloat{
        \centering
        \includegraphics[width=0.45\linewidth]{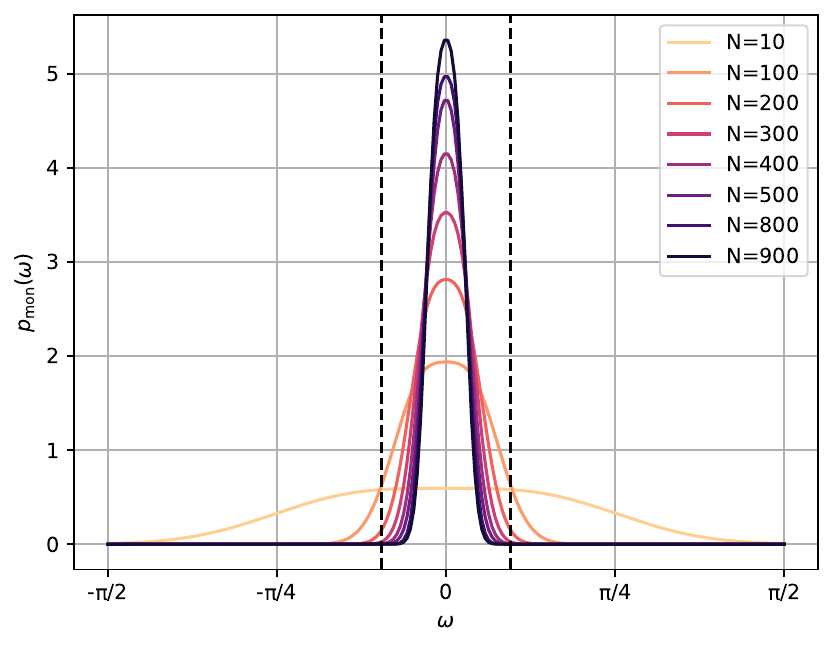}}
    % \hspace{0.1\linewidth}
    \hfill
    \subfloat{
        \centering
        \includegraphics[width=0.45\linewidth]{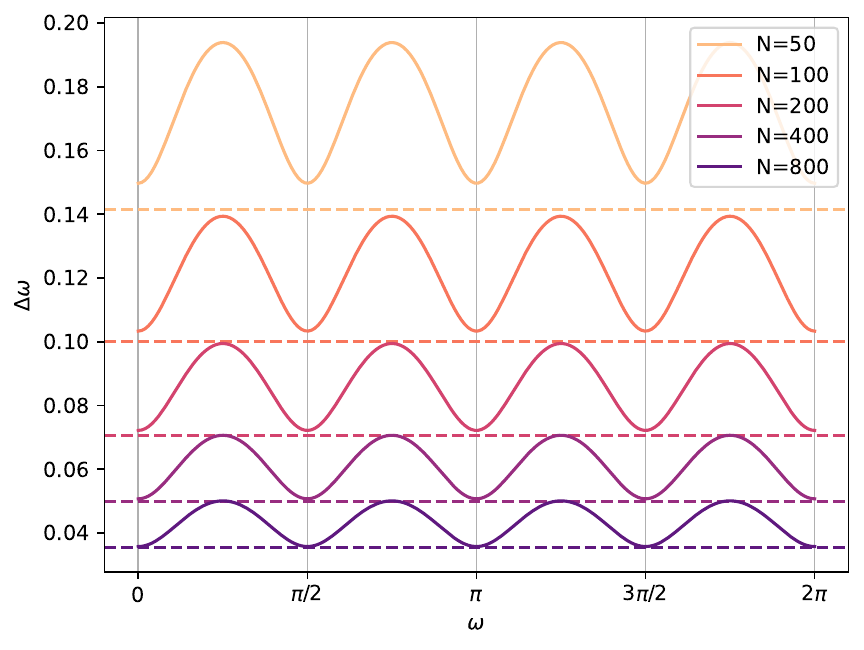}}
    \caption{Performance of the monitor qubits for the two tasks detailed in the main text. $\omega$ and $\hat{\omega}$ are respectively the rotation angle associated with the shuttling link, and its estimator computed from the measurement of $N$ monitor qubits. The cumulative dephasing and measurement noise is set to $\lambda=0.2\%$.
    Left: performance of task $(i)$, as the distribution of the angle $\omega$, post-selected according to a selection rule leading to an approximately $5\%$ rejection rate of the shuttling links. This amounts to cancelling the shuttle via snake surgery (Section \ref{sec:snake_surgery}) approximately $5\%$ of the time. The vertical dashed lines indicate the maximum angle $\omega_{\max}$ that the surface code can tolerate to guarantee fault-tolerance (if not using the complementary gap scheme of Section \ref{sec:complementary_gap}). Therefore, a logical error occurs when a qubit is shuttled through a post-selected shuttling link, yet the rotation angle $\omega$ is outside the window defined by the dashed lines. For $N=900$ monitor qubits, the probability of this occurring reaches $10^{-8}$.
    Right: performance of task $(ii)$, as the RMS error $\Delta \omega$ of the estimator $\hat{\omega}$ with respect to $\omega$. The dashed lines represent the corresponding Cramér-Rao bounds as $1/\sqrt{N}$ and confirm that even for a relatively low number of shots these asymptotic bounds are practically saturated near the rotation angles $\pm\pi/2$, $0$ and $\pm \pi$, \textit{i.e.}, we are optimally sensitive to these rotation angles. Dominant noise contributions as dephasing and readout errors merely shift the dashed lines above by a fraction of a percentage.}
    \label{fig:monitor_qubits}
\end{figure*}

The collected samples can also be split into batches of $n$ bits
in post-processing to perform several independent estimations
$\hat{\omega}_1, \hat{\omega}_2 \dots, \hat{\omega}_{N/n}$. For instance, for $n=200$  each of these yields an uncertainty $\Delta \omega \leq  0.1$ as illustrated in \cref{fig:monitor_qubits} (right).
This is still comfortably below $\omega_{\max}=0.3$.

\subsection{Complementary gap} \label{sec:complementary_gap}

The previous section described a method capable of detecting defective shuttling links via the use of additional monitor qubits. Nonetheless, one must remember that throughout the course of the computation, information about potential errors is already collected by ancilla qubits measuring stabilisers. Therefore, in addition to enabling key logical error suppression, these measurement outcomes can be put to use for the assessment of the quality of a shuttling link. The general idea simply is that, when a snake is shuttled near a defect, its data qubits may experience a higher physical error rate, yielding a higher number of stabiliser events. As we suppose that defects are purely phase-like, we will solely focus on $X$ stabilisers. In the present section, we formalise a criterion permitting us to distinguish with high probability between stabiliser signatures arising from the presence of a defect, or from normal circuit noise, and to dramatically decrease the resulting logical error rate.

Note however that stabilisers alone cannot be utilised to detect the most catastrophic kind of defects, namely defects inducing a $\pi$ rotation on all data qubits. These would indeed incur a logical failure with no stabiliser signature whatsoever. Nevertheless, detecting such large rotations is the task where monitor qubits excel. Thus, one must think of the method presented in this section as a second filter used on top of the monitor-qubit-based scheme, the joint use of both protocols eventually leading to an almost-perfect detection of all defects.

Additionally, one can note that using this scheme inevitably leads to data qubits experiencing a high rate of errors due to the defect, which is only detected \textit{a posteriori} through the stabiliser measurements. This is similar to the case of the previous section where monitor qubits are interlaced with data qubits.
This defect detection could at first appear pointless if the data qubits have already undergone the defect-induced error channel. However, we demonstrated in Section \ref{sec:snake_surgery} that such errors can be reversed via so-called snake surgery, as long as they are phase-like and have been \textit{a priori} identified.

The criterion we build for identifying defects from stabiliser measurements uses the recently-introduced concept of \textit{complementary gap} \cite{gidney2023yokedsurfacecodes}. This quantity measures the confidence with which the decoder decided on a given correction. For instance, imagine a scenario where a single stabiliser at the centre of a surface code renders a non-trivial measurement outcome. This syndrome can equally be explained by an error string terminating at either of the code's boundaries. There is therefore a $50\%$ chance of a logical failure. Since snake surgery grants us the effective ability to \textit{reverse time} whenever a high-risk event is detected, it is in this case more wise to do so by measuring out the head of the shuttled snake.

More generally, we denote by complementary gap the length difference between the shortest error strings explaining the syndrome but yielding opposite logical outcomes. For instance, in the previous example of a single non-trivial stabiliser measurement right at the centre of the code, the complementary gap is 0. Besides, as explained in \cite{gidney2023yokedsurfacecodes}, this quantity can be efficiently simulated by running the decoder with a switched boundary detector. Intuitively, when the physical error rate approaches the code's threshold \textit{e.g.} due to a defect, the obtained complementary gaps typically decrease. If the other noise sources are kept sufficiently low below threshold, it is thus only natural to assert that a defect occurred when the observed complementary gap $g$ is below a given value $g_{\min}$. As in the case of monitor qubits, our aim is to minimise the rate of false negatives while keeping the rate of false positives around 5-10$\%$. This guarantees that almost all defects are detected, while unnecessarily reverting the shuttle event in snake surgery 5-10$\%$ of the time. In general, the value of $g_{\min}$ permitting such a rejection rate depends on the code distance.

\begin{figure*}
    \centering
    \subfloat{
        \centering
        \includegraphics[width=0.48\linewidth]{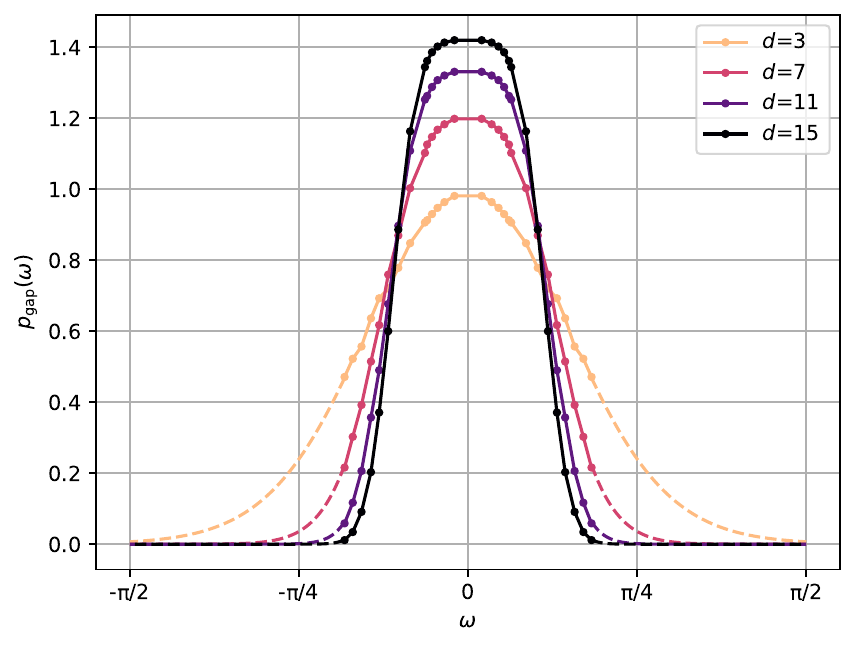}}
    % \hspace{0.1\linewidth}
    \hfill
    \subfloat{
        \centering
        \includegraphics[width=0.5\linewidth]{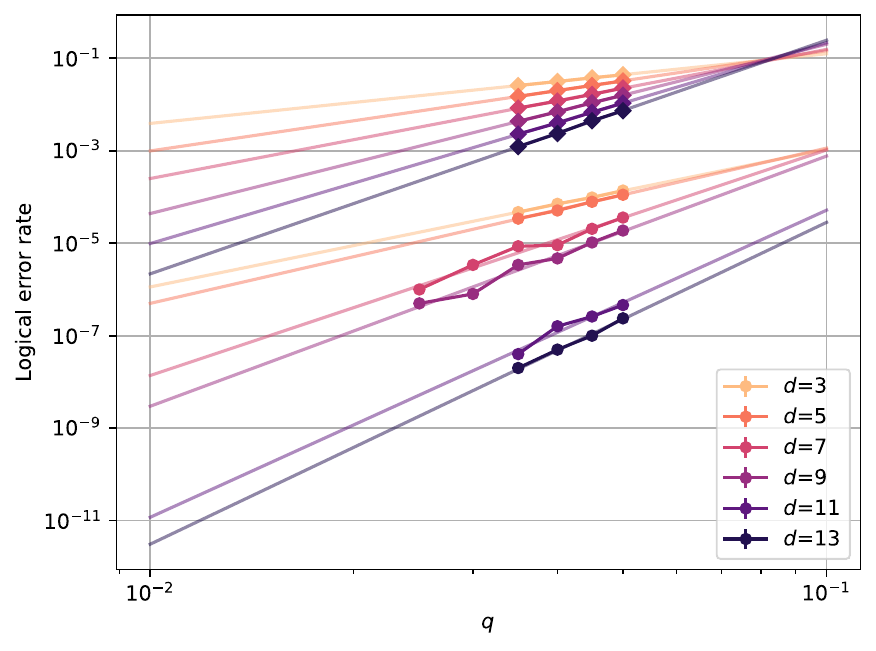}}
    \caption{Performance of the complementary-gap-based detection scheme for various code distances $d$ under circuit-level noise at $p=0.1\%$, with one round of additional defect-induced dephasing at a rate $q=\sin^2(\omega/2)$. Left: distribution of post-selected defect angles $\omega$, assuming a uniform initial distribution and according to a selection rule leading to the rejection of less than $5\%$ of the shuttling links. Additional filtering from monitor qubits is not implemented here. The tails of the distributions (dashed lines) are obtained by Gaussian extrapolation. Right: logical error rate after post-selection (circles), compared to the logical error rate without post-selection (diamonds). Additional linear regressions (in log scale) are plotted for both datasets.}
    \label{fig:complementary_gap}
\end{figure*}

In Fig. \ref{fig:complementary_gap} (left), we plot the distribution of post-selected angles, assuming an initial uniform distribution (therefore not using additional monitor qubits here). The selection rule we use is $g \geq g_{\min}=(d+1)/2$, which leads to the rejection of less than $5\%$ of the shuttling links (see Appendix \ref{app:complementary_gap}). One can observe that the distribution is far less peaked than the one obtained from the sole use of monitor qubits. This means that the present method is \textit{not} efficient at detecting defects. Rather, its strength lies in the significant logical error rate reductions it provides: even if a large defect angle $\omega$ passes the filters, the action of rejecting instances with small complementary gaps highly improves the decoder's performance, hence shrinking the rate of errors.
This contrasts with the monitor-qubit case, for which the filtering has no such impact. Therefore, we plot in Fig. \ref{fig:complementary_gap} (right) the logical error rate obtained after such post-selection (circles) and compare it to the error rate that would be obtained if our detection scheme was not utilised (diamonds). The strength of this protocol is made evident, showing up to five orders of magnitude improvements for the plotted distances and error rates. Given the tendency observed in the plot, we expect even further reductions for higher code distances (see Appendix \ref{app:complementary_gap}).

\subsection{Defect resilience}

Putting together the elements from all previous sections, one can estimate the resilience of our architecture to defects, as a function of the code distance $d$ and the defect rate $\rho$ \textit{per $d$ stabiliser cycles}. We additionally denote as $P_L$ the normal logical error rate one would observe with a distance-$d$ rotated surface code under circuit-level noise at $p=0.1\%$ (with no defect), again over $d$ stabiliser cycles.

When a snake is required to travel between two distant points in the device, its length is doubled and the head is detached from the tail which remains stabilised in place. The state of the shuttling path is then constantly monitored by the monitor qubits and stabiliser outcomes. If a defect is suspected, the head is destroyed and the shuttling event is restarted. We showed in Section \ref{sec:snake_surgery} that this procedure is fault-tolerant \textit{i.e.} the snake does \textit{not} pick up any error from the defect, as long they are purely phase-like (the error rate reduces to the normal logical error rate $P_L$). Therefore, the suspected main source of error corresponds to the case where a defect occurs yet is left undetected. In this situation, one expects that the detection strategies designed in the previous sections would catch most of the defects (\textit{i.e.} the distribution of post-selected angles is peaked around the origin), and that the leftover angles can be handled by normal error correction, enhanced by the rejection of small complementary gaps.

We rigorously demonstrate in Appendix \ref{app:defect_resistance} that the final logical error rate $\tilde{P}_L$ \textit{per $d$ stabiliser cycles} indeed reduces to the contribution of undetected defects, plus terms smaller than $P_L$. In particular, we show that, owing to the low rate of false positives that we set, the rejection of multiple shuttling paths until a non-defective one is found does not lead to any error accumulation. The defect-induced term in $\tilde{P}_L$ is of the form:
\begin{equation}
    \rho \times \int_{|\omega|=0}^{\pi} P(\omega)p_{\text{both}}(\omega) \mathrm{d}\omega
\end{equation}
where $\rho$ is the rate of defects, $P(\omega)$ is the probability of a logical error given that a defect characterised by an angle $\omega$ is left undetected, and $p_{\text{both}}(\omega)$ is the distribution of post-selected angles from both defect-detection techniques.

As the monitor qubits are never entangled with the data or ancilla qubits of the error correcting codes, it follows that, for each value of $\omega$, the monitoring and complementary gap strategies outcomes are uncorrelated. In other words, the detection strategies are conditionally independent given any defect angle $\omega$. Therefore one can write:
\begin{equation}
    p_{\text{both}}(\omega) = \frac{p_{\text{mon}}(\omega)p_{\text{gap}}(\omega)}{\int_{-\pi}^{\pi}p_{\text{mon}}(\omega)p_{\text{gap}}(\omega)\mathrm{d}\omega}
\end{equation}

Consequently, all quantities involved in the calculation of the final logical error rate $\tilde{P}_L$ have already been computed and analysed: $P(\omega)$ and $p_{\text{gap}}(\omega)$ in Fig. \ref{fig:complementary_gap}, and $p_{\text{mon}}(\omega)$ in Fig. \ref{fig:monitor_qubits} (left).

\begin{figure}
    \centering
    \includegraphics[width=\linewidth]{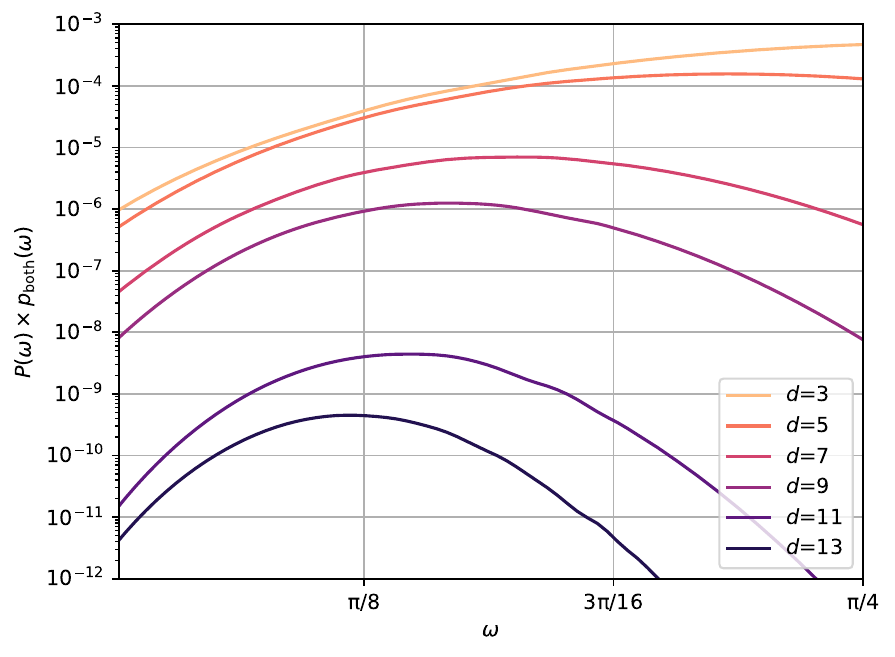}
    \caption{Probability that a defect of angle $\omega$ is undetected and triggers a logical error, plotted against $\omega$ and for various code distances $d$. We use one monitor qubit per data qubit, \textit{i.e.} $N=d^2$ monitor qubits.}
    \label{fig:Pw_p_both}
\end{figure}

In Fig. \ref{fig:Pw_p_both}, we thus plot the quantity $P(\omega)p_{\text{both}}(\omega)$ for various code distances, using the data and extrapolations from the previous sections. This product represents the probability that a defect of angle $\omega$ is undetected and triggers a logical error. We use one monitor qubit per data qubit, \textit{i.e.} $N=d^2$ monitor qubits. $P(\omega)$ is a growing function of the defect angle $\omega$ (higher angles yield higher error rates) while $p_{\text{both}}(\omega)$ is decreasing with $\omega$ (high angles are more likely to be detected by the monitoring schemes). Therefore, their product should feature a maximum, which is what we observe. Values of $\omega$ closer to 0 are not plotted as the effect of the circuit-level noise would become predominant and the linear extrapolations we used from Fig. \ref{fig:complementary_gap} (right) would break down.

Every code distance is associated with a given achievable logical error rate $P_L$ under circuit-level noise at $p=0.1\%$ and in the absence of defects. By computing the integrals under the curves, one can observe that for all code distances $d\leq 13$, these are smaller than the normal error rates $P_L$:
\begin{equation} \label{eq:Ptilde_lower_than_P}
    \int_{|\omega|=0}^{\pi} P(\omega)p_{\text{both}}(\omega) \mathrm{d}\omega \leq P_L
\end{equation}
The ratio between the left- and right-hand side of the inequality also decreases with $d$, suggesting that it would be satisfied for \textit{any} code distance.
The integrals are computed by making use of the extrapolations of the previous sections. As noted before, the linear extrapolations used for $P(\omega)$ are imperfect at small $\omega$, but this corresponds to minimal values of $P(\omega)$ (orders of magnitude lower than at its maximum), hence to a good approximation this has no impact on the final value of the integral.

Evidently the combined use of our defect detection and recovering strategies is extremely powerful, actually bringing the impact of defects below the normal logical error rate. One can now wonder to what extent both methods are necessary. On the one hand, only using monitor qubits does not seem to be sufficient, as we observed in Eq. \ref{Pm} that for a $30\times 30$ surface code using $900$ monitor qubits, the probability of observing an angle $\omega$ leading to a logical error was $P_m=10^{-8}$. If one aimed to reach error rates as low as $10^{-15}$, a defect rate $\rho=10^{-7}$ would thus be necessary to bring defect-induced noise as low as the normal logical error rate. This does not mean that monitor qubits must be overlooked however, as they are the only way to sense angles around $\pi$, which stabilisers cannot detect. On the other hand, the post-selection of high complementary gaps promises massive error rate reductions, with a post-selected error rate $P(\omega)$ orders of magnitude lower than the normal error rate $P_L$. Similarly to Eq. \ref{eq:Ptilde_lower_than_P}, one can show that:
\begin{equation}
    \int_{|\omega|=0}^{\pi/2} P(\omega)p_{\text{gap}}(\omega) \mathrm{d}\omega \leq P_L
\end{equation}
This means that, if angles closer to $\pi$ are excluded, the complementary gap method is sufficient to bring the defect-induced error rate below the normal logical error rate. Therefore, while both detection strategies appear necessary, one improvement one could think of is to reduce the number of monitor qubits so as to only reject angles around $\pi$. In all other cases, stabilisers aided by the post-selection of high complementary gaps would be able to lower the defect-induced error rate below the normal logical error rate.

Do note however that in all our derivations, we did assume a low rate of defects, such that two defects would not occur during the same shuttling event, and such that defect declarations would mostly arise from false positive detections (at a rate of 5-10$\%$). This is typically true if $\rho$ is much smaller than 5-10$\%$, \textit{i.e.} smaller than 1 defect every $10d$ stabiliser cycles on a given shuttling link.

\section{Universal quantum computation} \label{sec:univ_quant_comput}

In this section we demonstrate that universal quantum computation is possible on the snakes on a plane architecture, and give protocols for the implementation of a universal set of gates.

\subsection{Logical level connectivity}

One of the strengths of our proposal is that it boasts an all-to-all connectivity at the logical level. This is permitted by the expected fast and high-fidelity shuttling of silicon spin qubits, enabling the global movement of the snake-like logical qubits across the device. During these long shuttles, the preservation of a low logical error rate is guaranteed by the protocols of Section \ref{sec:noise_protection}. In particular, it relies on the rejection of a finite number of shuttles, whenever an abnormal event is detected by monitor qubits or stabiliser measurements. If such an abnormality is repeatedly detected, a defect can be declared, ultimately leading to the deactivation of the shuttling link. Additionally, some 2$\times$N filaments may be deemed defective even before the computation started, as the presence of fabrication defects will be inevitable in such large devices. Therefore, a finite proportion of the shuttling links will necessarily be unusable. Yet, we demonstrate in the following that logical-level all-to-all connectivity is still guaranteed as long as this proportion is below a given threshold.

In Fig. \ref{fig:general_layout}, snakes live in white tiles, which are connected through interaction edges (small edges where two white tiles are adjacent). Therefore, one can identify such tiles as the vertices of the logical-level connectivity graph, and the interactions edges as the edges. These form a (rotated) square lattice. In the worst case, one can assume that a defective shuttling link would lead to the entire deactivation of an white tile (vertex) or of an interaction edge (edge). In these cases, the maximum proportion of deactivated links that is tolerable to keep a fully connected lattice (at the scale of the entire device) is given by the side and bond percolation thresholds, which for the square lattice are respectively $60\%$ and $50\%$. Therefore, as long as the proportion of defective shuttling links is below $50\%$, all-to-all connectivity is guaranteed. However, the compute time may significantly increase when the number of disabled links approaches the percolation threshold. Indeed, the route a snake must follow to travel between two given points in the device may naturally be extended to avoid defective links.

\subsection{Universal set of gates}

The set of universal gates we advocate for is the classic set $\{CNOT,H,S,T\}$. One of the strengths of our architecture is that transversal gates can seamlessly be implemented via the interaction edges, where two logical snakes can be made adjacent. $CNOT$ gates can therefore be applied particularly efficiently, much faster than in more conventional static 2D setups where lattice surgery must be used.

\begin{figure}
    \centering
    \includegraphics[width=0.9\linewidth]{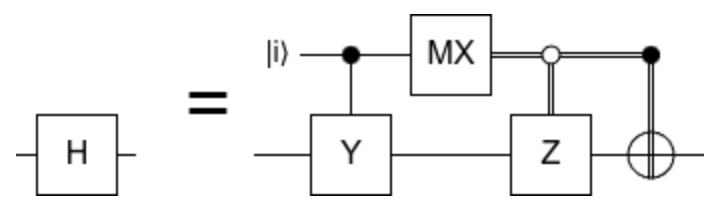}
    \caption{Implementation of a Hadamard gate via gate teleportation.}
    \label{fig:h_gate_teleportation}
\end{figure}

As for Hadamard gates, they can be applied transversally as well, although this induces a 90-degree rotation of the surface code. This rotation can be reversed by physically reshuffling the qubits through the junctions. This however incurs a relatively high time overhead and changes the orientation of the logical operators of the code, meaning $Z$ logical operators would now be parallel to the direction of shuttling, drastically reducing the protection against $Z$ noise. While there may exist alternative architectures or protocols that permit an efficient and safe implementation of such rotations, we leave this investigation for further research. Alternatively, one possible implementation of Hadamard gates via gate teleportation is given in Fig. \ref{fig:h_gate_teleportation}. Here the last two gates are just dependent on the classical outcome of the $MX$ measurement, which can be implemented by applying a transversal $H$ gate then measuring in the $Z$ basis (the subsequent rotation of the logical operators does not matter since they are directly measured). The $CY$ gate can be decomposed as $S.CNOT.S^\dagger$, where implementations for the $S$ and $T$ gates are given thereafter.

For these two gates, we adopt the conventional choice of an implementation via gate teleportation and additional magic state distillation. This process relies on the concatenation of the surface code with an additional code whose $S$ or $T$ gate is transversal. The entire procedure is based on the successive implementation of many $CNOT$ gates, which as explained before are not a bottleneck in our platform. One could even envision implementing the multi-target $CNOT$s appearing in the distillation circuits via quantum fan-out (as explained at the end of Section \ref{sec:snake_surgery}), which may greatly simplify them.

\subsection{Semi-transversal $CNOT$ gates} \label{sec:semi_transversal_gates}

% \subsubsection{Method description}

In order to apply a logical $CNOT$ gate transversally, two logical qubits have to be entirely shuttled through an interaction edge, which amounts to a total shuttling distance of $O(d^2)$. If one were to implement all transversal physical gates at once, this would imply $O(d^2)$ shuttling increments without stabiliser measurements, while these are normally performed in $O(d)$ shuttling increments (Figs. \ref{fig:surface_code_2xN} and \ref{fig:snake_movement}). This scaling becomes problematic when shuttling operations are noisy, leading to greater losses in fidelity.

\begin{figure}
    \centering
    \subfloat{
        \centering
        \includegraphics[width=0.45\linewidth]{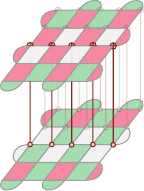}}
    % \hspace{0.1\linewidth}
    \hfill
    \subfloat{
        \centering
        \includegraphics[width=0.45\linewidth]{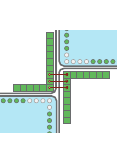}}
    \caption{Semi-transversal $CNOT$ gate. In this protocol, a logical $CNOT$ is implemented by applying transversal $CNOT$s in batches \textit{e.g.} row by row. $CNOT$s that are applied at the current round are represented in brighter brown, while $CNOT$s that have previously been applied are drawn in a faded brown. Some stabilisers cannot be measured at the current round due to their anti-commutation with certain $CNOT$s: these are drawn in light gray. Left panel: canonical 2D representation of a surface code. Right panel: snakes on a plane representation. }
    \label{fig:semi_tranversal}
\end{figure}

To address this problem, we can employ a semi-transversal gate protocol (Fig. \ref{fig:semi_tranversal}). The fundamental principle behind semi-transversal gates involves performing transversal operations in discrete batches while concurrently stabilising all physical qubits that are not actively involved in the interaction. By batching the transversal operations, it becomes possible to reduce the cumulative impact of shuttling noise. The number of transversal gates executed per batch can be adjusted based on the shuttling noise levels. Specifically, larger batches yield faster but noisier logical gates (the limiting case being the implementation of a normal transversal gate, with one batch of $d^2$ $CNOT$s). The length of the interaction edge can basically be set to the number of transversal operations per batch.

Because transversal $CNOT$s are not applied all at once, some $X$ (resp. $Z$ stabilisers) have a support that only partially overlaps with control (resp. target) qubits of already-implemented $CNOT$s. These cannot be measured as they would lead to a propagation of the errors. More precisely, let us denote as $\ket{\Psi}$ the state of the two-qubit logical state before applying the $CNOT$s. By definition, the application of any stabiliser $S$ should leave $\ket{\Psi}$ unchanged:
\begin{equation}
    S\ket{\Psi} = \ket{\Psi}
\end{equation}
When applying a unitary operation $U$ on $\ket{\Psi}$, only stabilisers that leave $U\ket{\Psi}$ unchanged can be measured; otherwise the state could randomly be projected outside the logical subspace:
\begin{equation} \label{eq:semi_transversal_condition}
    SU\ket{\Psi} = U\ket{\Psi}
\end{equation}
Now let us assume that $S$ is an $X$ stabiliser acting on four data qubits, indexed from 1 to 4, in the control logical qubit. Let us additionally assume that transversal $CNOT$s have only been applied between qubits 1 and 1', and 2 and 2' (where the $i'$ denotes qubits indices in the target logical qubit). It follows:
\begin{align*}
    SU\ket{\Psi} &= X_1X_2X_3X_4 CNOT_{1,1'} CNOT_{2,2'}\ket{\Psi} \\
                 &= USX'_1X'_2\ket{\Psi} \\
                 &= UX'_1X'_2\ket{\Psi}
\end{align*}
where the equality between the first and second lines follows from $CNOT$ and $X$ gates commutation rules. $X'_1X'_2$ however is not a stabiliser of the target logical qubit ($X'_1X'_2X'_3X'_4$ is). Therefore, Eq. \ref{eq:semi_transversal_condition} is not respected. With this logic, one can easily deduce that, at all times, all $X$ stabilisers on the target logical qubit and and all $Z$ stabilisers on the control logical qubits can be measured. However, $X$ (resp. $Z$) stabilisers whose support only partially overlaps with already-implemented $CNOT$s on the control (resp. target) logical qubits cannot be measured. This is represented in Fig. \ref{fig:semi_tranversal}.

% \subsubsection{Numerical evaluation of the performance}

% As explained previously, our semi-transversal scheme allows one to mitigate shuttling error accumulation. Indeed, a direct transversal implementation of two-qubit logical gates requires $O(d^2)$ shuttles with no stabiliser measurement. On the contrary, in the semi-transversal picture, the logical snakes can be stopped at any point for additional stabilisers measurements. While this comes at an apparent time overhead, it also intuitively promises a better control of the errors.

% In this section, we numerically estimate this time/performance trade-off. 

% \red{increase noise in the semi-transversal simulations because it takes longer $\rightarrow$ higher chances for error accumulation (not from shuttling)?}

\section{Discussion}

In this paper, we explore the practicality of shuttling logical qubits in a latticework structure. The structure is designed to be suitable for realisation with silicon-spin chip devices. Our approach leverages ideas presented in the 2$\times$N architecture of Ref.\,\cite{siegel_two_by_n_2024}. In that earlier piece of research, it was demonstrated that certain code classes could efficiently be embedded on a 2$\times$N array equipped with shuttling, and that universal quantum computation was possible. However, due to the linear structure of the device, the connectivity at the logical level was overly constrained, leading to relatively long run times. Additionally, single-point failures would cause the device to be split into two disconnected halves.

Here we have demonstrated that interfacing a large number of such structures together through junctions would solve the aforementioned problems. The proposed latticework arrangement employs linked 2$\times$N filaments along which logical qubits can travel. Our accompanying protocols enable long-range shuttling of logical qubits and indeed general fault-tolerant quantum computation through semi-transversal gate operations. 

Specifically, we first characterise the noise experienced by silicon spin qubits in our device, in the form of charge noise. We demonstrate that this error source can be ameliorated by the right choice of qubit encoding, namely a singlet-triplet encoding. As such a construction lives in a decoherence-free subspace, it is naturally first-order immune against phase errors. However, the sole use of this encoding would not be sufficient to protect the device against the most catastrophic noise source: the appearance of pin defects materialising near a shuttling track. This could incur a dramatic error surge yielding a logical failure. We therefore present a protocol that actively addresses the issue and provides a profound suppression of logical errors. The method involves the high-fidelity detection of defects, via the analysis of the collected stabiliser measurements and the use of additional monitor qubits. Armed with this capability, we show that one can revert a process that is suspected to have involved a defect-induced error. In particular, assuming that the defect-induced error mechanism is purely $Z$-like, we demonstrate that the errors it incurs can perfectly be annihilated, by effectively \textit{reversing time} and going back to the pre-defect version of the erroneous snake. Jointly using all above protocols, we demonstrate a tolerance to defects exceeding the tolerance to normal circuit-level noise, and for any code distance.

In addition to these noise-resistance schemes, we show that our proposed architecture is suited for universal quantum computation. Its first strength is the all-to-all connectivity at the logical level, enabled by long-distance shuttling. In particular, we note that a high proportion of defective edges is tolerable without the loss of  connectivity. The second strength of our current proposal is the possibility to interface logical qubits so as to perform transversal gates, which represents a considerable time reduction compared to more common lattice surgery protocols. In this context we describe a generalised \textit{semi-transversal} implementation of two-qubit logical gates. This scheme offers an opportunity to implement transversal gates in batches rather than all at once (which would be the case in a traditional transversal setup). 

We conclude that the use of monitored, fault tolerant snake-like logical qubits (`MF snake qubits') is a natural solution to realising scalable quantum computing in two-dimensional solid state architectures that support shuttling. The latticework structure on which the MF snakes move has a high degree of design flexibility, and its locally linear nature is compatible with near-term silicon manufacturing capabilities.

Our proposed architecture is not limited to surface code QEC. As explored in our previous work~\cite{siegel_two_by_n_2024}, many classes of qLDPC codes are a good fit for the 2$\times$N qubit layout and therefore also for our architecture. In fact, the addition of junctions allows one to reshuffle qubits in a much simpler way than in the previous work. Hence, one can further optimise the syndrome extraction circuits of the proposed codes, or even find novel classes of codes whose efficient implementation is enabled by the use of junctions. Since all of the mentioned classes of codes in~\cite{siegel_two_by_n_2024} are CSS codes, they all have a transversal CNOT gate that can be implemented in a similar semi-transversal fashion as proposed in Section~\ref{sec:semi_transversal_gates}. The same requirement holds -- one needs to be careful not to measure stabilisers for which only a partial set of data qubits in their support have undergone the semi-transversal operation. Note that performing a transversal CNOT between higher-rate codes implements a logical CNOT on each pair of respective logical qubits and generally that may be undesirable. 

We leave the adaption of snake-surgery protocol for higher-rate codes for future work. 

\section{Acknowledgements}
The authors are grateful to Zhu Sun for helpful conversations.
The numerical modelling involved in this study made use of the Quantum Exact Simulation Toolkit (QuEST),
and the recent development pyQuEST which permits the user to use Python as the interface front end. We are grateful to those who have contributed to these valuable tools. The authors acknowledge support from the EPSRC QCS Hub grant (agreement No. EP/T001062/1), EPSRC’s Robust and Reliable Quantum Computing (RoaRQ) project (EP/W032635/1), and the SEEQA project (EP/Y004655/1). B.K. thanks UKRI for the Future Leaders Fellowship
project titled Theory to Enable Practical Quantum Advantage (MR/Y015843/1).

\bibliography{ref}

\begin{thebibliography}{53}%
\makeatletter
\providecommand \@ifxundefined [1]{%
 \@ifx{#1\undefined}
}%
\providecommand \@ifnum [1]{%
 \ifnum #1\expandafter \@firstoftwo
 \else \expandafter \@secondoftwo
 \fi
}%
\providecommand \@ifx [1]{%
 \ifx #1\expandafter \@firstoftwo
 \else \expandafter \@secondoftwo
 \fi
}%
\providecommand \natexlab [1]{#1}%
\providecommand \enquote  [1]{``#1''}%
\providecommand \bibnamefont  [1]{#1}%
\providecommand \bibfnamefont [1]{#1}%
\providecommand \citenamefont [1]{#1}%
\providecommand \href@noop [0]{\@secondoftwo}%
\providecommand \href [0]{\begingroup \@sanitize@url \@href}%
\providecommand \@href[1]{\@@startlink{#1}\@@href}%
\providecommand \@@href[1]{\endgroup#1\@@endlink}%
\providecommand \@sanitize@url [0]{\catcode `\\12\catcode `\$12\catcode `\&12\catcode `\#12\catcode `\^12\catcode `\_12\catcode `\%12\relax}%
\providecommand \@@startlink[1]{}%
\providecommand \@@endlink[0]{}%
\providecommand \url  [0]{\begingroup\@sanitize@url \@url }%
\providecommand \@url [1]{\endgroup\@href {#1}{\urlprefix }}%
\providecommand \urlprefix  [0]{URL }%
\providecommand \Eprint [0]{\href }%
\providecommand \doibase [0]{https://doi.org/}%
\providecommand \selectlanguage [0]{\@gobble}%
\providecommand \bibinfo  [0]{\@secondoftwo}%
\providecommand \bibfield  [0]{\@secondoftwo}%
\providecommand \translation [1]{[#1]}%
\providecommand \BibitemOpen [0]{}%
\providecommand \bibitemStop [0]{}%
\providecommand \bibitemNoStop [0]{.\EOS\space}%
\providecommand \EOS [0]{\spacefactor3000\relax}%
\providecommand \BibitemShut  [1]{\csname bibitem#1\endcsname}%
\let\auto@bib@innerbib\@empty
%</preamble>
\bibitem [{\citenamefont {Fowler}\ \emph {et~al.}(2012)\citenamefont {Fowler}, \citenamefont {Mariantoni}, \citenamefont {Martinis},\ and\ \citenamefont {Cleland}}]{Fowler_2012}%
  \BibitemOpen
  \bibfield  {author} {\bibinfo {author} {\bibfnamefont {A.~G.}\ \bibnamefont {Fowler}}, \bibinfo {author} {\bibfnamefont {M.}~\bibnamefont {Mariantoni}}, \bibinfo {author} {\bibfnamefont {J.~M.}\ \bibnamefont {Martinis}},\ and\ \bibinfo {author} {\bibfnamefont {A.~N.}\ \bibnamefont {Cleland}},\ }\bibfield  {title} {\bibinfo {title} {Surface codes: Towards practical large-scale quantum computation},\ }\bibfield  {journal} {\bibinfo  {journal} {Physical Review A}\ }\textbf {\bibinfo {volume} {86}},\ \href {https://doi.org/10.1103/physreva.86.032324} {10.1103/physreva.86.032324} (\bibinfo {year} {2012})\BibitemShut {NoStop}%
\bibitem [{\citenamefont {Gidney}\ and\ \citenamefont {Eker{\aa}}(2021)}]{resource_estimation_shor_algo}%
  \BibitemOpen
  \bibfield  {author} {\bibinfo {author} {\bibfnamefont {C.}~\bibnamefont {Gidney}}\ and\ \bibinfo {author} {\bibfnamefont {M.}~\bibnamefont {Eker{\aa}}},\ }\bibfield  {title} {\bibinfo {title} {How to factor 2048 bit {RSA} integers in 8 hours using 20 million noisy qubits},\ }\href {https://doi.org/10.22331/q-2021-04-15-433} {\bibfield  {journal} {\bibinfo  {journal} {Quantum}\ }\textbf {\bibinfo {volume} {5}},\ \bibinfo {pages} {433} (\bibinfo {year} {2021})}\BibitemShut {NoStop}%
\bibitem [{\citenamefont {Bravyi}\ and\ \citenamefont {Kitaev}(1998)}]{Bravyi_Kitaev_1998}%
  \BibitemOpen
  \bibfield  {author} {\bibinfo {author} {\bibfnamefont {S.~B.}\ \bibnamefont {Bravyi}}\ and\ \bibinfo {author} {\bibfnamefont {A.~Y.}\ \bibnamefont {Kitaev}},\ }\href {https://doi.org/10.48550/ARXIV.QUANT-PH/9811052} {\bibinfo {title} {Quantum codes on a lattice with boundary}} (\bibinfo {year} {1998})\BibitemShut {NoStop}%
\bibitem [{\citenamefont {Kitaev}(1997)}]{Kitaev_1997}%
  \BibitemOpen
  \bibfield  {author} {\bibinfo {author} {\bibfnamefont {A.~Y.}\ \bibnamefont {Kitaev}},\ }\bibfield  {title} {\bibinfo {title} {Quantum computations: algorithms and error correction},\ }\href {https://doi.org/10.1070/rm1997v052n06abeh002155} {\bibfield  {journal} {\bibinfo  {journal} {Russian Mathematical Surveys}\ }\textbf {\bibinfo {volume} {52}},\ \bibinfo {pages} {1191} (\bibinfo {year} {1997})}\BibitemShut {NoStop}%
\bibitem [{\citenamefont {Li}\ \emph {et~al.}(2018)\citenamefont {Li}, \citenamefont {Petit}, \citenamefont {Franke}, \citenamefont {Dehollain}, \citenamefont {Helsen}, \citenamefont {Steudtner}, \citenamefont {Thomas}, \citenamefont {Yoscovits}, \citenamefont {Singh}, \citenamefont {Wehner}, \citenamefont {Vandersypen}, \citenamefont {Clarke},\ and\ \citenamefont {Veldhorst}}]{li_crossbar_2018}%
  \BibitemOpen
  \bibfield  {author} {\bibinfo {author} {\bibfnamefont {R.}~\bibnamefont {Li}}, \bibinfo {author} {\bibfnamefont {L.}~\bibnamefont {Petit}}, \bibinfo {author} {\bibfnamefont {D.~P.}\ \bibnamefont {Franke}}, \bibinfo {author} {\bibfnamefont {J.~P.}\ \bibnamefont {Dehollain}}, \bibinfo {author} {\bibfnamefont {J.}~\bibnamefont {Helsen}}, \bibinfo {author} {\bibfnamefont {M.}~\bibnamefont {Steudtner}}, \bibinfo {author} {\bibfnamefont {N.~K.}\ \bibnamefont {Thomas}}, \bibinfo {author} {\bibfnamefont {Z.~R.}\ \bibnamefont {Yoscovits}}, \bibinfo {author} {\bibfnamefont {K.~J.}\ \bibnamefont {Singh}}, \bibinfo {author} {\bibfnamefont {S.}~\bibnamefont {Wehner}}, \bibinfo {author} {\bibfnamefont {L.~M.~K.}\ \bibnamefont {Vandersypen}}, \bibinfo {author} {\bibfnamefont {J.~S.}\ \bibnamefont {Clarke}},\ and\ \bibinfo {author} {\bibfnamefont {M.}~\bibnamefont {Veldhorst}},\ }\bibfield  {title} {\bibinfo {title} {A crossbar network for silicon quantum dot qubits},\ }\href {https://doi.org/10.1126/sciadv.aar3960}
  {\bibfield  {journal} {\bibinfo  {journal} {Science Advances}\ }\textbf {\bibinfo {volume} {4}},\ \bibinfo {pages} {eaar3960} (\bibinfo {year} {2018})},\ \Eprint {https://arxiv.org/abs/https://www.science.org/doi/pdf/10.1126/sciadv.aar3960} {https://www.science.org/doi/pdf/10.1126/sciadv.aar3960} \BibitemShut {NoStop}%
\bibitem [{\citenamefont {Cai}\ \emph {et~al.}(2023)\citenamefont {Cai}, \citenamefont {Siegel},\ and\ \citenamefont {Benjamin}}]{cai_looped_pipelines_2023}%
  \BibitemOpen
  \bibfield  {author} {\bibinfo {author} {\bibfnamefont {Z.}~\bibnamefont {Cai}}, \bibinfo {author} {\bibfnamefont {A.}~\bibnamefont {Siegel}},\ and\ \bibinfo {author} {\bibfnamefont {S.}~\bibnamefont {Benjamin}},\ }\bibfield  {title} {\bibinfo {title} {Looped pipelines enabling effective 3d qubit lattices in a strictly 2d device},\ }\href {https://doi.org/10.1103/PRXQuantum.4.020345} {\bibfield  {journal} {\bibinfo  {journal} {PRX Quantum}\ }\textbf {\bibinfo {volume} {4}},\ \bibinfo {pages} {020345} (\bibinfo {year} {2023})}\BibitemShut {NoStop}%
\bibitem [{\citenamefont {Künne}\ \emph {et~al.}(2024)\citenamefont {Künne}, \citenamefont {Willmes}, \citenamefont {Oberländer}, \citenamefont {Gorjaew}, \citenamefont {Teske}, \citenamefont {Bhardwaj}, \citenamefont {Beer}, \citenamefont {Kammerloher}, \citenamefont {Otten}, \citenamefont {Seidler}, \citenamefont {Xue}, \citenamefont {Schreiber},\ and\ \citenamefont {Bluhm}}]{kunne_spinbus_2024}%
  \BibitemOpen
  \bibfield  {author} {\bibinfo {author} {\bibfnamefont {M.}~\bibnamefont {Künne}}, \bibinfo {author} {\bibfnamefont {A.}~\bibnamefont {Willmes}}, \bibinfo {author} {\bibfnamefont {M.}~\bibnamefont {Oberländer}}, \bibinfo {author} {\bibfnamefont {C.}~\bibnamefont {Gorjaew}}, \bibinfo {author} {\bibfnamefont {J.~D.}\ \bibnamefont {Teske}}, \bibinfo {author} {\bibfnamefont {H.}~\bibnamefont {Bhardwaj}}, \bibinfo {author} {\bibfnamefont {M.}~\bibnamefont {Beer}}, \bibinfo {author} {\bibfnamefont {E.}~\bibnamefont {Kammerloher}}, \bibinfo {author} {\bibfnamefont {R.}~\bibnamefont {Otten}}, \bibinfo {author} {\bibfnamefont {I.}~\bibnamefont {Seidler}}, \bibinfo {author} {\bibfnamefont {R.}~\bibnamefont {Xue}}, \bibinfo {author} {\bibfnamefont {L.~R.}\ \bibnamefont {Schreiber}},\ and\ \bibinfo {author} {\bibfnamefont {H.}~\bibnamefont {Bluhm}},\ }\bibfield  {title} {\bibinfo {title} {The {SpinBus} architecture for scaling spin qubits with electron shuttling},\ }\href {https://doi.org/10.1038/s41467-024-49182-4}
  {\bibfield  {journal} {\bibinfo  {journal} {Nature Communications}\ }\textbf {\bibinfo {volume} {15}},\ \bibinfo {pages} {4977} (\bibinfo {year} {2024})}\BibitemShut {NoStop}%
\bibitem [{\citenamefont {Siegel}\ \emph {et~al.}(2024)\citenamefont {Siegel}, \citenamefont {Strikis},\ and\ \citenamefont {Fogarty}}]{siegel_two_by_n_2024}%
  \BibitemOpen
  \bibfield  {author} {\bibinfo {author} {\bibfnamefont {A.}~\bibnamefont {Siegel}}, \bibinfo {author} {\bibfnamefont {A.}~\bibnamefont {Strikis}},\ and\ \bibinfo {author} {\bibfnamefont {M.}~\bibnamefont {Fogarty}},\ }\href@noop {} {\bibinfo {title} {Towards early fault tolerance on a 2$\times$n array of qubits equipped with shuttling}} (\bibinfo {year} {2024}),\ \Eprint {https://arxiv.org/abs/2402.12599} {arXiv:2402.12599 [quant-ph]} \BibitemShut {NoStop}%
\bibitem [{\citenamefont {Pataki}\ and\ \citenamefont {Pályi}(2024)}]{pataki_surface_code_crossbar_2024}%
  \BibitemOpen
  \bibfield  {author} {\bibinfo {author} {\bibfnamefont {D.}~\bibnamefont {Pataki}}\ and\ \bibinfo {author} {\bibfnamefont {A.}~\bibnamefont {Pályi}},\ }\href {https://arxiv.org/abs/2412.05425} {\bibinfo {title} {Compiling the surface code to crossbar spin qubit architectures}} (\bibinfo {year} {2024}),\ \Eprint {https://arxiv.org/abs/2412.05425} {arXiv:2412.05425 [cond-mat.mes-hall]} \BibitemShut {NoStop}%
\bibitem [{\citenamefont {Bluvstein}\ \emph {et~al.}(2023)\citenamefont {Bluvstein}, \citenamefont {Evered}, \citenamefont {Geim}, \citenamefont {Li}, \citenamefont {Zhou}, \citenamefont {Manovitz}, \citenamefont {Ebadi}, \citenamefont {Cain}, \citenamefont {Kalinowski}, \citenamefont {Hangleiter}, \citenamefont {Ataides}, \citenamefont {Maskara}, \citenamefont {Cong}, \citenamefont {Gao}, \citenamefont {Rodriguez}, \citenamefont {Karolyshyn}, \citenamefont {Semeghini}, \citenamefont {Gullans}, \citenamefont {Greiner}, \citenamefont {Vuleti\'{c}},\ and\ \citenamefont {Lukin}}]{Bluvstein_2023}%
  \BibitemOpen
  \bibfield  {author} {\bibinfo {author} {\bibfnamefont {D.}~\bibnamefont {Bluvstein}}, \bibinfo {author} {\bibfnamefont {S.~J.}\ \bibnamefont {Evered}}, \bibinfo {author} {\bibfnamefont {A.~A.}\ \bibnamefont {Geim}}, \bibinfo {author} {\bibfnamefont {S.~H.}\ \bibnamefont {Li}}, \bibinfo {author} {\bibfnamefont {H.}~\bibnamefont {Zhou}}, \bibinfo {author} {\bibfnamefont {T.}~\bibnamefont {Manovitz}}, \bibinfo {author} {\bibfnamefont {S.}~\bibnamefont {Ebadi}}, \bibinfo {author} {\bibfnamefont {M.}~\bibnamefont {Cain}}, \bibinfo {author} {\bibfnamefont {M.}~\bibnamefont {Kalinowski}}, \bibinfo {author} {\bibfnamefont {D.}~\bibnamefont {Hangleiter}}, \bibinfo {author} {\bibfnamefont {J.~P.~B.}\ \bibnamefont {Ataides}}, \bibinfo {author} {\bibfnamefont {N.}~\bibnamefont {Maskara}}, \bibinfo {author} {\bibfnamefont {I.}~\bibnamefont {Cong}}, \bibinfo {author} {\bibfnamefont {X.}~\bibnamefont {Gao}}, \bibinfo {author} {\bibfnamefont {P.~S.}\ \bibnamefont {Rodriguez}}, \bibinfo {author} {\bibfnamefont
  {T.}~\bibnamefont {Karolyshyn}}, \bibinfo {author} {\bibfnamefont {G.}~\bibnamefont {Semeghini}}, \bibinfo {author} {\bibfnamefont {M.~J.}\ \bibnamefont {Gullans}}, \bibinfo {author} {\bibfnamefont {M.}~\bibnamefont {Greiner}}, \bibinfo {author} {\bibfnamefont {V.}~\bibnamefont {Vuleti\'{c}}},\ and\ \bibinfo {author} {\bibfnamefont {M.~D.}\ \bibnamefont {Lukin}},\ }\bibfield  {title} {\bibinfo {title} {Logical quantum processor based on reconfigurable atom arrays},\ }\bibfield  {journal} {\bibinfo  {journal} {Nature}\ }\href {https://doi.org/10.1038/s41586-023-06927-3} {10.1038/s41586-023-06927-3} (\bibinfo {year} {2023})\BibitemShut {NoStop}%
\bibitem [{\citenamefont {Pino}\ \emph {et~al.}(2021)\citenamefont {Pino}, \citenamefont {Dreiling}, \citenamefont {Figgatt}, \citenamefont {Gaebler}, \citenamefont {Moses}, \citenamefont {Allman}, \citenamefont {Baldwin}, \citenamefont {Foss-Feig}, \citenamefont {Hayes}, \citenamefont {Mayer}, \citenamefont {Ryan-Anderson},\ and\ \citenamefont {Neyenhuis}}]{Pino_2021}%
  \BibitemOpen
  \bibfield  {author} {\bibinfo {author} {\bibfnamefont {J.~M.}\ \bibnamefont {Pino}}, \bibinfo {author} {\bibfnamefont {J.~M.}\ \bibnamefont {Dreiling}}, \bibinfo {author} {\bibfnamefont {C.}~\bibnamefont {Figgatt}}, \bibinfo {author} {\bibfnamefont {J.~P.}\ \bibnamefont {Gaebler}}, \bibinfo {author} {\bibfnamefont {S.~A.}\ \bibnamefont {Moses}}, \bibinfo {author} {\bibfnamefont {M.~S.}\ \bibnamefont {Allman}}, \bibinfo {author} {\bibfnamefont {C.~H.}\ \bibnamefont {Baldwin}}, \bibinfo {author} {\bibfnamefont {M.}~\bibnamefont {Foss-Feig}}, \bibinfo {author} {\bibfnamefont {D.}~\bibnamefont {Hayes}}, \bibinfo {author} {\bibfnamefont {K.}~\bibnamefont {Mayer}}, \bibinfo {author} {\bibfnamefont {C.}~\bibnamefont {Ryan-Anderson}},\ and\ \bibinfo {author} {\bibfnamefont {B.}~\bibnamefont {Neyenhuis}},\ }\bibfield  {title} {\bibinfo {title} {Demonstration of the trapped-ion quantum ccd computer architecture},\ }\href {https://doi.org/10.1038/s41586-021-03318-4} {\bibfield  {journal} {\bibinfo  {journal} {Nature}\
  }\textbf {\bibinfo {volume} {592}},\ \bibinfo {pages} {209} (\bibinfo {year} {2021})}\BibitemShut {NoStop}%
\bibitem [{\citenamefont {Langrock}\ \emph {et~al.}(2023)\citenamefont {Langrock}, \citenamefont {Krzywda}, \citenamefont {Focke}, \citenamefont {Seidler}, \citenamefont {Schreiber},\ and\ \citenamefont {Cywi\ifmmode~\acute{n}\else \'{n}\fi{}ski}}]{langrockBlueprintScalableSpin2023}%
  \BibitemOpen
  \bibfield  {author} {\bibinfo {author} {\bibfnamefont {V.}~\bibnamefont {Langrock}}, \bibinfo {author} {\bibfnamefont {J.~A.}\ \bibnamefont {Krzywda}}, \bibinfo {author} {\bibfnamefont {N.}~\bibnamefont {Focke}}, \bibinfo {author} {\bibfnamefont {I.}~\bibnamefont {Seidler}}, \bibinfo {author} {\bibfnamefont {L.~R.}\ \bibnamefont {Schreiber}},\ and\ \bibinfo {author} {\bibfnamefont {L.}~\bibnamefont {Cywi\ifmmode~\acute{n}\else \'{n}\fi{}ski}},\ }\bibfield  {title} {\bibinfo {title} {Blueprint of a scalable spin qubit shuttle device for coherent mid-range qubit transfer in disordered ${\text{si/sige/sio}}_{2}$},\ }\href {https://doi.org/10.1103/PRXQuantum.4.020305} {\bibfield  {journal} {\bibinfo  {journal} {PRX Quantum}\ }\textbf {\bibinfo {volume} {4}},\ \bibinfo {pages} {020305} (\bibinfo {year} {2023})}\BibitemShut {NoStop}%
\bibitem [{\citenamefont {Struck}\ \emph {et~al.}(2020)\citenamefont {Struck}, \citenamefont {Hollmann}, \citenamefont {Schauer}, \citenamefont {Fedorets}, \citenamefont {Schmidbauer}, \citenamefont {Sawano}, \citenamefont {Riemann}, \citenamefont {Abrosimov}, \citenamefont {Cywinski}, \citenamefont {Bougeard},\ and\ \citenamefont {Schreiber}}]{Struck_2020}%
  \BibitemOpen
  \bibfield  {author} {\bibinfo {author} {\bibfnamefont {T.}~\bibnamefont {Struck}}, \bibinfo {author} {\bibfnamefont {A.}~\bibnamefont {Hollmann}}, \bibinfo {author} {\bibfnamefont {F.}~\bibnamefont {Schauer}}, \bibinfo {author} {\bibfnamefont {O.}~\bibnamefont {Fedorets}}, \bibinfo {author} {\bibfnamefont {A.}~\bibnamefont {Schmidbauer}}, \bibinfo {author} {\bibfnamefont {K.}~\bibnamefont {Sawano}}, \bibinfo {author} {\bibfnamefont {H.}~\bibnamefont {Riemann}}, \bibinfo {author} {\bibfnamefont {N.~V.}\ \bibnamefont {Abrosimov}}, \bibinfo {author} {\bibfnamefont {L.}~\bibnamefont {Cywinski}}, \bibinfo {author} {\bibfnamefont {D.}~\bibnamefont {Bougeard}},\ and\ \bibinfo {author} {\bibfnamefont {L.~R.}\ \bibnamefont {Schreiber}},\ }\bibfield  {title} {\bibinfo {title} {Low-frequency spin qubit energy splitting noise in highly purified 28si/sige},\ }\bibfield  {journal} {\bibinfo  {journal} {npj Quantum Information}\ }\textbf {\bibinfo {volume} {6}},\ \href {https://doi.org/10.1038/s41534-020-0276-2}
  {10.1038/s41534-020-0276-2} (\bibinfo {year} {2020})\BibitemShut {NoStop}%
\bibitem [{\citenamefont {{Gonzalez-Zalba}}\ \emph {et~al.}(2021)\citenamefont {{Gonzalez-Zalba}}, \citenamefont {{de Franceschi}}, \citenamefont {Charbon}, \citenamefont {Meunier}, \citenamefont {Vinet},\ and\ \citenamefont {Dzurak}}]{gonzalez-zalbaScalingSiliconbasedQuantum2021}%
  \BibitemOpen
  \bibfield  {author} {\bibinfo {author} {\bibfnamefont {M.~F.}\ \bibnamefont {{Gonzalez-Zalba}}}, \bibinfo {author} {\bibfnamefont {S.}~\bibnamefont {{de Franceschi}}}, \bibinfo {author} {\bibfnamefont {E.}~\bibnamefont {Charbon}}, \bibinfo {author} {\bibfnamefont {T.}~\bibnamefont {Meunier}}, \bibinfo {author} {\bibfnamefont {M.}~\bibnamefont {Vinet}},\ and\ \bibinfo {author} {\bibfnamefont {A.~S.}\ \bibnamefont {Dzurak}},\ }\bibfield  {title} {\bibinfo {title} {Scaling silicon-based quantum computing using {{CMOS}} technology},\ }\href {https://doi.org/10.1038/s41928-021-00681-y} {\bibfield  {journal} {\bibinfo  {journal} {Nature Electronics}\ }\textbf {\bibinfo {volume} {4}},\ \bibinfo {pages} {872} (\bibinfo {year} {2021})}\BibitemShut {NoStop}%
\bibitem [{\citenamefont {Maurand}\ \emph {et~al.}(2016)\citenamefont {Maurand}, \citenamefont {Jehl}, \citenamefont {Kotekar-Patil}, \citenamefont {Corna}, \citenamefont {Bohuslavskyi}, \citenamefont {Lavi\'{e}ville}, \citenamefont {Hutin}, \citenamefont {Barraud}, \citenamefont {Vinet}, \citenamefont {Sanquer},\ and\ \citenamefont {De~Franceschi}}]{Maurand_2016}%
  \BibitemOpen
  \bibfield  {author} {\bibinfo {author} {\bibfnamefont {R.}~\bibnamefont {Maurand}}, \bibinfo {author} {\bibfnamefont {X.}~\bibnamefont {Jehl}}, \bibinfo {author} {\bibfnamefont {D.}~\bibnamefont {Kotekar-Patil}}, \bibinfo {author} {\bibfnamefont {A.}~\bibnamefont {Corna}}, \bibinfo {author} {\bibfnamefont {H.}~\bibnamefont {Bohuslavskyi}}, \bibinfo {author} {\bibfnamefont {R.}~\bibnamefont {Lavi\'{e}ville}}, \bibinfo {author} {\bibfnamefont {L.}~\bibnamefont {Hutin}}, \bibinfo {author} {\bibfnamefont {S.}~\bibnamefont {Barraud}}, \bibinfo {author} {\bibfnamefont {M.}~\bibnamefont {Vinet}}, \bibinfo {author} {\bibfnamefont {M.}~\bibnamefont {Sanquer}},\ and\ \bibinfo {author} {\bibfnamefont {S.}~\bibnamefont {De~Franceschi}},\ }\bibfield  {title} {\bibinfo {title} {A cmos silicon spin qubit},\ }\bibfield  {journal} {\bibinfo  {journal} {Nature Communications}\ }\textbf {\bibinfo {volume} {7}},\ \href {https://doi.org/10.1038/ncomms13575} {10.1038/ncomms13575} (\bibinfo {year} {2016})\BibitemShut {NoStop}%
\bibitem [{\citenamefont {Zwerver}\ \emph {et~al.}(2022)\citenamefont {Zwerver}, \citenamefont {Kr\"{a}henmann}, \citenamefont {Watson}, \citenamefont {Lampert}, \citenamefont {George}, \citenamefont {Pillarisetty}, \citenamefont {Bojarski}, \citenamefont {Amin}, \citenamefont {Amitonov}, \citenamefont {Boter}, \citenamefont {Caudillo}, \citenamefont {Correas-Serrano}, \citenamefont {Dehollain}, \citenamefont {Droulers}, \citenamefont {Henry}, \citenamefont {Kotlyar}, \citenamefont {Lodari}, \citenamefont {L\"{u}thi}, \citenamefont {Michalak}, \citenamefont {Mueller}, \citenamefont {Neyens}, \citenamefont {Roberts}, \citenamefont {Samkharadze}, \citenamefont {Zheng}, \citenamefont {Zietz}, \citenamefont {Scappucci}, \citenamefont {Veldhorst}, \citenamefont {Vandersypen},\ and\ \citenamefont {Clarke}}]{Zwerver_2022}%
  \BibitemOpen
  \bibfield  {author} {\bibinfo {author} {\bibfnamefont {A.~M.~J.}\ \bibnamefont {Zwerver}}, \bibinfo {author} {\bibfnamefont {T.}~\bibnamefont {Kr\"{a}henmann}}, \bibinfo {author} {\bibfnamefont {T.~F.}\ \bibnamefont {Watson}}, \bibinfo {author} {\bibfnamefont {L.}~\bibnamefont {Lampert}}, \bibinfo {author} {\bibfnamefont {H.~C.}\ \bibnamefont {George}}, \bibinfo {author} {\bibfnamefont {R.}~\bibnamefont {Pillarisetty}}, \bibinfo {author} {\bibfnamefont {S.~A.}\ \bibnamefont {Bojarski}}, \bibinfo {author} {\bibfnamefont {P.}~\bibnamefont {Amin}}, \bibinfo {author} {\bibfnamefont {S.~V.}\ \bibnamefont {Amitonov}}, \bibinfo {author} {\bibfnamefont {J.~M.}\ \bibnamefont {Boter}}, \bibinfo {author} {\bibfnamefont {R.}~\bibnamefont {Caudillo}}, \bibinfo {author} {\bibfnamefont {D.}~\bibnamefont {Correas-Serrano}}, \bibinfo {author} {\bibfnamefont {J.~P.}\ \bibnamefont {Dehollain}}, \bibinfo {author} {\bibfnamefont {G.}~\bibnamefont {Droulers}}, \bibinfo {author} {\bibfnamefont {E.~M.}\ \bibnamefont {Henry}},
  \bibinfo {author} {\bibfnamefont {R.}~\bibnamefont {Kotlyar}}, \bibinfo {author} {\bibfnamefont {M.}~\bibnamefont {Lodari}}, \bibinfo {author} {\bibfnamefont {F.}~\bibnamefont {L\"{u}thi}}, \bibinfo {author} {\bibfnamefont {D.~J.}\ \bibnamefont {Michalak}}, \bibinfo {author} {\bibfnamefont {B.~K.}\ \bibnamefont {Mueller}}, \bibinfo {author} {\bibfnamefont {S.}~\bibnamefont {Neyens}}, \bibinfo {author} {\bibfnamefont {J.}~\bibnamefont {Roberts}}, \bibinfo {author} {\bibfnamefont {N.}~\bibnamefont {Samkharadze}}, \bibinfo {author} {\bibfnamefont {G.}~\bibnamefont {Zheng}}, \bibinfo {author} {\bibfnamefont {O.~K.}\ \bibnamefont {Zietz}}, \bibinfo {author} {\bibfnamefont {G.}~\bibnamefont {Scappucci}}, \bibinfo {author} {\bibfnamefont {M.}~\bibnamefont {Veldhorst}}, \bibinfo {author} {\bibfnamefont {L.~M.~K.}\ \bibnamefont {Vandersypen}},\ and\ \bibinfo {author} {\bibfnamefont {J.~S.}\ \bibnamefont {Clarke}},\ }\bibfield  {title} {\bibinfo {title} {Qubits made by advanced semiconductor manufacturing},\ }\href
  {https://doi.org/10.1038/s41928-022-00727-9} {\bibfield  {journal} {\bibinfo  {journal} {Nature Electronics}\ }\textbf {\bibinfo {volume} {5}},\ \bibinfo {pages} {184} (\bibinfo {year} {2022})}\BibitemShut {NoStop}%
\bibitem [{\citenamefont {Xue}\ \emph {et~al.}(2021)\citenamefont {Xue}, \citenamefont {Patra}, \citenamefont {van Dijk}, \citenamefont {Samkharadze}, \citenamefont {Subramanian}, \citenamefont {Corna}, \citenamefont {Paquelet~Wuetz}, \citenamefont {Jeon}, \citenamefont {Sheikh}, \citenamefont {Juarez-Hernandez}, \citenamefont {Esparza}, \citenamefont {Rampurawala}, \citenamefont {Carlton}, \citenamefont {Ravikumar}, \citenamefont {Nieva}, \citenamefont {Kim}, \citenamefont {Lee}, \citenamefont {Sammak}, \citenamefont {Scappucci}, \citenamefont {Veldhorst}, \citenamefont {Sebastiano}, \citenamefont {Babaie}, \citenamefont {Pellerano}, \citenamefont {Charbon},\ and\ \citenamefont {Vandersypen}}]{XueXiao2021Ccco}%
  \BibitemOpen
  \bibfield  {author} {\bibinfo {author} {\bibfnamefont {X.}~\bibnamefont {Xue}}, \bibinfo {author} {\bibfnamefont {B.}~\bibnamefont {Patra}}, \bibinfo {author} {\bibfnamefont {J.~P.~G.}\ \bibnamefont {van Dijk}}, \bibinfo {author} {\bibfnamefont {N.}~\bibnamefont {Samkharadze}}, \bibinfo {author} {\bibfnamefont {S.}~\bibnamefont {Subramanian}}, \bibinfo {author} {\bibfnamefont {A.}~\bibnamefont {Corna}}, \bibinfo {author} {\bibfnamefont {B.}~\bibnamefont {Paquelet~Wuetz}}, \bibinfo {author} {\bibfnamefont {C.}~\bibnamefont {Jeon}}, \bibinfo {author} {\bibfnamefont {F.}~\bibnamefont {Sheikh}}, \bibinfo {author} {\bibfnamefont {E.}~\bibnamefont {Juarez-Hernandez}}, \bibinfo {author} {\bibfnamefont {B.~P.}\ \bibnamefont {Esparza}}, \bibinfo {author} {\bibfnamefont {H.}~\bibnamefont {Rampurawala}}, \bibinfo {author} {\bibfnamefont {B.}~\bibnamefont {Carlton}}, \bibinfo {author} {\bibfnamefont {S.}~\bibnamefont {Ravikumar}}, \bibinfo {author} {\bibfnamefont {C.}~\bibnamefont {Nieva}}, \bibinfo {author}
  {\bibfnamefont {S.}~\bibnamefont {Kim}}, \bibinfo {author} {\bibfnamefont {H.-J.}\ \bibnamefont {Lee}}, \bibinfo {author} {\bibfnamefont {A.}~\bibnamefont {Sammak}}, \bibinfo {author} {\bibfnamefont {G.}~\bibnamefont {Scappucci}}, \bibinfo {author} {\bibfnamefont {M.}~\bibnamefont {Veldhorst}}, \bibinfo {author} {\bibfnamefont {F.}~\bibnamefont {Sebastiano}}, \bibinfo {author} {\bibfnamefont {M.}~\bibnamefont {Babaie}}, \bibinfo {author} {\bibfnamefont {S.}~\bibnamefont {Pellerano}}, \bibinfo {author} {\bibfnamefont {E.}~\bibnamefont {Charbon}},\ and\ \bibinfo {author} {\bibfnamefont {L.~M.~K.}\ \bibnamefont {Vandersypen}},\ }\bibfield  {title} {\bibinfo {title} {Cmos-based cryogenic control of silicon quantum circuits},\ }\href@noop {} {\bibfield  {journal} {\bibinfo  {journal} {Nature (London)}\ }\textbf {\bibinfo {volume} {593}},\ \bibinfo {pages} {205} (\bibinfo {year} {2021})}\BibitemShut {NoStop}%
\bibitem [{\citenamefont {Ruffino}\ \emph {et~al.}(2022)\citenamefont {Ruffino}, \citenamefont {Yang}, \citenamefont {Michniewicz}, \citenamefont {Peng}, \citenamefont {Charbon},\ and\ \citenamefont {Gonzalez-Zalba}}]{ruffino2021integrated}%
  \BibitemOpen
  \bibfield  {author} {\bibinfo {author} {\bibfnamefont {A.}~\bibnamefont {Ruffino}}, \bibinfo {author} {\bibfnamefont {T.-Y.}\ \bibnamefont {Yang}}, \bibinfo {author} {\bibfnamefont {J.}~\bibnamefont {Michniewicz}}, \bibinfo {author} {\bibfnamefont {Y.}~\bibnamefont {Peng}}, \bibinfo {author} {\bibfnamefont {E.}~\bibnamefont {Charbon}},\ and\ \bibinfo {author} {\bibfnamefont {M.~F.}\ \bibnamefont {Gonzalez-Zalba}},\ }\bibfield  {title} {\bibinfo {title} {A cryo-cmos chip that integrates silicon quantum dots and multiplexed dispersive readout electronics},\ }\href {https://doi.org/10.1038/s41928-021-00687-6} {\bibfield  {journal} {\bibinfo  {journal} {Nature electronics}\ }\textbf {\bibinfo {volume} {5}},\ \bibinfo {pages} {53} (\bibinfo {year} {2022})}\BibitemShut {NoStop}%
\bibitem [{\citenamefont {Philips}\ \emph {et~al.}(2022)\citenamefont {Philips}, \citenamefont {Madzik}, \citenamefont {Amitonov}, \citenamefont {de~Snoo}, \citenamefont {Russ}, \citenamefont {Kalhor}, \citenamefont {Volk}, \citenamefont {Lawrie}, \citenamefont {Brousse}, \citenamefont {Tryputen}, \citenamefont {Wuetz}, \citenamefont {Sammak}, \citenamefont {Veldhorst}, \citenamefont {Scappucci},\ and\ \citenamefont {Vandersypen}}]{Philips_2022}%
  \BibitemOpen
  \bibfield  {author} {\bibinfo {author} {\bibfnamefont {S.~G.~J.}\ \bibnamefont {Philips}}, \bibinfo {author} {\bibfnamefont {M.~T.}\ \bibnamefont {Madzik}}, \bibinfo {author} {\bibfnamefont {S.~V.}\ \bibnamefont {Amitonov}}, \bibinfo {author} {\bibfnamefont {S.~L.}\ \bibnamefont {de~Snoo}}, \bibinfo {author} {\bibfnamefont {M.}~\bibnamefont {Russ}}, \bibinfo {author} {\bibfnamefont {N.}~\bibnamefont {Kalhor}}, \bibinfo {author} {\bibfnamefont {C.}~\bibnamefont {Volk}}, \bibinfo {author} {\bibfnamefont {W.~I.~L.}\ \bibnamefont {Lawrie}}, \bibinfo {author} {\bibfnamefont {D.}~\bibnamefont {Brousse}}, \bibinfo {author} {\bibfnamefont {L.}~\bibnamefont {Tryputen}}, \bibinfo {author} {\bibfnamefont {B.~P.}\ \bibnamefont {Wuetz}}, \bibinfo {author} {\bibfnamefont {A.}~\bibnamefont {Sammak}}, \bibinfo {author} {\bibfnamefont {M.}~\bibnamefont {Veldhorst}}, \bibinfo {author} {\bibfnamefont {G.}~\bibnamefont {Scappucci}},\ and\ \bibinfo {author} {\bibfnamefont {L.~M.~K.}\ \bibnamefont {Vandersypen}},\ }\bibfield
  {title} {\bibinfo {title} {Universal control of a six-qubit quantum processor in silicon},\ }\href {https://doi.org/10.1038/s41586-022-05117-x} {\bibfield  {journal} {\bibinfo  {journal} {Nature}\ }\textbf {\bibinfo {volume} {609}},\ \bibinfo {pages} {919} (\bibinfo {year} {2022})}\BibitemShut {NoStop}%
\bibitem [{\citenamefont {Takeda}\ \emph {et~al.}(2022)\citenamefont {Takeda}, \citenamefont {Noiri}, \citenamefont {Nakajima}, \citenamefont {Kobayashi},\ and\ \citenamefont {Tarucha}}]{Takeda_2022}%
  \BibitemOpen
  \bibfield  {author} {\bibinfo {author} {\bibfnamefont {K.}~\bibnamefont {Takeda}}, \bibinfo {author} {\bibfnamefont {A.}~\bibnamefont {Noiri}}, \bibinfo {author} {\bibfnamefont {T.}~\bibnamefont {Nakajima}}, \bibinfo {author} {\bibfnamefont {T.}~\bibnamefont {Kobayashi}},\ and\ \bibinfo {author} {\bibfnamefont {S.}~\bibnamefont {Tarucha}},\ }\bibfield  {title} {\bibinfo {title} {Quantum error correction with silicon spin qubits},\ }\href {https://doi.org/10.1038/s41586-022-04986-6} {\bibfield  {journal} {\bibinfo  {journal} {Nature}\ }\textbf {\bibinfo {volume} {608}},\ \bibinfo {pages} {682} (\bibinfo {year} {2022})}\BibitemShut {NoStop}%
\bibitem [{\citenamefont {Xue}\ \emph {et~al.}(2022)\citenamefont {Xue}, \citenamefont {Russ}, \citenamefont {Samkharadze}, \citenamefont {Undseth}, \citenamefont {Sammak}, \citenamefont {Scappucci},\ and\ \citenamefont {Vandersypen}}]{Xue_2022}%
  \BibitemOpen
  \bibfield  {author} {\bibinfo {author} {\bibfnamefont {X.}~\bibnamefont {Xue}}, \bibinfo {author} {\bibfnamefont {M.}~\bibnamefont {Russ}}, \bibinfo {author} {\bibfnamefont {N.}~\bibnamefont {Samkharadze}}, \bibinfo {author} {\bibfnamefont {B.}~\bibnamefont {Undseth}}, \bibinfo {author} {\bibfnamefont {A.}~\bibnamefont {Sammak}}, \bibinfo {author} {\bibfnamefont {G.}~\bibnamefont {Scappucci}},\ and\ \bibinfo {author} {\bibfnamefont {L.~M.~K.}\ \bibnamefont {Vandersypen}},\ }\bibfield  {title} {\bibinfo {title} {Quantum logic with spin qubits crossing the surface code threshold},\ }\href {https://doi.org/10.1038/s41586-021-04273-w} {\bibfield  {journal} {\bibinfo  {journal} {Nature}\ }\textbf {\bibinfo {volume} {601}},\ \bibinfo {pages} {343} (\bibinfo {year} {2022})}\BibitemShut {NoStop}%
\bibitem [{\citenamefont {Noiri}\ \emph {et~al.}(2022)\citenamefont {Noiri}, \citenamefont {Takeda}, \citenamefont {Nakajima}, \citenamefont {Kobayashi}, \citenamefont {Sammak}, \citenamefont {Scappucci},\ and\ \citenamefont {Tarucha}}]{Noiri_2022}%
  \BibitemOpen
  \bibfield  {author} {\bibinfo {author} {\bibfnamefont {A.}~\bibnamefont {Noiri}}, \bibinfo {author} {\bibfnamefont {K.}~\bibnamefont {Takeda}}, \bibinfo {author} {\bibfnamefont {T.}~\bibnamefont {Nakajima}}, \bibinfo {author} {\bibfnamefont {T.}~\bibnamefont {Kobayashi}}, \bibinfo {author} {\bibfnamefont {A.}~\bibnamefont {Sammak}}, \bibinfo {author} {\bibfnamefont {G.}~\bibnamefont {Scappucci}},\ and\ \bibinfo {author} {\bibfnamefont {S.}~\bibnamefont {Tarucha}},\ }\bibfield  {title} {\bibinfo {title} {Fast universal quantum gate above the fault-tolerance threshold in silicon},\ }\href {https://doi.org/10.1038/s41586-021-04182-y} {\bibfield  {journal} {\bibinfo  {journal} {Nature}\ }\textbf {\bibinfo {volume} {601}},\ \bibinfo {pages} {338} (\bibinfo {year} {2022})}\BibitemShut {NoStop}%
\bibitem [{\citenamefont {Mills}\ \emph {et~al.}(2022)\citenamefont {Mills}, \citenamefont {Guinn}, \citenamefont {Gullans}, \citenamefont {Sigillito}, \citenamefont {Feldman}, \citenamefont {Nielsen},\ and\ \citenamefont {Petta}}]{Mills_2022}%
  \BibitemOpen
  \bibfield  {author} {\bibinfo {author} {\bibfnamefont {A.~R.}\ \bibnamefont {Mills}}, \bibinfo {author} {\bibfnamefont {C.~R.}\ \bibnamefont {Guinn}}, \bibinfo {author} {\bibfnamefont {M.~J.}\ \bibnamefont {Gullans}}, \bibinfo {author} {\bibfnamefont {A.~J.}\ \bibnamefont {Sigillito}}, \bibinfo {author} {\bibfnamefont {M.~M.}\ \bibnamefont {Feldman}}, \bibinfo {author} {\bibfnamefont {E.}~\bibnamefont {Nielsen}},\ and\ \bibinfo {author} {\bibfnamefont {J.~R.}\ \bibnamefont {Petta}},\ }\bibfield  {title} {\bibinfo {title} {Two-qubit silicon quantum processor with operation fidelity exceeding $99\%$},\ }\bibfield  {journal} {\bibinfo  {journal} {Science Advances}\ }\textbf {\bibinfo {volume} {8}},\ \href {https://doi.org/10.1126/sciadv.abn5130} {10.1126/sciadv.abn5130} (\bibinfo {year} {2022})\BibitemShut {NoStop}%
\bibitem [{\citenamefont {Steinacker}\ \emph {et~al.}(2024)\citenamefont {Steinacker}, \citenamefont {Stuyck}, \citenamefont {Lim}, \citenamefont {Tanttu}, \citenamefont {Feng}, \citenamefont {Nickl}, \citenamefont {Serrano}, \citenamefont {Candido}, \citenamefont {Cifuentes}, \citenamefont {Hudson}, \citenamefont {Chan}, \citenamefont {Kubicek}, \citenamefont {Jussot}, \citenamefont {Canvel}, \citenamefont {Beyne}, \citenamefont {Shimura}, \citenamefont {Loo}, \citenamefont {Godfrin}, \citenamefont {Raes}, \citenamefont {Baudot}, \citenamefont {Wan}, \citenamefont {Laucht}, \citenamefont {Yang}, \citenamefont {Saraiva}, \citenamefont {Escott}, \citenamefont {Greve},\ and\ \citenamefont {Dzurak}}]{steinacker2024300mmfoundrysilicon}%
  \BibitemOpen
  \bibfield  {author} {\bibinfo {author} {\bibfnamefont {P.}~\bibnamefont {Steinacker}}, \bibinfo {author} {\bibfnamefont {N.~D.}\ \bibnamefont {Stuyck}}, \bibinfo {author} {\bibfnamefont {W.~H.}\ \bibnamefont {Lim}}, \bibinfo {author} {\bibfnamefont {T.}~\bibnamefont {Tanttu}}, \bibinfo {author} {\bibfnamefont {M.}~\bibnamefont {Feng}}, \bibinfo {author} {\bibfnamefont {A.}~\bibnamefont {Nickl}}, \bibinfo {author} {\bibfnamefont {S.}~\bibnamefont {Serrano}}, \bibinfo {author} {\bibfnamefont {M.}~\bibnamefont {Candido}}, \bibinfo {author} {\bibfnamefont {J.~D.}\ \bibnamefont {Cifuentes}}, \bibinfo {author} {\bibfnamefont {F.~E.}\ \bibnamefont {Hudson}}, \bibinfo {author} {\bibfnamefont {K.~W.}\ \bibnamefont {Chan}}, \bibinfo {author} {\bibfnamefont {S.}~\bibnamefont {Kubicek}}, \bibinfo {author} {\bibfnamefont {J.}~\bibnamefont {Jussot}}, \bibinfo {author} {\bibfnamefont {Y.}~\bibnamefont {Canvel}}, \bibinfo {author} {\bibfnamefont {S.}~\bibnamefont {Beyne}}, \bibinfo {author} {\bibfnamefont {Y.}~\bibnamefont
  {Shimura}}, \bibinfo {author} {\bibfnamefont {R.}~\bibnamefont {Loo}}, \bibinfo {author} {\bibfnamefont {C.}~\bibnamefont {Godfrin}}, \bibinfo {author} {\bibfnamefont {B.}~\bibnamefont {Raes}}, \bibinfo {author} {\bibfnamefont {S.}~\bibnamefont {Baudot}}, \bibinfo {author} {\bibfnamefont {D.}~\bibnamefont {Wan}}, \bibinfo {author} {\bibfnamefont {A.}~\bibnamefont {Laucht}}, \bibinfo {author} {\bibfnamefont {C.~H.}\ \bibnamefont {Yang}}, \bibinfo {author} {\bibfnamefont {A.}~\bibnamefont {Saraiva}}, \bibinfo {author} {\bibfnamefont {C.~C.}\ \bibnamefont {Escott}}, \bibinfo {author} {\bibfnamefont {K.~D.}\ \bibnamefont {Greve}},\ and\ \bibinfo {author} {\bibfnamefont {A.~S.}\ \bibnamefont {Dzurak}},\ }\bibfield  {title} {\bibinfo {title} {A 300 mm foundry silicon spin qubit unit cell exceeding 99\% fidelity in all operations},\ }\href {https://arxiv.org/abs/2410.15590} {\bibfield  {journal} {\bibinfo  {journal} {arxiv:2410.15590}\ } (\bibinfo {year} {2024})}\BibitemShut {NoStop}%
\bibitem [{\citenamefont {Yoneda}\ \emph {et~al.}(2018)\citenamefont {Yoneda}, \citenamefont {Takeda}, \citenamefont {Otsuka}, \citenamefont {Nakajima}, \citenamefont {Delbecq}, \citenamefont {Allison}, \citenamefont {Honda}, \citenamefont {Kodera}, \citenamefont {Oda}, \citenamefont {Hoshi}, \citenamefont {Usami}, \citenamefont {Itoh},\ and\ \citenamefont {Tarucha}}]{yonedaQuantumdotSpinQubit2018}%
  \BibitemOpen
  \bibfield  {author} {\bibinfo {author} {\bibfnamefont {J.}~\bibnamefont {Yoneda}}, \bibinfo {author} {\bibfnamefont {K.}~\bibnamefont {Takeda}}, \bibinfo {author} {\bibfnamefont {T.}~\bibnamefont {Otsuka}}, \bibinfo {author} {\bibfnamefont {T.}~\bibnamefont {Nakajima}}, \bibinfo {author} {\bibfnamefont {M.~R.}\ \bibnamefont {Delbecq}}, \bibinfo {author} {\bibfnamefont {G.}~\bibnamefont {Allison}}, \bibinfo {author} {\bibfnamefont {T.}~\bibnamefont {Honda}}, \bibinfo {author} {\bibfnamefont {T.}~\bibnamefont {Kodera}}, \bibinfo {author} {\bibfnamefont {S.}~\bibnamefont {Oda}}, \bibinfo {author} {\bibfnamefont {Y.}~\bibnamefont {Hoshi}}, \bibinfo {author} {\bibfnamefont {N.}~\bibnamefont {Usami}}, \bibinfo {author} {\bibfnamefont {K.~M.}\ \bibnamefont {Itoh}},\ and\ \bibinfo {author} {\bibfnamefont {S.}~\bibnamefont {Tarucha}},\ }\bibfield  {title} {\bibinfo {title} {A quantum-dot spin qubit with coherence limited by charge noise and fidelity higher than 99.9\%},\ }\href
  {https://doi.org/10.1038/s41565-017-0014-x} {\bibfield  {journal} {\bibinfo  {journal} {Nature Nanotechnology}\ }\textbf {\bibinfo {volume} {13}},\ \bibinfo {pages} {102} (\bibinfo {year} {2018})}\BibitemShut {NoStop}%
\bibitem [{\citenamefont {Hutin}\ \emph {et~al.}(2019)\citenamefont {Hutin}, \citenamefont {Bertrand}, \citenamefont {Chanrion}, \citenamefont {Bohuslavskyi}, \citenamefont {Ansaloni}, \citenamefont {Yang}, \citenamefont {Michniewicz}, \citenamefont {Niegemann}, \citenamefont {Spence}, \citenamefont {Lundberg}, \citenamefont {Chatterjee}, \citenamefont {Crippa}, \citenamefont {Li}, \citenamefont {Maurand}, \citenamefont {Jehl}, \citenamefont {Sanquer}, \citenamefont {Gonzalez-Zalba}, \citenamefont {Kuemmeth}, \citenamefont {Niquet}, \citenamefont {Franceschi}, \citenamefont {Urdampilleta}, \citenamefont {Meunier},\ and\ \citenamefont {Vinet}}]{hutin2019gate}%
  \BibitemOpen
  \bibfield  {author} {\bibinfo {author} {\bibfnamefont {L.}~\bibnamefont {Hutin}}, \bibinfo {author} {\bibfnamefont {B.}~\bibnamefont {Bertrand}}, \bibinfo {author} {\bibfnamefont {E.}~\bibnamefont {Chanrion}}, \bibinfo {author} {\bibfnamefont {H.}~\bibnamefont {Bohuslavskyi}}, \bibinfo {author} {\bibfnamefont {F.}~\bibnamefont {Ansaloni}}, \bibinfo {author} {\bibfnamefont {T.~Y.}\ \bibnamefont {Yang}}, \bibinfo {author} {\bibfnamefont {J.}~\bibnamefont {Michniewicz}}, \bibinfo {author} {\bibfnamefont {D.~J.}\ \bibnamefont {Niegemann}}, \bibinfo {author} {\bibfnamefont {C.}~\bibnamefont {Spence}}, \bibinfo {author} {\bibfnamefont {T.}~\bibnamefont {Lundberg}}, \bibinfo {author} {\bibfnamefont {A.}~\bibnamefont {Chatterjee}}, \bibinfo {author} {\bibfnamefont {A.}~\bibnamefont {Crippa}}, \bibinfo {author} {\bibfnamefont {J.}~\bibnamefont {Li}}, \bibinfo {author} {\bibfnamefont {R.}~\bibnamefont {Maurand}}, \bibinfo {author} {\bibfnamefont {X.}~\bibnamefont {Jehl}}, \bibinfo {author} {\bibfnamefont
  {M.}~\bibnamefont {Sanquer}}, \bibinfo {author} {\bibfnamefont {M.~F.}\ \bibnamefont {Gonzalez-Zalba}}, \bibinfo {author} {\bibfnamefont {F.}~\bibnamefont {Kuemmeth}}, \bibinfo {author} {\bibfnamefont {Y.~M.}\ \bibnamefont {Niquet}}, \bibinfo {author} {\bibfnamefont {S.~D.}\ \bibnamefont {Franceschi}}, \bibinfo {author} {\bibfnamefont {M.}~\bibnamefont {Urdampilleta}}, \bibinfo {author} {\bibfnamefont {T.}~\bibnamefont {Meunier}},\ and\ \bibinfo {author} {\bibfnamefont {M.}~\bibnamefont {Vinet}},\ }\href@noop {} {\bibinfo {title} {Gate reflectometry for probing charge and spin states in linear si mos split-gate arrays}} (\bibinfo {year} {2019}),\ \Eprint {https://arxiv.org/abs/1912.10884} {arXiv:1912.10884 [cond-mat.mes-hall]} \BibitemShut {NoStop}%
\bibitem [{\citenamefont {Smet}\ \emph {et~al.}(2024)\citenamefont {Smet}, \citenamefont {Matsumoto}, \citenamefont {Zwerver}, \citenamefont {Tryputen}, \citenamefont {de~Snoo}, \citenamefont {Amitonov}, \citenamefont {Sammak}, \citenamefont {Samkharadze}, \citenamefont {\"{O}nder G\"{u}l}, \citenamefont {Wasserman}, \citenamefont {Rimbach-Russ}, \citenamefont {Scappucci},\ and\ \citenamefont {Vandersypen}}]{desmet2024highfidelitysinglespinshuttlingsilicon}%
  \BibitemOpen
  \bibfield  {author} {\bibinfo {author} {\bibfnamefont {M.~D.}\ \bibnamefont {Smet}}, \bibinfo {author} {\bibfnamefont {Y.}~\bibnamefont {Matsumoto}}, \bibinfo {author} {\bibfnamefont {A.-M.~J.}\ \bibnamefont {Zwerver}}, \bibinfo {author} {\bibfnamefont {L.}~\bibnamefont {Tryputen}}, \bibinfo {author} {\bibfnamefont {S.~L.}\ \bibnamefont {de~Snoo}}, \bibinfo {author} {\bibfnamefont {S.~V.}\ \bibnamefont {Amitonov}}, \bibinfo {author} {\bibfnamefont {A.}~\bibnamefont {Sammak}}, \bibinfo {author} {\bibfnamefont {N.}~\bibnamefont {Samkharadze}}, \bibinfo {author} {\bibnamefont {\"{O}nder G\"{u}l}}, \bibinfo {author} {\bibfnamefont {R.~N.~M.}\ \bibnamefont {Wasserman}}, \bibinfo {author} {\bibfnamefont {M.}~\bibnamefont {Rimbach-Russ}}, \bibinfo {author} {\bibfnamefont {G.}~\bibnamefont {Scappucci}},\ and\ \bibinfo {author} {\bibfnamefont {L.~M.~K.}\ \bibnamefont {Vandersypen}},\ }\href {https://arxiv.org/abs/2406.07267} {\bibinfo {title} {High-fidelity single-spin shuttling in silicon}} (\bibinfo {year}
  {2024}),\ \Eprint {https://arxiv.org/abs/2406.07267} {arXiv:2406.07267 [cond-mat.mes-hall]} \BibitemShut {NoStop}%
\bibitem [{\citenamefont {Yoneda}\ \emph {et~al.}(2021)\citenamefont {Yoneda}, \citenamefont {Huang}, \citenamefont {Feng}, \citenamefont {Yang}, \citenamefont {Chan}, \citenamefont {Tanttu}, \citenamefont {Gilbert}, \citenamefont {Leon}, \citenamefont {Hudson}, \citenamefont {Itoh}, \citenamefont {Morello}, \citenamefont {Bartlett}, \citenamefont {Laucht}, \citenamefont {Saraiva},\ and\ \citenamefont {Dzurak}}]{yonedaCoherentSpinQubit2021}%
  \BibitemOpen
  \bibfield  {author} {\bibinfo {author} {\bibfnamefont {J.}~\bibnamefont {Yoneda}}, \bibinfo {author} {\bibfnamefont {W.}~\bibnamefont {Huang}}, \bibinfo {author} {\bibfnamefont {M.}~\bibnamefont {Feng}}, \bibinfo {author} {\bibfnamefont {C.~H.}\ \bibnamefont {Yang}}, \bibinfo {author} {\bibfnamefont {K.~W.}\ \bibnamefont {Chan}}, \bibinfo {author} {\bibfnamefont {T.}~\bibnamefont {Tanttu}}, \bibinfo {author} {\bibfnamefont {W.}~\bibnamefont {Gilbert}}, \bibinfo {author} {\bibfnamefont {R.~C.~C.}\ \bibnamefont {Leon}}, \bibinfo {author} {\bibfnamefont {F.~E.}\ \bibnamefont {Hudson}}, \bibinfo {author} {\bibfnamefont {K.~M.}\ \bibnamefont {Itoh}}, \bibinfo {author} {\bibfnamefont {A.}~\bibnamefont {Morello}}, \bibinfo {author} {\bibfnamefont {S.~D.}\ \bibnamefont {Bartlett}}, \bibinfo {author} {\bibfnamefont {A.}~\bibnamefont {Laucht}}, \bibinfo {author} {\bibfnamefont {A.}~\bibnamefont {Saraiva}},\ and\ \bibinfo {author} {\bibfnamefont {A.~S.}\ \bibnamefont {Dzurak}},\ }\bibfield  {title} {\bibinfo {title}
  {Coherent spin qubit transport in silicon},\ }\href {https://doi.org/10.1038/s41467-021-24371-7} {\bibfield  {journal} {\bibinfo  {journal} {Nature Communications}\ }\textbf {\bibinfo {volume} {12}},\ \bibinfo {pages} {4114} (\bibinfo {year} {2021})}\BibitemShut {NoStop}%
\bibitem [{\citenamefont {Seidler}\ \emph {et~al.}(2021)\citenamefont {Seidler}, \citenamefont {Struck}, \citenamefont {Xue}, \citenamefont {Focke}, \citenamefont {Trellenkamp}, \citenamefont {Bluhm},\ and\ \citenamefont {Schreiber}}]{seidlerConveyormodeSingleelectronShuttling2021}%
  \BibitemOpen
  \bibfield  {author} {\bibinfo {author} {\bibfnamefont {I.}~\bibnamefont {Seidler}}, \bibinfo {author} {\bibfnamefont {T.}~\bibnamefont {Struck}}, \bibinfo {author} {\bibfnamefont {R.}~\bibnamefont {Xue}}, \bibinfo {author} {\bibfnamefont {N.}~\bibnamefont {Focke}}, \bibinfo {author} {\bibfnamefont {S.}~\bibnamefont {Trellenkamp}}, \bibinfo {author} {\bibfnamefont {H.}~\bibnamefont {Bluhm}},\ and\ \bibinfo {author} {\bibfnamefont {L.~R.}\ \bibnamefont {Schreiber}},\ }\bibfield  {title} {\bibinfo {title} {Conveyor-mode single-electron shuttling in {{Si}}/{{SiGe}} for a scalable quantum computing architecture},\ }\href {http://arxiv.org/abs/2108.00879} {\bibfield  {journal} {\bibinfo  {journal} {arXiv:2108.00879 [cond-mat, physics:quant-ph]}\ } (\bibinfo {year} {2021})}\BibitemShut {NoStop}%
\bibitem [{\citenamefont {Maune}\ \emph {et~al.}(2012)\citenamefont {Maune}, \citenamefont {Borselli}, \citenamefont {Huang}, \citenamefont {Ladd}, \citenamefont {Deelman}, \citenamefont {Holabird}, \citenamefont {Kiselev}, \citenamefont {Alvarado-Rodriguez}, \citenamefont {Ross}, \citenamefont {Schmitz}, \citenamefont {Sokolich}, \citenamefont {Watson}, \citenamefont {Gyure},\ and\ \citenamefont {Hunter}}]{MAUNEB.M2012Csoi}%
  \BibitemOpen
  \bibfield  {author} {\bibinfo {author} {\bibfnamefont {B.~M.}\ \bibnamefont {Maune}}, \bibinfo {author} {\bibfnamefont {M.~G.}\ \bibnamefont {Borselli}}, \bibinfo {author} {\bibfnamefont {B.}~\bibnamefont {Huang}}, \bibinfo {author} {\bibfnamefont {T.~D.}\ \bibnamefont {Ladd}}, \bibinfo {author} {\bibfnamefont {P.~W.}\ \bibnamefont {Deelman}}, \bibinfo {author} {\bibfnamefont {K.~S.}\ \bibnamefont {Holabird}}, \bibinfo {author} {\bibfnamefont {A.~A.}\ \bibnamefont {Kiselev}}, \bibinfo {author} {\bibfnamefont {I.}~\bibnamefont {Alvarado-Rodriguez}}, \bibinfo {author} {\bibfnamefont {R.~S.}\ \bibnamefont {Ross}}, \bibinfo {author} {\bibfnamefont {A.~E.}\ \bibnamefont {Schmitz}}, \bibinfo {author} {\bibfnamefont {M.}~\bibnamefont {Sokolich}}, \bibinfo {author} {\bibfnamefont {C.~A.}\ \bibnamefont {Watson}}, \bibinfo {author} {\bibfnamefont {M.~F.}\ \bibnamefont {Gyure}},\ and\ \bibinfo {author} {\bibfnamefont {A.~T.}\ \bibnamefont {Hunter}},\ }\bibfield  {title} {\bibinfo {title} {Coherent singlet-triplet
  oscillations in a silicon-based double quantum dot},\ }\href {https://api.semanticscholar.org/CorpusID:4385331} {\bibfield  {journal} {\bibinfo  {journal} {Nature}\ }\textbf {\bibinfo {volume} {481}},\ \bibinfo {pages} {344} (\bibinfo {year} {2012})}\BibitemShut {NoStop}%
\bibitem [{\citenamefont {Zheng}\ \emph {et~al.}(2019)\citenamefont {Zheng}, \citenamefont {Samkharadze}, \citenamefont {Noordam}, \citenamefont {Kalhor}, \citenamefont {Brousse}, \citenamefont {Sammak}, \citenamefont {Scappucci},\ and\ \citenamefont {Vandersypen}}]{Zheng_2019}%
  \BibitemOpen
  \bibfield  {author} {\bibinfo {author} {\bibfnamefont {G.}~\bibnamefont {Zheng}}, \bibinfo {author} {\bibfnamefont {N.}~\bibnamefont {Samkharadze}}, \bibinfo {author} {\bibfnamefont {M.~L.}\ \bibnamefont {Noordam}}, \bibinfo {author} {\bibfnamefont {N.}~\bibnamefont {Kalhor}}, \bibinfo {author} {\bibfnamefont {D.}~\bibnamefont {Brousse}}, \bibinfo {author} {\bibfnamefont {A.}~\bibnamefont {Sammak}}, \bibinfo {author} {\bibfnamefont {G.}~\bibnamefont {Scappucci}},\ and\ \bibinfo {author} {\bibfnamefont {L.~M.~K.}\ \bibnamefont {Vandersypen}},\ }\bibfield  {title} {\bibinfo {title} {Rapid gate-based spin read-out in silicon using an on-chip resonator},\ }\href {https://doi.org/10.1038/s41565-019-0488-9} {\bibfield  {journal} {\bibinfo  {journal} {Nature Nanotechnology}\ }\textbf {\bibinfo {volume} {14}},\ \bibinfo {pages} {742} (\bibinfo {year} {2019})}\BibitemShut {NoStop}%
\bibitem [{\citenamefont {Takeda}\ \emph {et~al.}(2024)\citenamefont {Takeda}, \citenamefont {Noiri}, \citenamefont {Nakajima}, \citenamefont {Camenzind}, \citenamefont {Kobayashi}, \citenamefont {Sammak}, \citenamefont {Scappucci},\ and\ \citenamefont {Tarucha}}]{Takeda_2024}%
  \BibitemOpen
  \bibfield  {author} {\bibinfo {author} {\bibfnamefont {K.}~\bibnamefont {Takeda}}, \bibinfo {author} {\bibfnamefont {A.}~\bibnamefont {Noiri}}, \bibinfo {author} {\bibfnamefont {T.}~\bibnamefont {Nakajima}}, \bibinfo {author} {\bibfnamefont {L.~C.}\ \bibnamefont {Camenzind}}, \bibinfo {author} {\bibfnamefont {T.}~\bibnamefont {Kobayashi}}, \bibinfo {author} {\bibfnamefont {A.}~\bibnamefont {Sammak}}, \bibinfo {author} {\bibfnamefont {G.}~\bibnamefont {Scappucci}},\ and\ \bibinfo {author} {\bibfnamefont {S.}~\bibnamefont {Tarucha}},\ }\bibfield  {title} {\bibinfo {title} {Rapid single-shot parity spin readout in a silicon double quantum dot with fidelity exceeding $99\%$},\ }\href {https://doi.org/https://doi.org/10.1038/s41534-024-00813-0} {\bibfield  {journal} {\bibinfo  {journal} {npj quantum information}\ }\textbf {\bibinfo {volume} {10}},\ \bibinfo {pages} {22} (\bibinfo {year} {2024})}\BibitemShut {NoStop}%
\bibitem [{\citenamefont {Gidney}\ and\ \citenamefont {Ekerå}(2021)}]{Gidney_2021}%
  \BibitemOpen
  \bibfield  {author} {\bibinfo {author} {\bibfnamefont {C.}~\bibnamefont {Gidney}}\ and\ \bibinfo {author} {\bibfnamefont {M.}~\bibnamefont {Ekerå}},\ }\bibfield  {title} {\bibinfo {title} {How to factor 2048 bit rsa integers in 8 hours using 20 million noisy qubits},\ }\href {https://doi.org/10.22331/q-2021-04-15-433} {\bibfield  {journal} {\bibinfo  {journal} {Quantum}\ }\textbf {\bibinfo {volume} {5}},\ \bibinfo {pages} {433} (\bibinfo {year} {2021})}\BibitemShut {NoStop}%
\bibitem [{\citenamefont {Stano}\ and\ \citenamefont {Loss}(2022)}]{stano_review_2022}%
  \BibitemOpen
  \bibfield  {author} {\bibinfo {author} {\bibfnamefont {P.}~\bibnamefont {Stano}}\ and\ \bibinfo {author} {\bibfnamefont {D.}~\bibnamefont {Loss}},\ }\bibfield  {title} {\bibinfo {title} {Review of performance metrics of spin qubits in gated semiconducting nanostructures},\ }\href {https://doi.org/10.1038/s42254-022-00484-w} {\bibfield  {journal} {\bibinfo  {journal} {Nature Reviews Physics}\ }\textbf {\bibinfo {volume} {4}},\ \bibinfo {pages} {672} (\bibinfo {year} {2022})}\BibitemShut {NoStop}%
\bibitem [{\citenamefont {K{\c e}pa}\ \emph {et~al.}(2023)\citenamefont {K{\c e}pa}, \citenamefont {Focke}, \citenamefont {Cywi{\'n}ski},\ and\ \citenamefont {Krzywda}}]{kepa_2023}%
  \BibitemOpen
  \bibfield  {author} {\bibinfo {author} {\bibfnamefont {M.}~\bibnamefont {K{\c e}pa}}, \bibinfo {author} {\bibfnamefont {N.}~\bibnamefont {Focke}}, \bibinfo {author} {\bibfnamefont {{\L}.}~\bibnamefont {Cywi{\'n}ski}},\ and\ \bibinfo {author} {\bibfnamefont {J.~A.}\ \bibnamefont {Krzywda}},\ }\bibfield  {title} {\bibinfo {title} {{Simulation of 1/f charge noise affecting a quantum dot in a Si/SiGe structure}},\ }\href {https://doi.org/10.1063/5.0151029} {\bibfield  {journal} {\bibinfo  {journal} {Applied Physics Letters}\ }\textbf {\bibinfo {volume} {123}},\ \bibinfo {pages} {034005} (\bibinfo {year} {2023})}\BibitemShut {NoStop}%
\bibitem [{\citenamefont {Elsayed}\ \emph {et~al.}(2024)\citenamefont {Elsayed}, \citenamefont {Shehata}, \citenamefont {Godfrin}, \citenamefont {Kubicek}, \citenamefont {Massar}, \citenamefont {Canvel}, \citenamefont {Jussot}, \citenamefont {Simion}, \citenamefont {Mongillo}, \citenamefont {Wan}, \citenamefont {Govoreanu}, \citenamefont {Radu}, \citenamefont {Li}, \citenamefont {Van~Dorpe},\ and\ \citenamefont {De~Greve}}]{elsayed_low_2024}%
  \BibitemOpen
  \bibfield  {author} {\bibinfo {author} {\bibfnamefont {A.}~\bibnamefont {Elsayed}}, \bibinfo {author} {\bibfnamefont {M.~M.~K.}\ \bibnamefont {Shehata}}, \bibinfo {author} {\bibfnamefont {C.}~\bibnamefont {Godfrin}}, \bibinfo {author} {\bibfnamefont {S.}~\bibnamefont {Kubicek}}, \bibinfo {author} {\bibfnamefont {S.}~\bibnamefont {Massar}}, \bibinfo {author} {\bibfnamefont {Y.}~\bibnamefont {Canvel}}, \bibinfo {author} {\bibfnamefont {J.}~\bibnamefont {Jussot}}, \bibinfo {author} {\bibfnamefont {G.}~\bibnamefont {Simion}}, \bibinfo {author} {\bibfnamefont {M.}~\bibnamefont {Mongillo}}, \bibinfo {author} {\bibfnamefont {D.}~\bibnamefont {Wan}}, \bibinfo {author} {\bibfnamefont {B.}~\bibnamefont {Govoreanu}}, \bibinfo {author} {\bibfnamefont {I.~P.}\ \bibnamefont {Radu}}, \bibinfo {author} {\bibfnamefont {R.}~\bibnamefont {Li}}, \bibinfo {author} {\bibfnamefont {P.}~\bibnamefont {Van~Dorpe}},\ and\ \bibinfo {author} {\bibfnamefont {K.}~\bibnamefont {De~Greve}},\ }\bibfield  {title} {\bibinfo {title} {Low
  charge noise quantum dots with industrial {CMOS} manufacturing},\ }\href {https://doi.org/10.1038/s41534-024-00864-3} {\bibfield  {journal} {\bibinfo  {journal} {npj Quantum Information}\ }\textbf {\bibinfo {volume} {10}},\ \bibinfo {pages} {1} (\bibinfo {year} {2024})}\BibitemShut {NoStop}%
\bibitem [{\citenamefont {Yoneda}\ \emph {et~al.}(2023)\citenamefont {Yoneda}, \citenamefont {Rojas-Arias}, \citenamefont {Stano}, \citenamefont {Takeda}, \citenamefont {Noiri}, \citenamefont {Nakajima}, \citenamefont {Loss},\ and\ \citenamefont {Tarucha}}]{yoneda_noise_correlation_2023}%
  \BibitemOpen
  \bibfield  {author} {\bibinfo {author} {\bibfnamefont {J.}~\bibnamefont {Yoneda}}, \bibinfo {author} {\bibfnamefont {J.~S.}\ \bibnamefont {Rojas-Arias}}, \bibinfo {author} {\bibfnamefont {P.}~\bibnamefont {Stano}}, \bibinfo {author} {\bibfnamefont {K.}~\bibnamefont {Takeda}}, \bibinfo {author} {\bibfnamefont {A.}~\bibnamefont {Noiri}}, \bibinfo {author} {\bibfnamefont {T.}~\bibnamefont {Nakajima}}, \bibinfo {author} {\bibfnamefont {D.}~\bibnamefont {Loss}},\ and\ \bibinfo {author} {\bibfnamefont {S.}~\bibnamefont {Tarucha}},\ }\bibfield  {title} {\bibinfo {title} {Noise-correlation spectrum for a pair of spin qubits in silicon},\ }\href {https://doi.org/10.1038/s41567-023-02238-6} {\bibfield  {journal} {\bibinfo  {journal} {Nature Physics}\ }\textbf {\bibinfo {volume} {19}},\ \bibinfo {pages} {1793} (\bibinfo {year} {2023})}\BibitemShut {NoStop}%
\bibitem [{\citenamefont {Jnane}\ and\ \citenamefont {Benjamin}(2024)}]{jnane_ab_initio_2024}%
  \BibitemOpen
  \bibfield  {author} {\bibinfo {author} {\bibfnamefont {H.}~\bibnamefont {Jnane}}\ and\ \bibinfo {author} {\bibfnamefont {S.~C.}\ \bibnamefont {Benjamin}},\ }\href {https://arxiv.org/abs/2403.00191} {\bibinfo {title} {Ab initio modelling of quantum dot qubits: Coupling, gate dynamics and robustness versus charge noise}} (\bibinfo {year} {2024}),\ \Eprint {https://arxiv.org/abs/2403.00191} {arXiv:2403.00191 [cond-mat.mes-hall]} \BibitemShut {NoStop}%
\bibitem [{\citenamefont {Shehata}\ \emph {et~al.}(2023)\citenamefont {Shehata}, \citenamefont {Simion}, \citenamefont {Li}, \citenamefont {Mohiyaddin}, \citenamefont {Wan}, \citenamefont {Mongillo}, \citenamefont {Govoreanu}, \citenamefont {Radu}, \citenamefont {De~Greve},\ and\ \citenamefont {Van~Dorpe}}]{shehata_2023}%
  \BibitemOpen
  \bibfield  {author} {\bibinfo {author} {\bibfnamefont {M.~M. E.~K.}\ \bibnamefont {Shehata}}, \bibinfo {author} {\bibfnamefont {G.}~\bibnamefont {Simion}}, \bibinfo {author} {\bibfnamefont {R.}~\bibnamefont {Li}}, \bibinfo {author} {\bibfnamefont {F.~A.}\ \bibnamefont {Mohiyaddin}}, \bibinfo {author} {\bibfnamefont {D.}~\bibnamefont {Wan}}, \bibinfo {author} {\bibfnamefont {M.}~\bibnamefont {Mongillo}}, \bibinfo {author} {\bibfnamefont {B.}~\bibnamefont {Govoreanu}}, \bibinfo {author} {\bibfnamefont {I.}~\bibnamefont {Radu}}, \bibinfo {author} {\bibfnamefont {K.}~\bibnamefont {De~Greve}},\ and\ \bibinfo {author} {\bibfnamefont {P.}~\bibnamefont {Van~Dorpe}},\ }\bibfield  {title} {\bibinfo {title} {Modeling semiconductor spin qubits and their charge noise environment for quantum gate fidelity estimation},\ }\href {https://doi.org/10.1103/PhysRevB.108.045305} {\bibfield  {journal} {\bibinfo  {journal} {Phys. Rev. B}\ }\textbf {\bibinfo {volume} {108}},\ \bibinfo {pages} {045305} (\bibinfo {year}
  {2023})}\BibitemShut {NoStop}%
\bibitem [{\citenamefont {Culcer}\ \emph {et~al.}(2009)\citenamefont {Culcer}, \citenamefont {Hu},\ and\ \citenamefont {Das~Sarma}}]{culcer_2009}%
  \BibitemOpen
  \bibfield  {author} {\bibinfo {author} {\bibfnamefont {D.}~\bibnamefont {Culcer}}, \bibinfo {author} {\bibfnamefont {X.}~\bibnamefont {Hu}},\ and\ \bibinfo {author} {\bibfnamefont {S.}~\bibnamefont {Das~Sarma}},\ }\bibfield  {title} {\bibinfo {title} {{Dephasing of Si spin qubits due to charge noise}},\ }\href {https://doi.org/10.1063/1.3194778} {\bibfield  {journal} {\bibinfo  {journal} {Applied Physics Letters}\ }\textbf {\bibinfo {volume} {95}},\ \bibinfo {pages} {073102} (\bibinfo {year} {2009})},\ \Eprint {https://arxiv.org/abs/https://pubs.aip.org/aip/apl/article-pdf/doi/10.1063/1.3194778/14105785/073102\_1\_online.pdf} {https://pubs.aip.org/aip/apl/article-pdf/doi/10.1063/1.3194778/14105785/073102\_1\_online.pdf} \BibitemShut {NoStop}%
\bibitem [{\citenamefont {Rojas-Arias}\ \emph {et~al.}(2023)\citenamefont {Rojas-Arias}, \citenamefont {Noiri}, \citenamefont {Stano}, \citenamefont {Nakajima}, \citenamefont {Yoneda}, \citenamefont {Takeda}, \citenamefont {Kobayashi}, \citenamefont {Sammak}, \citenamefont {Scappucci}, \citenamefont {Loss},\ and\ \citenamefont {Tarucha}}]{rojas_spatial_correlations_2023}%
  \BibitemOpen
  \bibfield  {author} {\bibinfo {author} {\bibfnamefont {J.}~\bibnamefont {Rojas-Arias}}, \bibinfo {author} {\bibfnamefont {A.}~\bibnamefont {Noiri}}, \bibinfo {author} {\bibfnamefont {P.}~\bibnamefont {Stano}}, \bibinfo {author} {\bibfnamefont {T.}~\bibnamefont {Nakajima}}, \bibinfo {author} {\bibfnamefont {J.}~\bibnamefont {Yoneda}}, \bibinfo {author} {\bibfnamefont {K.}~\bibnamefont {Takeda}}, \bibinfo {author} {\bibfnamefont {T.}~\bibnamefont {Kobayashi}}, \bibinfo {author} {\bibfnamefont {A.}~\bibnamefont {Sammak}}, \bibinfo {author} {\bibfnamefont {G.}~\bibnamefont {Scappucci}}, \bibinfo {author} {\bibfnamefont {D.}~\bibnamefont {Loss}},\ and\ \bibinfo {author} {\bibfnamefont {S.}~\bibnamefont {Tarucha}},\ }\bibfield  {title} {\bibinfo {title} {Spatial noise correlations beyond nearest neighbors in ${}^{28}\mathrm{Si}/$si-ge spin qubits},\ }\href {https://doi.org/10.1103/PhysRevApplied.20.054024} {\bibfield  {journal} {\bibinfo  {journal} {Phys. Rev. Appl.}\ }\textbf {\bibinfo {volume} {20}},\ \bibinfo
  {pages} {054024} (\bibinfo {year} {2023})}\BibitemShut {NoStop}%
\bibitem [{\citenamefont {Mehmandoost}\ and\ \citenamefont {Dobrovitski}(2024)}]{mehmandoost_decoherence_2024}%
  \BibitemOpen
  \bibfield  {author} {\bibinfo {author} {\bibfnamefont {M.}~\bibnamefont {Mehmandoost}}\ and\ \bibinfo {author} {\bibfnamefont {V.~V.}\ \bibnamefont {Dobrovitski}},\ }\bibfield  {title} {\bibinfo {title} {Decoherence induced by a sparse bath of two-level fluctuators: Peculiar features of $1/f$ noise in high-quality qubits},\ }\href {https://doi.org/10.1103/PhysRevResearch.6.033175} {\bibfield  {journal} {\bibinfo  {journal} {Phys. Rev. Res.}\ }\textbf {\bibinfo {volume} {6}},\ \bibinfo {pages} {033175} (\bibinfo {year} {2024})}\BibitemShut {NoStop}%
\bibitem [{\citenamefont {Loss}\ and\ \citenamefont {DiVincenzo}(1998)}]{lossQuantumComputationQuantum1998}%
  \BibitemOpen
  \bibfield  {author} {\bibinfo {author} {\bibfnamefont {D.}~\bibnamefont {Loss}}\ and\ \bibinfo {author} {\bibfnamefont {D.~P.}\ \bibnamefont {DiVincenzo}},\ }\bibfield  {title} {\bibinfo {title} {Quantum computation with quantum dots},\ }\href {https://doi.org/10.1103/PhysRevA.57.120} {\bibfield  {journal} {\bibinfo  {journal} {Physical Review A}\ }\textbf {\bibinfo {volume} {57}},\ \bibinfo {pages} {120} (\bibinfo {year} {1998})}\BibitemShut {NoStop}%
\bibitem [{\citenamefont {Burkard}\ \emph {et~al.}(2021)\citenamefont {Burkard}, \citenamefont {Ladd}, \citenamefont {Nichol}, \citenamefont {Pan},\ and\ \citenamefont {Petta}}]{burkardSemiconductorSpinQubits2021}%
  \BibitemOpen
  \bibfield  {author} {\bibinfo {author} {\bibfnamefont {G.}~\bibnamefont {Burkard}}, \bibinfo {author} {\bibfnamefont {T.~D.}\ \bibnamefont {Ladd}}, \bibinfo {author} {\bibfnamefont {J.~M.}\ \bibnamefont {Nichol}}, \bibinfo {author} {\bibfnamefont {A.}~\bibnamefont {Pan}},\ and\ \bibinfo {author} {\bibfnamefont {J.~R.}\ \bibnamefont {Petta}},\ }\bibfield  {title} {\bibinfo {title} {Semiconductor {{Spin Qubits}}},\ }\href {http://arxiv.org/abs/2112.08863} {\bibfield  {journal} {\bibinfo  {journal} {arXiv:2112.08863 [cond-mat, physics:physics, physics:quant-ph]}\ } (\bibinfo {year} {2021})}\BibitemShut {NoStop}%
\bibitem [{\citenamefont {Levy}(2002)}]{levy_st_2002}%
  \BibitemOpen
  \bibfield  {author} {\bibinfo {author} {\bibfnamefont {J.}~\bibnamefont {Levy}},\ }\bibfield  {title} {\bibinfo {title} {Universal quantum computation with spin-$1/2$ pairs and heisenberg exchange},\ }\href {https://doi.org/10.1103/PhysRevLett.89.147902} {\bibfield  {journal} {\bibinfo  {journal} {Phys. Rev. Lett.}\ }\textbf {\bibinfo {volume} {89}},\ \bibinfo {pages} {147902} (\bibinfo {year} {2002})}\BibitemShut {NoStop}%
\bibitem [{\citenamefont {DiVincenzo}\ \emph {et~al.}(2000)\citenamefont {DiVincenzo}, \citenamefont {Bacon}, \citenamefont {Kempe}, \citenamefont {Burkard},\ and\ \citenamefont {Whaley}}]{divincenzo_universal_2000}%
  \BibitemOpen
  \bibfield  {author} {\bibinfo {author} {\bibfnamefont {D.~P.}\ \bibnamefont {DiVincenzo}}, \bibinfo {author} {\bibfnamefont {D.}~\bibnamefont {Bacon}}, \bibinfo {author} {\bibfnamefont {J.}~\bibnamefont {Kempe}}, \bibinfo {author} {\bibfnamefont {G.}~\bibnamefont {Burkard}},\ and\ \bibinfo {author} {\bibfnamefont {K.~B.}\ \bibnamefont {Whaley}},\ }\bibfield  {title} {\bibinfo {title} {Universal quantum computation with the exchange interaction},\ }\href {https://doi.org/10.1038/35042541} {\bibfield  {journal} {\bibinfo  {journal} {Nature}\ }\textbf {\bibinfo {volume} {408}},\ \bibinfo {pages} {339} (\bibinfo {year} {2000})}\BibitemShut {NoStop}%
\bibitem [{\citenamefont {Mokeev}\ \emph {et~al.}(2024)\citenamefont {Mokeev}, \citenamefont {Zhang},\ and\ \citenamefont {Dobrovitski}}]{mokeev2024modelingdecoherencefidelityenhancement}%
  \BibitemOpen
  \bibfield  {author} {\bibinfo {author} {\bibfnamefont {A.~S.}\ \bibnamefont {Mokeev}}, \bibinfo {author} {\bibfnamefont {Y.-N.}\ \bibnamefont {Zhang}},\ and\ \bibinfo {author} {\bibfnamefont {V.~V.}\ \bibnamefont {Dobrovitski}},\ }\href {https://arxiv.org/abs/2409.04404} {\bibinfo {title} {Modeling of decoherence and fidelity enhancement during transport of entangled qubits}} (\bibinfo {year} {2024}),\ \Eprint {https://arxiv.org/abs/2409.04404} {arXiv:2409.04404 [cond-mat.mes-hall]} \BibitemShut {NoStop}%
\bibitem [{\citenamefont {Horsman}\ \emph {et~al.}(2012)\citenamefont {Horsman}, \citenamefont {Fowler}, \citenamefont {Devitt},\ and\ \citenamefont {Meter}}]{Horsman_2012}%
  \BibitemOpen
  \bibfield  {author} {\bibinfo {author} {\bibfnamefont {D.}~\bibnamefont {Horsman}}, \bibinfo {author} {\bibfnamefont {A.~G.}\ \bibnamefont {Fowler}}, \bibinfo {author} {\bibfnamefont {S.}~\bibnamefont {Devitt}},\ and\ \bibinfo {author} {\bibfnamefont {R.~V.}\ \bibnamefont {Meter}},\ }\bibfield  {title} {\bibinfo {title} {Surface code quantum computing by lattice surgery},\ }\href {https://doi.org/10.1088/1367-2630/14/12/123011} {\bibfield  {journal} {\bibinfo  {journal} {New Journal of Physics}\ }\textbf {\bibinfo {volume} {14}},\ \bibinfo {pages} {123011} (\bibinfo {year} {2012})}\BibitemShut {NoStop}%
\bibitem [{\citenamefont {Gokhale}\ \emph {et~al.}(2021)\citenamefont {Gokhale}, \citenamefont {Koretsky}, \citenamefont {Huang}, \citenamefont {Majumder}, \citenamefont {Drucker}, \citenamefont {Brown},\ and\ \citenamefont {Chong}}]{fan_out}%
  \BibitemOpen
  \bibfield  {author} {\bibinfo {author} {\bibfnamefont {P.}~\bibnamefont {Gokhale}}, \bibinfo {author} {\bibfnamefont {S.}~\bibnamefont {Koretsky}}, \bibinfo {author} {\bibfnamefont {S.}~\bibnamefont {Huang}}, \bibinfo {author} {\bibfnamefont {S.}~\bibnamefont {Majumder}}, \bibinfo {author} {\bibfnamefont {A.}~\bibnamefont {Drucker}}, \bibinfo {author} {\bibfnamefont {K.~R.}\ \bibnamefont {Brown}},\ and\ \bibinfo {author} {\bibfnamefont {F.~T.}\ \bibnamefont {Chong}},\ }\bibfield  {title} {\bibinfo {title} {Quantum fan-out: Circuit optimizations and technology modeling},\ }in\ \href {https://doi.org/10.1109/QCE52317.2021.00045} {\emph {\bibinfo {booktitle} {2021 IEEE International Conference on Quantum Computing and Engineering (QCE)}}}\ (\bibinfo {year} {2021})\ pp.\ \bibinfo {pages} {276--290}\BibitemShut {NoStop}%
\bibitem [{\citenamefont {Gidney}\ \emph {et~al.}(2023)\citenamefont {Gidney}, \citenamefont {Newman}, \citenamefont {Brooks},\ and\ \citenamefont {Jones}}]{gidney2023yokedsurfacecodes}%
  \BibitemOpen
  \bibfield  {author} {\bibinfo {author} {\bibfnamefont {C.}~\bibnamefont {Gidney}}, \bibinfo {author} {\bibfnamefont {M.}~\bibnamefont {Newman}}, \bibinfo {author} {\bibfnamefont {P.}~\bibnamefont {Brooks}},\ and\ \bibinfo {author} {\bibfnamefont {C.}~\bibnamefont {Jones}},\ }\href {https://arxiv.org/abs/2312.04522} {\bibinfo {title} {Yoked surface codes}} (\bibinfo {year} {2023}),\ \Eprint {https://arxiv.org/abs/2312.04522} {arXiv:2312.04522 [quant-ph]} \BibitemShut {NoStop}%
\bibitem [{\citenamefont {Wallman}\ and\ \citenamefont {Emerson}(2016)}]{wallmanNoiseTailoringScalable2016}%
  \BibitemOpen
  \bibfield  {author} {\bibinfo {author} {\bibfnamefont {J.~J.}\ \bibnamefont {Wallman}}\ and\ \bibinfo {author} {\bibfnamefont {J.}~\bibnamefont {Emerson}},\ }\bibfield  {title} {\bibinfo {title} {Noise tailoring for scalable quantum computation via randomized compiling},\ }\href {https://doi.org/10.1103/PhysRevA.94.052325} {\bibfield  {journal} {\bibinfo  {journal} {Physical Review A}\ }\textbf {\bibinfo {volume} {94}},\ \bibinfo {pages} {052325} (\bibinfo {year} {2016})}\BibitemShut {NoStop}%
\bibitem [{\citenamefont {T{\'o}th}\ and\ \citenamefont {Apellaniz}(2014)}]{toth2014quantum}%
  \BibitemOpen
  \bibfield  {author} {\bibinfo {author} {\bibfnamefont {G.}~\bibnamefont {T{\'o}th}}\ and\ \bibinfo {author} {\bibfnamefont {I.}~\bibnamefont {Apellaniz}},\ }\bibfield  {title} {\bibinfo {title} {Quantum metrology from a quantum information science perspective},\ }\href@noop {} {\bibfield  {journal} {\bibinfo  {journal} {Journal of Physics A: Mathematical and Theoretical}\ }\textbf {\bibinfo {volume} {47}},\ \bibinfo {pages} {424006} (\bibinfo {year} {2014})}\BibitemShut {NoStop}%
\bibitem [{\citenamefont {Koczor}\ \emph {et~al.}(2020)\citenamefont {Koczor}, \citenamefont {Endo}, \citenamefont {Jones}, \citenamefont {Matsuzaki},\ and\ \citenamefont {Benjamin}}]{koczor2020variational}%
  \BibitemOpen
  \bibfield  {author} {\bibinfo {author} {\bibfnamefont {B.}~\bibnamefont {Koczor}}, \bibinfo {author} {\bibfnamefont {S.}~\bibnamefont {Endo}}, \bibinfo {author} {\bibfnamefont {T.}~\bibnamefont {Jones}}, \bibinfo {author} {\bibfnamefont {Y.}~\bibnamefont {Matsuzaki}},\ and\ \bibinfo {author} {\bibfnamefont {S.~C.}\ \bibnamefont {Benjamin}},\ }\bibfield  {title} {\bibinfo {title} {Variational-state quantum metrology},\ }\href@noop {} {\bibfield  {journal} {\bibinfo  {journal} {New Journal of Physics}\ }\textbf {\bibinfo {volume} {22}},\ \bibinfo {pages} {083038} (\bibinfo {year} {2020})}\BibitemShut {NoStop}%
\end{thebibliography}%

\appendix

\section{Alternative latticework structures} \label{app:alternative_latticework}

\begin{figure}
    \centering
    \includegraphics[width=\linewidth]{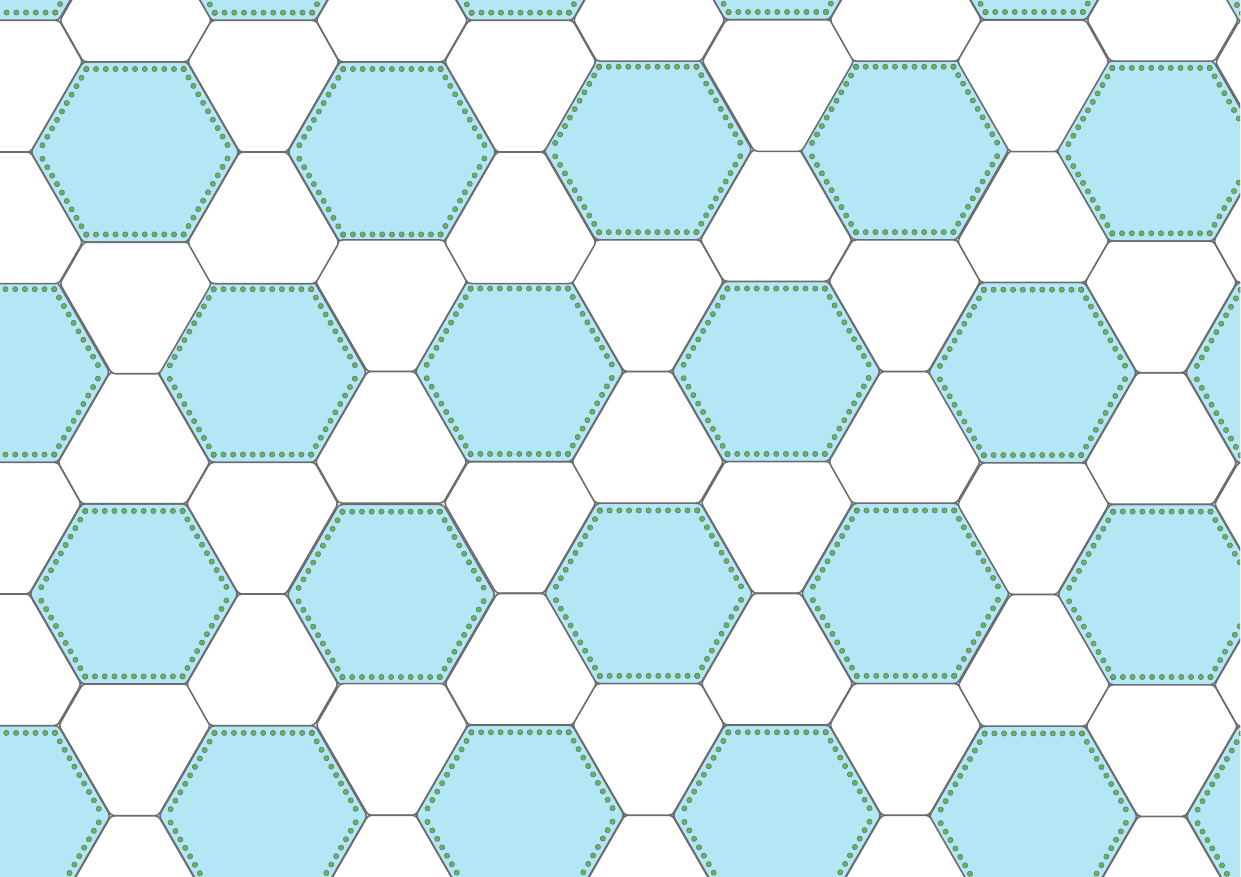}
    \caption{Hexagonal lattice using three-way junctions.}
    \label{fig:hexagonal_lattice}
\end{figure}

\begin{figure}
    \centering
    \includegraphics[width=\linewidth]{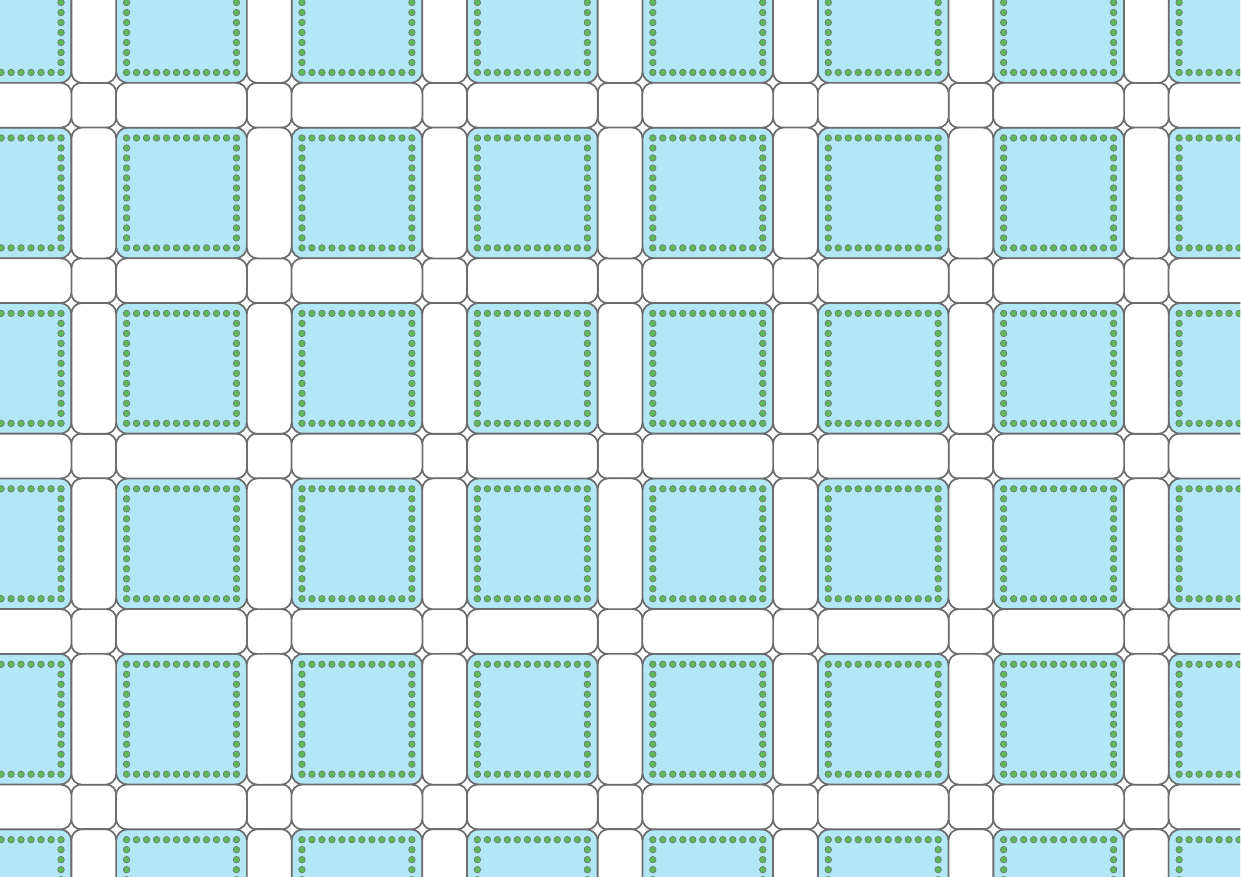}
    \caption{Rectangular lattice using four-way junctions.}
    \label{fig:rectangular_lattice}
\end{figure}

We here give two examples of arrangements of the 2$\times$N devices, other than the one studied in the main text (Fig. \ref{fig:general_layout}). These may allow for different connectivity opportunities at the logical level. Fig. \ref{fig:hexagonal_lattice} represents a hexagonal lattice, using three-way junctions only as in the main text. Fig. \ref{fig:rectangular_lattice} is a rectangular lattice, made of four-way junctions.

\section{\label{app:shuttling_infidelity} Shuttling infidelity for different encodings}

\begin{figure}
    \centering
    \includegraphics[width=\linewidth]{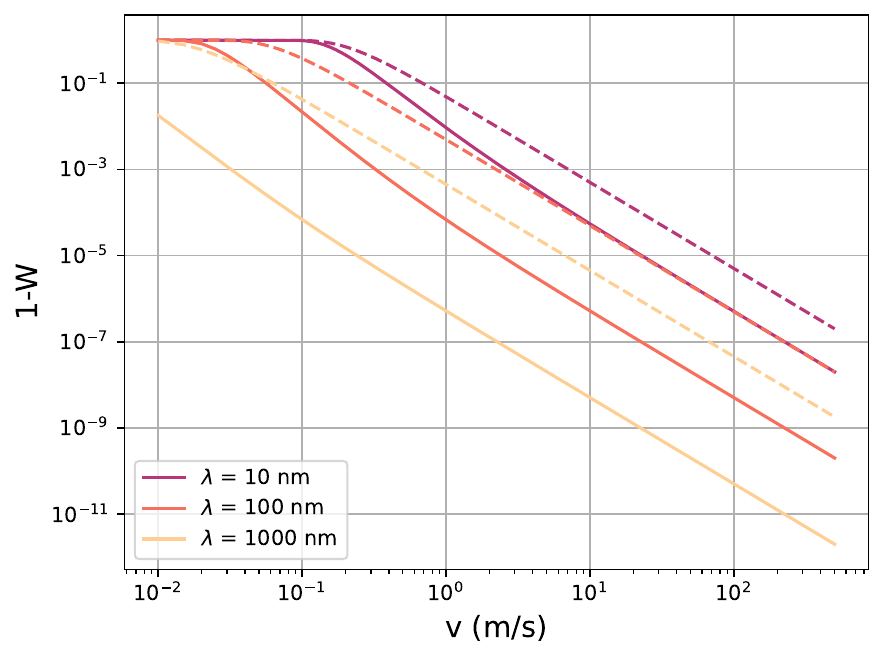}
    \caption{Evolution of the shuttling infidelity as a function of speed for different values of the correlation length $\lambda$ with $d = 100$ nm and $\tau = 20 \; \mu$s. The solid and dashed lines correspond to the ST and LD encodings respectively.}
    \label{fig:shuttling_infidelity_corr_length}
\end{figure}

\begin{figure}
    \centering
    \includegraphics[width=\linewidth]{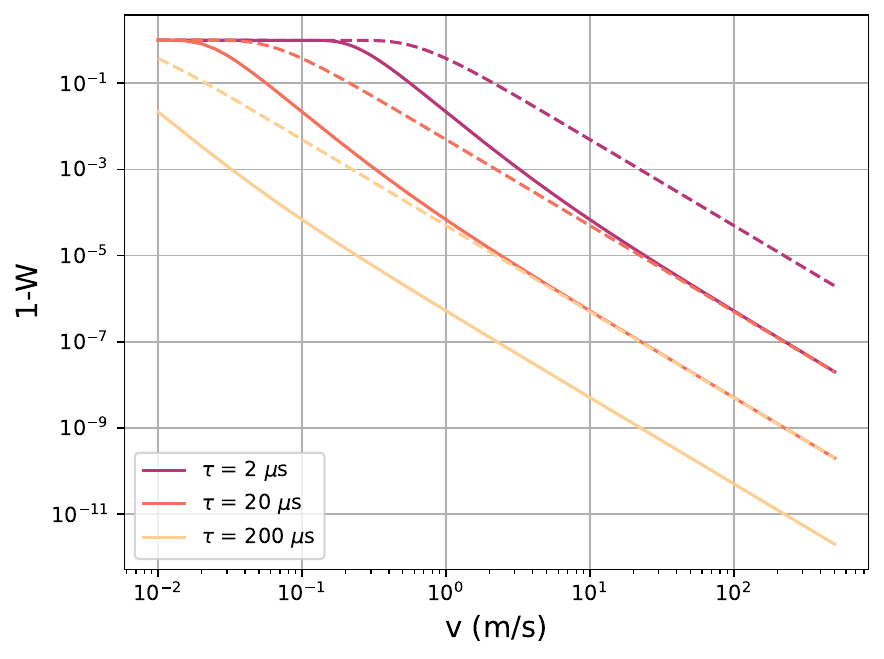}
    \caption{Evolution of the shuttling infidelity as a function of speed for different values of the correlation time $\tau$ with $d = \lambda = 100$ nm. The solid and dashed lines correspond to the ST and LD encodings respectively.}
    \label{fig:shuttling_infidelity_corr_time}
\end{figure}

In the main text, we plotted the evolution of the infidelity with respect to speed for different values of the distance between the ST qubits. In order to confirm our understanding, we study here the evolution of the shuttling infidelity for different values of the correlation length $\lambda$ in \cref{fig:shuttling_infidelity_corr_length} and correlation time $\tau$ and \cref{fig:shuttling_infidelity_corr_time}. 

In both plots, we find that the ST encoding outperforms the LD encoding and that the infidelity decreases with the shuttling speed, which has been explained in the main text. In addition, we find that as $\lambda$ or $\tau$ increases, the infidelity decreases. This is consistent with the fact that by increasing these quantities we reduce the spatial and time fluctuations, ultimately creating a smoother shuttling track for both encodings. 
More precisely, by increasing $\lambda$, the protection power of the ST encoding in \cref{fig:shuttling_infidelity_corr_length} is improved for any speed $v$, as seen by the growing difference in infidelity between solid and dashed lines of the same colour. This comes from the fact that the difference in experienced magnetic field between the two electrons becomes smaller as the correlation length increases. As for \cref{fig:shuttling_infidelity_corr_time}, the protection is only increased up to a certain speed due to the fact that the delay between the electrons decreases with $v$. After this threshold    the fidelity improvement is limited by the correlation length. 

\section{Snake surgery: recovering from dramatic phase errors after transversal gates} \label{app:snake_surgery}

We here show that the protocol described in Section \ref{sec:snake_surgery} can also be applied when the logical snakes are interacted \textit{before} a defective shuttling link was detected.

In order to bring two logical snakes closer together such that they can interact, they are first grown then split in halves, as explained in Section \ref{sec:snake_surgery}. This renders the states:
\begin{align}
    \ket{\psi} &= \alpha\ket{++} + \beta\ket{--} \\
    \ket{\psi'} &= \alpha'\ket{++} + \beta'\ket{--}
\end{align}
where $\ket{\psi}$ and $\ket{\psi'}$ are the snakes states before the growth and split. The heads are then shuttled while the tails are stabilised in place (which is protected against scratches as stabiliser cycles only requires $O(d)$ shuttling increments).

The application of transversal $CNOT$s to the heads brings the system to the following state:
\begin{align*}
    \ket{\Psi} &= (\alpha\ket{+} + \beta\ket{-}) \ket{0} (\alpha\ket{++} + \beta\ket{--}) \\
               &+ (\alpha\ket{+} - \beta\ket{-}) \ket{1} (\alpha\ket{++} - \beta\ket{--})
\end{align*}
where the four qubits are respectively the tail and head of the first logical qubit, then the tail and head of the second logical qubit.

One first case corresponds to a successful interaction, where no defect was detected during the shuttle. In this case the tails are measured in the $Z$ basis, which projects $\ket{\Psi}$ onto:
\begin{align*}
    \ket{\Psi} &\rightarrow \frac{1}{2} [(\alpha + (-1)^{m_1}\beta) \ket{0} (\alpha'\ket{+}+(-1)^{m_2}\beta'\ket{-}) \\
               &+ (\alpha - (-1)^{m_1}\beta) \ket{1} (\alpha'\ket{+}-(-1)^{m_2}\beta'\ket{-})]
\end{align*}
where $m_1$ and $m_2$ are the measurement outcomes of the tail measurements. By applying Pauli corrections in the form $X_1^{m_1}X_2^{m_2}$, one can obtain the desired final state \textit{i.e.}
\begin{equation}
    \ket{\Psi} \rightarrow CNOT_{12}[(\alpha\ket{+} + \beta\ket{-})(\alpha'\ket{+} + \beta'\ket{-})]
\end{equation}

Now let us assume that the interaction was deemed unsuccessful due to the appearance of a defect during the shuttle. This translates into catastrophic $Z$ errors affecting the heads. In this case, they are measured in the $Z$ basis, rendering the state:
\begin{align*}
    \ket{\Psi} &\rightarrow (\alpha\ket{+} + (-1)^{m_1}\beta\ket{-})(\alpha'\ket{+} + (-1)^{m_1+m_2}\beta'\ket{-})
\end{align*}
By this time applying Pauli corrections in the form $X_1^{m_1}X_2^{m_1+m_2}$, one can recover the initial two-logical-qubit state:
\begin{equation}
    \ket{\Psi} \rightarrow (\alpha\ket{+} + \beta\ket{-})(\alpha'\ket{+} + \beta'\ket{-})
\end{equation}

\section{QEC simulations details} \label{qec_sim_details}

In all QEC simulations in this paper, we only focus on $Z$ errors as they are the main error mechanism at play in our system. This is possible since the surface code is a CSS code, hence $X$ and $Z$ errors can be analysed separately. Therefore, we evaluate the $X$ logical error rate of a distance-$d$ surface code over $d$ rounds of stabiliser measurements, via $N_\text{runs}$ runs of Monte Carlo simulations. $N_\text{runs}$ ranges between 10,000 and 1,000,000 depending on the target logical error rates.

In all simulations, circuit-level noise of strength $p$ is adopted, \textit{i.e.} depolarising channels after every initialisation and single- or two-qubit gate, as well as a classical flip of every measurement outcome, all of strength $p$. Assuming high-fidelity shuttling, we showed in \cite{siegel_two_by_n_2024} that shuttling noise had a minimal impact on logical error rates compared to gate noise, thus we neglect it here. Additional noise processes described in further details in the relevant sections may be added as well.

The syndrome these errors create is decoded via Minimum Weight Perfect Matching \cite{Fowler_2012}. The initial errors and the correction are then added up to determine if the $X$ logical operator value was flipped.

\section{Surface code threshold in the presence of a defect} \label{app:defect_threshold}

When the entire length of a snake is shuttled near a defect (\textit{i.e.} for a long shuttle), all data qubits experience \textit{one round} of increased error. More specifically, we assume that this defect-induced noise is pure $Z$, and for convenience of QEC simulations, we suppose that it is a dephasing channel of strength $q$.
If the defect was instead inducing single-qubit coherent rotations of angle $\omega$, one could just twirl them \cite{wallmanNoiseTailoringScalable2016}, which would indeed generate a dephasing channel of strength $q=\sin^2(\omega/2)$.

\begin{figure}
    \centering
    \includegraphics[width=\linewidth]{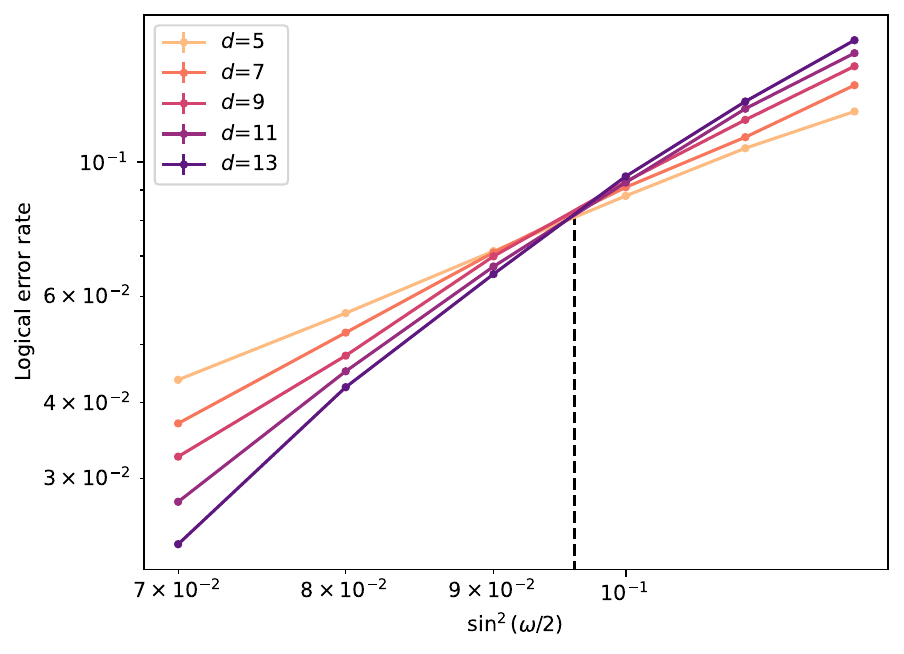}
    \caption{Defect-induced error threshold. Ancilla and data qubits are subject to circuit-level error of strength $p=0.1\%$. Data qubits additionally experience one round of high errors of strength $\sin^2(\omega/2)$, owing to a shuttling event in the vicinity of a defect. The black dashed line indicates the value of the threshold for such error mechanism.}
    \label{fig:defect_threshold}
\end{figure}

With this assumption and normal circuit-level noise (Appendix \ref{qec_sim_details}), we obtain the plot of Fig. \ref{fig:defect_threshold}. From it, one can deduce a threshold value $q_{th}=0.95\%$, corresponding to a threshold angle $\omega_{th}=0.62$.

\section{Monitor qubits\label{app:monitor}}

\subsection{Parameter estimation task}
A dominant source of error in spin qubits is the fluctuation of the internal Rabi frequency of the qubits by a variable amount $\omega$ via the Hamiltonian as $\mathcal{H} = \omega Z/2$, where $Z$ is the Pauli $Z$ operator. In such a scenario the channel $\Phi (\cdot)  = U (\cdot) U^\dagger$ is unitary and acts on each qubit as a $Z$ rotation gate by an unknown angle as $U=e^{-i \omega  Z/2}$.

\textbf{Fundamental precision limits }--- The fundamental limits can be understood in terms of the Cramér-Rao bound which is a lower bound on the variance of the estimator $\hat{\omega}$ as
\begin{equation}
	(\Delta \omega)^2 := \mathrm{Var}[\hat{\omega}] \geq (\nu F_c)^{-1},
\end{equation}
where $\nu$ is the number of times the shuttling event is repeated.
Here the classical Fisher information quantifies the sensitivity of the classical probability distribution $p(q|\omega)$ of measuring the bitstring $q$ given a particular 
rotation angle $\omega$ as
\begin{equation}\label{eq:classical_fisher}
	 F_c =	\sum_{q} p(q|\omega) \left(  \frac{\partial \ln p(q|\omega) }{\partial \omega} \right)^2.
\end{equation}
Given a fixed input state $\rho$, the classical Fisher information is maximised when the best possible measurement basis is used, by mapping the standard basis measurements using the unitary $\mathcal{U}$ and
in this case the quantum Fisher Information is obtained which is therefore a general upper bound as $F_c \leq F_Q$, please refer to \cite{toth2014quantum} for more details.

\textbf{Optimal setting with unentangled monitor qubits }--- We first consider the simple scenario when all $N$ sensing qubits are initialised in a separable pure state.
Given the channel $\Phi$ is assumed to be a unitary $Z$ rotation, an optimal initial state
that maximises the quantum Fisher information is the $\ket{+}^{\otimes N}$ state and the optimal measurement
that maximises the classical Fisher information, and thus the achievable precision, is performed in the eigenbasis of the $X$ Pauli operator. Due to the additivity of the quantum and classical
Fisher information one has 
\begin{equation*}
	F_c[\ket{+}^{\otimes N}] = N F_c[\ket{+}] = N,	
\end{equation*}
where the second equation follows from $F_c[\ket{+}] = 1$ via the probability distribution

\begin{align}
	p(q{=}0|\omega)  &= \cos^2(\frac{\omega}{2}) \\
	p(q{=}1|\omega)  &= \sin^2(\frac{\omega}{2}) 
\end{align}
 by substituting into \cref{eq:classical_fisher}.
Therefore, the Cramér-Rao bound for this scenario states that the uncertainty $\Delta \omega$
in learning the rotation angle $\omega$ is
lower bounded as $\Delta \omega \geq [\nu N]^{-1/2}$.

The CR bound is saturated asymptotically by the maximum likelihood estimator but the actual
precision may deviate in case of relatively few samples as presently we assume 
as few as $N=200$ samples. 
Given the likelihood function
\begin{equation}
	\mathcal{L} (\omega) = \cos(\frac{\omega}{2})^{2 m_0}   \times \sin(\frac{\omega}{2})^{2(N-m_0)}
\end{equation}
where $m_0$ denotes the number of times the $0$ outcome was observed,  
one can show that the the maximum likelihood estimator is the inverse cosine function
of the estimated $Z$ coordinate of the Bloch sphere as
\begin{equation}\label{eq:estimator}
	\hat{\omega} = \arccos\left( \frac{2\hat{m}_0}{N} -1 \right).
\end{equation}
Since $\omega$ is estimated from a finite number of shots, our estimates
can only take up $N$ different discrete values as $\hat{\omega} \in \{\omega_k\}$
via
\begin{equation}\label{eq:prob_dens}
	\omega_k = \arccos\left( \frac{2 k}{N} -1 \right), \quad 0 \leq k \leq N.
\end{equation}
The discrete probability density of $\omega_k$ can readily be computed given $\hat{m}_0$ follows a discrete binomial distribution $B(N,p)$, where we denote:
\begin{equation} \label{eq:monitor_noiseless_p}
    p := p(q{=}0|\omega) = \cos^2(\omega/2)
\end{equation}
This yields the probability $\mathbb{P}(\omega_k) = \mathbb{P}_{\text{binom}}(k, \omega)$ as the discrete probability density function of the binomial distribution.
Furthermore, we compute explicitly the precision (Root Mean Squared Error) $\Delta \omega$ via
\begin{equation}
	\Delta \omega^2	 = \sum_{k=1}^N \mathbb{P}_{\text{binom}}(k,\omega) \left(\omega_k- \omega \right)^2.
\end{equation}

\subsection{Effect of dephasing and readout errors}

So far we analysed the precision of ideal state preparation and measurement and now argue that the present scheme is particularly robust against the dominant noise sources in solid-state devices as dephasing and readout errors.

\noindent\textbf{Effect of dephasing noise on precision }--- A dominant noise contribution is expected to be due to fluctuations of $\omega$ faster than the characteristic shuttling times in which case the average effect can be modelled as a dephasing channel. Given that the unitary $Z$ rotation and dephasing channels commute, this can be viewed as first dephasing the initial state as $\rho_{deph}= \mathcal{D}[\ket{+}\bra{+}]^{\otimes N}$, to which a unitary rotation is then applied via $\Phi$. In this case the information gain is guaranteed to be lower given that the quantum Fisher information, whose inverse lower bounds the achievable precision, is maximised only by pure states: $F_Q[\rho_{deph}] < F_Q[\rho] = N$ where $\rho$ is the optimal pure initial state we detailed above. In particular, dephasing with probability $\lambda$ affects our individual single qubit probe states $\rho_{sing} = \ket{+}\bra{+} $ as
\begin{equation}\label{eq:glob_dep}
	\rho' = (1-\lambda) \rho_{sing} + \lambda Z\rho_{sing}Z = (1-2\lambda) \rho_{sing} + 2 \lambda \frac{\openone}{2}.
\end{equation}
Above in the second equation we identify the effect of dephasing on our particular probe state with an equivalent global depolarising noise channel, for which we can analytically solve the QFI  (using Eq.~(76) in \cite{toth2014quantum}) as
\begin{equation}
	F_Q[\rho'] = \frac{1 - 4 \lambda}{1-\frac{3 \lambda }{2}}
	+ \frac{4 \lambda ^2}{1-\frac{3 \lambda }{2}}
	\geq 1- \frac{5}{2} \lambda.
\end{equation} 
Therefore, the QFI of the $N$-qubit separable probe state can be computed due to the additivity of the QFI over separable states as $F_Q[\rho_{deph}]  = N F_Q[\rho'] \geq N (1-\frac{5}{2} \lambda) $. Given a sub-surface-code threshold $\lambda < 0.7\%$ we see that dephasing merely decreases the QFI by a factor $(1-\frac{5}{2} \lambda)$ which is only fraction of a percentage. This leads to an increased Cramer-Rao bound, \textit{i.e.}, the dashed lines in \cref{fig:monitor_qubits}(right) are shifted up via a multiplication by the factor $(1-\frac{5}{2} \lambda)^{-1/2}$ due to dephasing. This robustness is thanks to the robustness of the QFI of separable states against decoherence which was studied in detail in \cite{koczor2020variational} ---  in contrast, maximally entangled GHZ states are exponentially deteriorated as a function of $N$.

Furthermore, probe states have been designed that are optimal against non-unitary channels $\Phi$~\cite{koczor2020variational} and, in particular, the information gain can be maximised
and improved by a constant factor when using entangled squeezed Dicke states as detailed in ref.~\cite{koczor2020variational}.
This, however, requires the monitor qubits be prepared in a non-trivial entangled state
and due to hardware noise in the state preparation gates, may not be beneficial and therefore our separable states offer a slightly suboptimal but simple alternative.

\noindent \textbf{Effect of measurement errors on precision }--- For our optimal protocol we assumed perfect projective measurements of the monitor qubits in the relevant bases. However, real measurements are characterised by general POVMs rather than by projections. We illustrate this via a simple readout-error model where entries in the bitstring $b$, as measurement outcomes of the $N$ probe qubits, undergo random and independent bit flips. More specifically, we assume that the $i^{th}$ bit is flipped as $0 \rightarrow 1$ with probability $\alpha_i^{+}$ and as
$1 \rightarrow 0$ with probability $\alpha_i^{-}$.
Therefore, the projective POVM $\{\ketbra{0}{0},\ketbra{1}{1}\}$ on the $i^{th}$ qubit
is replaced by the POVM 
\begin{equation*}
\left\{ (1{-}\alpha_i^{+}) \ketbra{0}{0}+ \alpha_i^{-} \ketbra{1}{1},
	\quad
	\alpha_i^{+} \ketbra{0}{0}+(1{-}\alpha_i^{-}) \ketbra{1}{1}
	\right\}
\end{equation*}

For the specific readout-error model when $\alpha_i^{+}=\alpha_i^{-}=:\alpha_i$, \textit{i.e.}, the probability is the same for flipping a 1 as flipping a 0, we can easily see that the POVM is identical to performing ideal projective measurements on a dephased probe state 
$\rho_{deph}= \mathcal{D}[\ket{+}\bra{+}]^{\otimes N}$ where effective dephasing happens with probability $\alpha_i$.
Therefore, exactly the same conclusions apply to readout noise as to dephasing noise: as long as the probabilities $\alpha_i$ 
are consistent with the surface code threshold, the Cramer-Rao bounds are only increased by a negligible amount due to readout or dephasing errors.

\noindent\textbf{Noisy probability distributions} ---
From Eq. \ref{eq:glob_dep}, one can easily update Eq. \ref{eq:monitor_noiseless_p} to include the effect of dephasing and measurement noise by setting:
\begin{align}
    p' &= (1-2\lambda)p + \lambda \label{eq:monitor_noisy_p} \\
       &= (1-2\lambda)\cos^2(\omega/2) + \lambda
\end{align}
All probability distributions (binomial laws) can thereon be computed with this new parameter. In all the following plots, we assume an equal level of measurement and dephasing noise at $0.1\%$, leading to a total noise strength $\lambda=0.2\%$.

\subsection{Task $(i)$: defect detection}

After explaining the generalities of monitor qubits, we can now delve into the implementation of the two tasks monitor qubits are utilised for. The first one is detecting with maximum probability if a defect occurred, regardless of the value of its angle.
More precisely, let us denote by $\omega_{\max}$ the maximum rotation angle the surface code can tolerate to guarantee sufficient error suppression at the chosen code distance, and $\hat{\omega}_{\max}$ the value of the estimator above which a catastrophic event is declared. As explained in the main text, we want to minimise the rate of false negatives while keeping the rate of false positives around 5-10$\%$. Both these parameters are controlled by the choice of $\hat{\omega}_{\max}$. Besides, we set $\omega_{\max}=\omega_{th}/2=0.3$ as per Appendix \ref{app:defect_threshold}.

Additionally, we assume that defects (of any angle) are rare and occur at a rate $\rho$, and that the distribution of defect angles $p(\omega)$ is uniform in $[-\pi,\pi]$, \textit{i.e.}:
\begin{equation}
    \forall \omega\in[-\pi,\pi]~~p(\omega)=1/2\pi
\end{equation}

\begin{figure}
    \centering
    \includegraphics[width=0.9\linewidth]{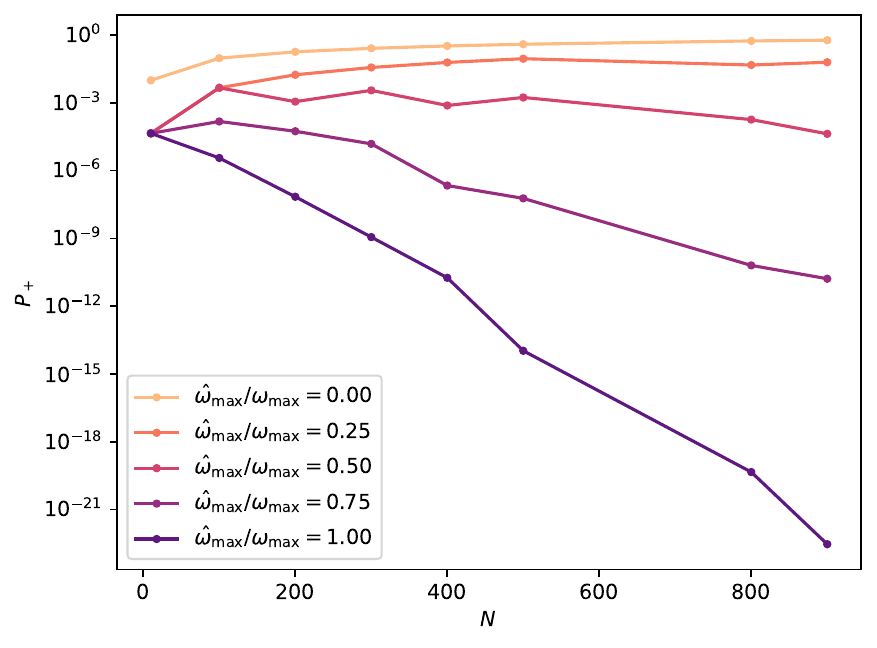}
    \caption{Probability of a false positive, \textit{i.e.} declaring that a defect occurred on a non-defective link. See main text for parameters explanation.}
    \label{fig:monitor_P+}
\end{figure}

\noindent\textbf{Rate of false positives }--- The rate $P_{+}$ of false positives is given by:
\begin{align*}
    P_+ &= \rho \times \mathbb{P}(|\hat{\omega}|>\hat{\omega}_{\max} ~|~ |\omega|<\omega_{\max}) \\
        &+ (1-\rho) \times \mathbb{P}(|\hat{\omega}|>\hat{\omega}_{\max} ~|~ \omega=0)
\end{align*}
When $\rho$ is small, one can typically neglect the first term of the equation and assume $1-\rho\sim1$, leading to:
\begin{align}
    P_+ &= \mathbb{P}(|\hat{\omega}|>\hat{\omega}_{\max} ~|~ \omega=0) \\
        &= \sum_{k=0}^{k_{\max}-1} \mathbb{P}_{\text{binom}}(k,0)
\end{align}
where, from Eq. \ref{eq:prob_dens}, $k_{\max} = \ceil{N \cos^2(\hat{\omega}_{\max}/2)}$. The binomial distributions are then computed using Eq. \ref{eq:monitor_noisy_p}. The result is plotted it in Fig. \ref{fig:monitor_P+} against the number $N$ of monitor qubits.
As expected, $\hat{\omega}_{\max}$ controls the rate $P_+$ of false positives. When $\hat{\omega}_{\max}$ is close to 0, there is a high chance that the estimator $\hat{\omega}$ falls outside the interval $[-\hat{\omega}_{\max},\hat{\omega}_{\max}]$, even when $\omega=0$. This translates into a high rejection rate, approaching 1. In our case, we are interested in a rejection rate around $5-10\%$, which corresponds to the salmon plot, defined by $\hat{\omega}_{\max} / \omega_{\max}=0.25$. We will thus set $\hat{\omega}_{\max} = 0.075$ in the following.\\

\noindent\textbf{Distribution of post-selected angles }--- The next step is determining the distribution $p_{\text{mon}}(\omega)$ of post-selected angles, according to the choice of $\hat{\omega}_{\max}$. This quantity is given by:
\begin{align*}
    p_{\text{mon}}(\omega) &= p(\omega ~|~ |\hat{\omega}|\leq\hat{\omega}_{\max}) \\
        &= \frac{p(|\hat{\omega}|\leq\hat{\omega}_{\max} ~|~ \omega) p(\omega)}{p(|\hat{\omega}|\leq\hat{\omega}_{\max})}  \\
        &= \frac{ p(|\hat{\omega}|\leq\hat{\omega}_{\max} ~|~ \omega) }{\int_{|\omega|=0}^\pi p(|\hat{\omega}|\leq\hat{\omega}_{\max} ~|~ \omega) \mathrm{d}\omega}
\end{align*}
where in the last equation we used that the original distribution $p(\omega)$ of defect rotation angles is uniform. Finally, one gets:
\begin{equation}
    p_{\text{mon}}(\omega) = \frac{1}{2} \frac{\sum_{k=k_{\max}}^{N} \mathbb{P}_{\text{binom}}(k,\omega)}{\sum_{k=k_{\max}}^{N} \int_{\omega=0}^\pi \mathbb{P}_{\text{binom}}(k,\omega) \mathrm{d}\omega}
\end{equation}
where the integration between 0 and $\pi$ and the factor 2 arise from symmetry arguments. The result is plotted in Fig. \ref{fig:monitor_qubits} (left). As expected, increasing the number of monitor qubits shrinks the probability distribution around 0. The black dashed lines indicate the value of $\pm \omega_{\max}$: when $\omega$ is outside this window, the chosen code distance does not suffice to adequately lower the logical error rate.\\

\noindent\textbf{Rate of false negatives }--- The probability of $\omega$ falling outside this window is the rate of false negatives, which can lead a logical failure. It is expressed as:
\begin{align}
    P_- &= \mathbb{P}(|\omega|>\omega_{\max} ~|~ |\hat{\omega}| \leq \hat{\omega}_{\max}) \\
        &= 2\int_{\omega=0}^{\pi} p_{\text{mon}}(\omega) \mathrm{d}\omega
\end{align}
Using $N=900$ monitor qubits, meaning that we interleave one monitor qubit per data qubit in a $30\times 30$ surface code, one obtains $P_-=1.4\times 10^{-8}$. This is still computed for a total dephasing and measurement noise strength $\lambda=0.2\%$ and $\hat{\omega}_{\max}=\omega_{\max}/4=0.075$. Therefore, by rejection $5\%$ of the shuttling events, the logical error rate due to defects is brought down to $1.4\times 10^{-8}\times\rho$.

\subsection{Task $(ii)$: estimating large rotation angles}
We now focus on task $(ii)$, \textit{i.e.} the precise evaluation of the angle $\omega$. For this purpose, we devise an estimator that can accurately approximate $\omega$ in the full range $(-\pi, \pi)$, such that we are optimally sensitive to the angles $0, \pm \pi/2, \pm \pi$. We therefore initialise the state $\rho$ of the monitor qubits as $\ket{-\pi/4}^{\otimes N}$, where $\ket{-\pi/4} := Rz(-\pi/4) \ket{+}$.
Measuring half of them in the eigenbasis of the Pauli $X$ operator and the other half in the $Y$ operator results in the probability distribution
\begin{align*}
	p_X(0|\omega)  &= \cos(\frac{\omega-\pi/4}{2})^2, \quad p_X(1|\omega)  = \sin(\frac{\omega-\pi/4}{2})^2 \\
	p_Y(0|\omega)  &= \cos(\frac{\omega-3\pi/4}{2})^2, \quad p_Y(1|\omega)  = \sin(\frac{\omega-3\pi/4}{2})^2.
\end{align*}
We then use the following estimator
\begin{equation}
	\hat{\omega} = \arctan2\left( \hat{x}, \hat{y} \right) +\pi/4,
\end{equation}
which outperforms the maximum likelihood estimator as it makes use of the fact that the true $x$ and $y$ coordinates satisfy $x^2+y^2=1$. Here $\hat{x}$ and $\hat{y}$ are estimators of the $x$ and $y$ coordinates of the Bloch sphere, \textit{e.g.} $\hat{x} := \frac{2\hat{x}_0}{N/2} -1$ is the number of times $\hat{x}_0$ the $X$ probes resulted in the $0$ outcome.

As before the estimator $\hat{\omega}$ can only take up discrete values as
\begin{equation}
	\omega_{kl} = \arctan2\left( \frac{2 k}{N/2} -1, \frac{2 l}{N/2} -1 \right) +\pi/4, 
\end{equation}
such that $0 \leq k,l \leq N/2$. We can then explicitly compute the discrete probability density function given both $\hat{x}$ and $\hat{y}$
follow binomial distributions as $B(N/2,p_x)$ and $B(N/2,p_y)$ where we denote as $p_x$ and $p_y$ the probabilities $p_X(0|\omega)$ and $p_Y(0|\omega)$ defined above. Denoting the discrete probability density functions of the two binomial distributions as $\mathbb{P}_{\text{binom},X}(k)$
and $\mathbb{P}_{\text{binom},Y}(l)$, the noiseless probability of the individual $\omega_{kl}$ estimates is given as 
\begin{equation} \label{eq:monitor_2pi_prob}
	\mathbb{P}(\omega_{kl}) = \mathbb{P}_{\text{binom},X}(k) \times \mathbb{P}_{\text{binom},Y}(l).
\end{equation}
In the presence of noise, one can easily verify that the arguments of the previous sections still apply. In particular, dephasing and measurement noise are still equivalent, whether the initial state is $\ket{+}^{\otimes N}$ or $\ket{-\pi/4}^{\otimes N}$. Besides, the noisy probability distribution can be derived from Eqs. \ref{eq:monitor_noisy_p} and \ref{eq:monitor_2pi_prob}, by simply updating the values of $p_x$ and $p_y$ as:
\begin{align}
    p_x' &= (1-2\lambda)p_x + q \\
    p_y' &= (1-2\lambda)p_y + q
\end{align}

\subsection{Test non-unitarity by estimating purity}

Unitary rotations can be estimated and then corrected by inverse rotations, however, significant non-unitary noise above threshold can potentially be detrimental. For this reason, we estimate resource requirements for estimating whether the purity of the shuttled qubits
remains nearly pure.

One approach for estimating the fidelity is by applying a generalised SWAP test to the qubits in which case we split the $N$ sensing qubits into batches of $n$ qubits and for each batch of qubits we apply a SWAP test to estimate
$f=\tr[\rho^n] = 2p_0-1$   from the probability $p_0$ of measuring $0$ on the ancilla qubit.
The qubits in each batch are initialised identically but can be initialised in different states, potentially random states, across the different batches.

We have overall $M=N/n$ bits of information to estimate $f$ via the
estimator  $\hat{p}_0  = \hat{m}_0/M$ where $\hat{m}_0$ is the number of times $0$
was observed. 
In particular, we obtain the estimator $\hat{f} =2 \hat{m}_0/M -1 $
and the variance
\begin{align*}
	\mathrm{Var}[\hat{f}] &= \mathrm{Var}[2 \hat{m}_0/M - 1 ] = 4 \mathrm{Var}[\hat{m}_0]/M^2  \\
	& = 4 p(1-p)/M = \frac{1-f^2}{M}
\end{align*}
where we used that $\hat{m}_0$ is binomially distributed and
in the last equation we substituted $p = 1/2 - f/2$.

\section{Complementary gap} \label{app:complementary_gap}

One of our defect-detection schemes is based on the computation of the complementary gap corresponding to the measured syndrome. This quantity is given by the length difference between the shortest error strings explaining the syndrome but yielding opposite logical outcomes. Low complementary gaps mean that the decoder's decision cannot be fully trusted.

Similarly to the monitor-qubit case, our aim is to establish a rejection criterion minimising the rate of false negatives (not detecting a defect) while keeping the rate of false positives around $5\%$ (so as to keep the impact on the compute time moderate). Our selection rule is of the form $g \geq g_{\min}$, where $g$ is the measured complementary gap and $g_{\min}$ is a distance-dependent parameter. This means that whenever a gap lower than $g_{\min}$ is measured, the shuttling event is aborted and restarted via snake surgery (Section \ref{sec:snake_surgery}). \\

\noindent\textbf{Rate of false positives }--- One can readily adapt the equations of Appendix \ref{app:monitor} and write:
\begin{align*}
    P_+ &= \rho \times \mathbb{P}(g<g_{\min} ~|~ |\omega|<\omega_{\max}) \\
    &+ (1-\rho) \times \mathbb{P}(g<g_{\min} ~|~ \omega=0)
\end{align*}
from which it follows
\begin{equation}
    P_+ \approx \mathbb{P}(g<g_{\min} ~|~ \omega=0)
\end{equation}
owing to the assumed low rate of defects (much lower than $5\%$).

\begin{figure}
    \centering
    \includegraphics[width=\linewidth]{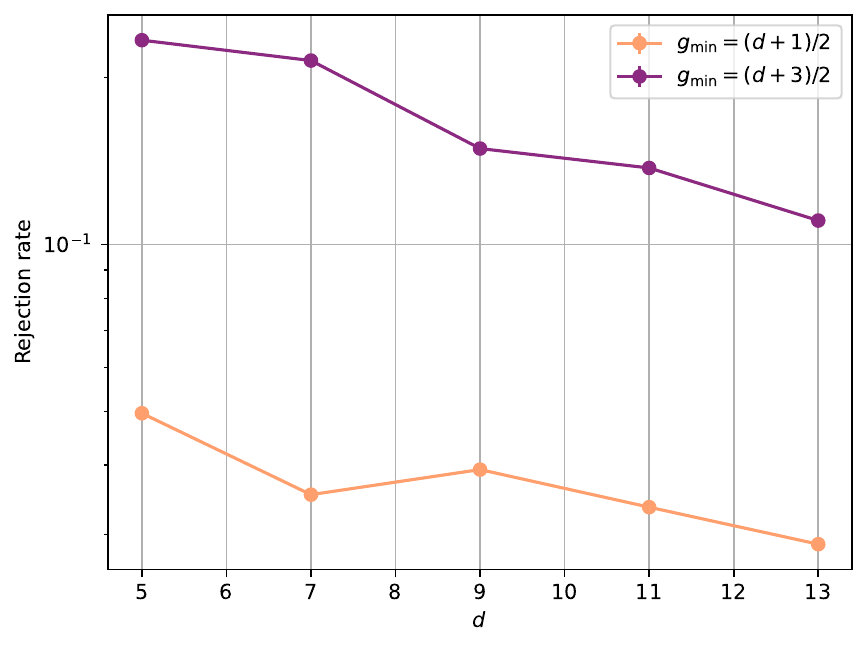}
    \caption{Rejection rate corresponding to the selection rule $g \geq g_{\min}$, where $g$ is the measured complementary gap and $g_{\min}$ is a distance-dependent parameter. The simulations are run under circuit-level noise at a rate $p=0.1\%$.}
    \label{fig:complementary_gap_055_rejection_rate}
\end{figure}

In Fig. \ref{fig:complementary_gap_055_rejection_rate}, we thus plot the rejection rates obtained for two values of $g_{\min}$ by Monte-Carlo simulations, assuming circuit-level noise at $p=0.1\%$ and no defect (see simulation details in Appendix \ref{qec_sim_details}). We observe that setting $g_{\min}=(d+3)/2$ leads to a rejection rate that is higher than our desired tolerance (around 5-10$\%$). In contrast, reducing it to $g_{\min}=(d+1)/2$ yields a rejection rate of 3-5$\%$, at least for the plotted code distances. This is thus the criterion we will adopt in the rest of the paper. For large code distances \textit{e.g.} $d=30$, $g_{\min}$ may be increased to $(d+3)/2$. Finally, note that this choice is highly dependent on the physical error rate $p$.\\

\noindent\textbf{Distribution of post-selected angles }--- The distribution of post-selected angles can again be easily obtained by adapting the equations derived for monitor qubits:
\begin{align*}
    p_{\text{gap}}(\omega) &= p(\omega ~|~ g \geq g_{\min}) \\
        &= \frac{ p(g \geq g_{\min} ~|~ \omega) }{\int_{|\omega|=0}^\pi p(g \geq g_{\min} ~|~ \omega) \mathrm{d}\omega}
\end{align*}
$p(g \geq g_{\min} ~|~ \omega)$ is the acceptance probability given a certain defect angle $\omega$. This quantity is obtained by Monte-Carlo simulations and the results are plotted in Fig. \ref{fig:complementary_gap} (left).\\

\noindent\textbf{Resulting logical error rate }--- The main difference between the complementary-gap and the monitor-qubit schemes is that the former additionally shrinks the logical error rate of the post-selected instances. This is because rejecting small complementary gaps highly improves the confidence of the decoder thus enhances its performance. We show such error reduction in Fig. \ref{fig:complementary_gap} (right). The data is obtained with the usual procedure described in Appendix \ref{qec_sim_details}, the only subtlety being than for every instance of the Monte-Carlo simulation, the code is reset and new errors are injected until a complementary gap higher than $g_{\min}$ is observed.

\section{Defect resistance} \label{app:defect_resistance}

Here, we rigorously estimate the logical error rate \textit{per $d$ stabiliser cycles} $\tilde{P}_L$ associated with the displacement of a snake from an initial location to a final location in the device. In particular, we aim to express it as a function of $P_L$, the static logical error rate per $d$ stabiliser cycles. This $P_L$ is the quantity that is normally used as a target error rate to run a predefined algorithm, generally ranging between $10^{-10}$ and $10^{-15}$. However, our architecture extensively makes use of long shuttles, with possible shuttling failures and rerouting due to defects, thereby potentially modifying the raw value of the error rate. In particular, we use a shuttling link deactivation rate of $10\%$, which is set by hyperparameters in the defect detection protocols based on monitor qubits and complementary gap.

For this purpose, we assume that, to travel from its initial to its final location, a snake has to attempt a collection of paths, indexed by $i$. 
We then define the following events:
\begin{itemize}
    \item log: a logical error occurred
    \item $R_i$: the $i$-th path is rejected due to the detection of a defect
    \item $\bar{R}_i$: the $i$-th path is accepted
\end{itemize}

The probability of logical error occurring on any of these paths is:
\begin{align*}
    \tilde{P}_L &= \sum_{i \geq 1} \mathbb{P}\Big(\text{log} \,| \bigcap_{j=1}^{i-1}R_j \cap \bar{R}_i\Big) \times \mathbb{P}\Big(\bigcap_{j=1}^{i-1}R_j \cap \bar{R}_i\Big) \\
                &= \sum_{i \geq 1} \left((i-1) \mathbb{P}(\text{log} | R_1) + \mathbb{P}(\text{log} | \bar{R}_1)\right) \times 0.1^{i-1}\times 0.9
\end{align*}
In the first equation, we are decomposing the probability of a logical failure depending on the number of attempted paths (trying $i$ paths amounts to rejecting $i-1$ paths and accepting the $i$-th one). Then, when $i$ paths are tried, a logical error can occur at any of the $i-1$ rejections, which is $\mathbb{P}(\text{log} | R_1)$; or it can occur on the accepted path, which is $\mathbb{P}(\text{log} | \bar{R}_1)$. Note that we are here assuming that, for low enough error rates, the probability of two logical errors is negligible. The final factors corresponds to the probability of having $i-1$ rejected paths and 1 accepted path.

Using geometric series formulae, one obtains:
\begin{equation}
    \tilde{P}_L = \frac{1}{9} \mathbb{P}(\text{log} | R_1) + \mathbb{P}(\text{log} | \bar{R}_1)
\end{equation}

Now, $\mathbb{P}(\text{log} | R_1)$ corresponds to the probability of a logical error given that a defect was detected on the chosen shuttling path. This is handled by the snake surgery protocol of Section \ref{sec:snake_surgery}, where it was shown that the subsequent error rate was that of a normal lattice surgery \textit{i.e.} of $d$ rounds of stabiliser measurements on a static snake. In other words:
\begin{equation} \label{eq:log_R1_PL}
    \mathbb{P}(\text{log} | R_1) = P_L
\end{equation}
Notably, due to the factor 1/9, attempting multiple paths does not increase the logical error rate. This stems from the relatively low rate of false positives we set.

$\mathbb{P}(\text{log} | \bar{R}_1)$ however is slightly more complicated to evaluate. It quantifies the probability of a logical failure given that the path was accepted. The main source of error in this case arises from the no-detection of a defect on the accepted path. This probability is thus conditional on the defect rate $\rho$ (per $d$ stabiliser cycles) and the performance of the detection strategies. Using the notations from the previous sections, one can write:
\begin{equation*}
    \mathbb{P}(\text{log} \cap \bar{R}_1) = \int_{|\omega|=0}^{\pi} p(\text{log} \cap \omega \cap \bar{M} \cap \bar{G}) \mathrm{d}\omega
\end{equation*}
where $M$ and $G$ are events respectively indicating that the monitor qubits or complementary gap rejected a given path. $\bar{R}_1= \bar{M} \cap \bar{G}$ as a path is only accepted if it passed both the monitoring and complementary gap filters. Additionally, we decomposed the probability depending on the defect angle $\omega$. Then:
\begin{align*}
    \mathbb{P}(\text{log} \cap \bar{R}_1) = \int_{|\omega|=0}^{\pi} \mathbb{P}(\text{log} &\,|\, \omega \cap \bar{M} \cap \bar{G}) \times p(\omega \,|\, \bar{M} \cap \bar{G}) \\
                &\times \mathbb{P}(\bar{M} \cap \bar{G}) \mathrm{d}\omega
\end{align*}
The first factor is the probability of a logical failure on a shuttling link that passed all detection tests and that induces a rotation by an angle $\omega$. The second factor is the distribution of post-selected angles $\omega$ after both filtering protocols. Finally, the third term is the probability of accepting a path \textit{i.e.} 0.9.

The second term contains two contributions, conditional on the presence of a defect or not:
\begin{equation*}
    p(\omega | \bar{M} \cap \bar{G}) = (1-\rho) \delta(\omega) + \rho p_{\text{both}}(\omega)
\end{equation*}
Here, $\delta$ is the Dirac function, indicating that the absence of a defect corresponds to a phase shift $\omega=0$. $p_{\text{both}}$ is the probability density of post-selected angles by both detection techniques in the presence of a defect.

Eventually, one gets:
\begin{align}
    \mathbb{P}(\text{log} &\cap \bar{R}_1) = 0.9 \times (1-\rho) \times \mathbb{P}(\text{log} \,|\, \omega=0 \cap \bar{M} \cap \bar{G}) \label{eq:log_r1_term1} \\
        &+ 0.9 \times \rho \times \int_{|\omega|=0}^{\pi} \mathbb{P}(\text{log} \,|\, \omega \cap \bar{M} \cap \bar{G}) p_{\text{both}}(\omega) \mathrm{d}\omega \label{eq:log_r1_term2}
\end{align}

The probability in the first term (Eq. \ref{eq:log_r1_term1}) is the probability of a logical error given that a snake travelled through a non-defective shuttling link that was correctly declared as non-defective. This probability is significantly smaller than $P_L$, as the rejection stemming from the complementary gap scheme highly improves the resulting error rates (Fig. \ref{fig:complementary_gap} right):
\begin{equation} \label{eq:log_no_def_MG}
    \mathbb{P}(\text{log} \,|\, \omega=0 \cap \bar{M} \cap \bar{G}) \ll P_L
\end{equation}
The probability appearing in the second term (Eq. \ref{eq:log_r1_term2}) however corresponds to the error-prone case of an undetected defect, and is pondered by the distribution $p_{\text{both}}$ of post-selected defect angles $\omega$. Defining 
\begin{equation*}
    P(\omega)=\mathbb{P}(\text{log} \,|\, \omega \cap \bar{M} \cap \bar{G}),
\end{equation*}
one can finally write the final form of $\tilde{P}_L$:
\begin{equation}
    \tilde{P}_L = \frac{1}{9}P_L + \rho \times \int_{|\omega|=0}^{\pi} P(\omega)p_{\text{both}}(\omega) \mathrm{d}\omega
\end{equation}
The first term is the probability that a logical error occurs while rejecting shuttling links via snake surgery, and the second term describes the impact of an undetected defect.

\end{document}